\documentclass[12pt,a4paper]{article}

\usepackage[parfill]{parskip}

\usepackage{graphicx}
\usepackage{xcolor}
\usepackage{multicol}
\usepackage{newtxtext,newtxmath}
\usepackage{hyperref}
\usepackage{wrapfig}
\usepackage{units}
\usepackage{paralist}
\usepackage{sidecap}
\usepackage{floatrow}

\newcommand{\hd}[1]{{\bf\sc\underline{#1}}}
\newcommand{\lya}{Ly$\alpha$}
\newcommand{\about}{$\sim\!\!$~} 
\newcommand{\kms}{\,km\,s$^{-1}$}
\newcommand{\ha}{H$\alpha$}
\newcommand{\hb}{H$\beta$}
\newcommand{\sii}{$[$SII$]$}

\begin{document}

\begin{figure}
\vspace{0.0cm}
\scalebox{0.72}[0.72]{\includegraphics{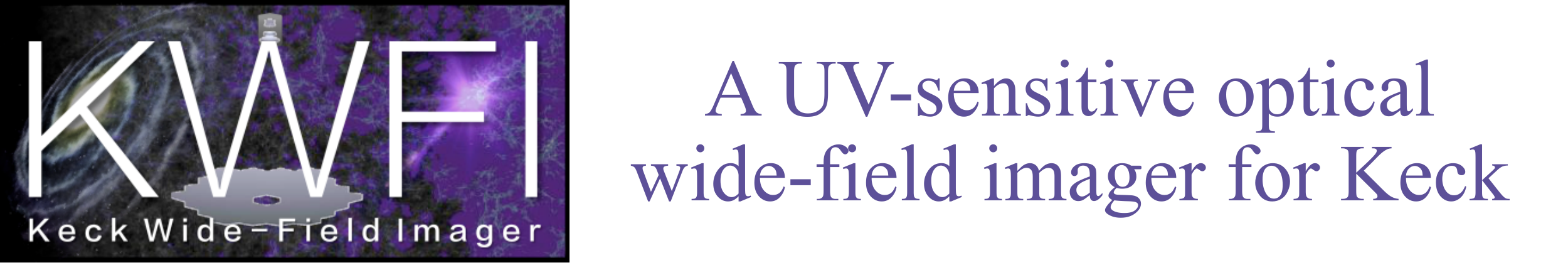}}
\end{figure}

{\bf \Large Science Cases for the Keck Wide-Field Imager}

\small{{\bf KWFI Team:} J.\,Cooke (Swinburne; PI), R.\,Dekany (Caltech/COO; Project Manager), R.\,Bertz (Caltech/COO; Deputy Project Manager), P.\,Gillingham (AAO+Macquarie; Lead Optical Engineer), M.\,Radovan (UCO; Lead Mechanical Engineer), R.\,Smith (Caltech/COO; Lead Detector Engineer), S.\,Krishnan (Swinburne/FoF; Lead FEX Engineer), R.\,Seikel (Swinburne/ADACS: Lead Software Engineer), G.\,Poole (Swinburne/ADACS), T.\,Travouillon (ANU/AITC), J.\,Fucik (Caltech/COO), H.\,Kaptan (Swinburne/FoF), C.\,Webster (Swinburne/FoF), A.\,Delacroix (Caltech/COO), J.\,Hurley (Swinburne/ADACS), S.\,Salaheen (Swinburne/ADACS), J.\,Cantos (Swinburne/ADACS), C.\,Steidel (Caltech), J.\,Brodie (Swinburne), M.\,Brown (Caltech),  N.\,Suzuki (Kavli IPMU, UC Berkeley), B.\,T.\,Bolin (Caltech/IPAC), J.\,Burchett (UCSC/NMSU),  M.\,Cowley (QUT/UQ),  D.\,Fisher (Swinburne), R.\,Foley (UCSC), G.\,Foran (Swinburne), K.\,Glazebrook (Swinburne), G.\,Kacprzak (Swinburne), R.\,Margutti (UC Berkeley), B.\,Mobasher (UCR), A.\,M\"{o}ller (Swinburne), J.\,Mould (Swinburne),  A.\,Rest (STScI), J.\,Rhodes (JPL/Caltech), R.\,M.\,Rich (UCLA), L.\,Wang (TAMU), I.\,Wold (NASA/GSFC), J.\,Zhang (Swinburne)}

{\bf Science authors:} Charlotte Angus (NBI/DARK), Katie Auchettl (Univ.Melbourne/UCSC), John Bally (Univ.Colorado), Bryce Bolin (Caltech), Sarah Brough (UNSW, Sydney), Joseph N.\,Burchett (NMSU), Jeff Cooke (Swinburne), Ryan Foley (UCSC), Garry Foran (Swinburne), Duncan Forbes (Swinburne), Jonah Gannon (Swinburne), Ryosuke Hirai (Monash), Glenn G. Kacprzak (Swinburne), Raffaella Margutti (UC Berkeley), Cristina Mart{\'i}nez-Lombilla (UNSW, Sydney), Uro\v{s} Me\v{s}tri\'{c} (INAF), Anais M\"{o}ller (Swinburne), Armin Rest (Space Telescope Institute), Jason Rhodes (JPL/Caltech), R.\,Michael Rich (UCLA), Fabian Sch\"ussler (IRFU, CEA Paris-Saclay), Richard Wainscoat (Univ.Hawaii), Josh Walawender (W.M.Keck Observatory), Isak Wold (NASA/GSFC), Jielai Zhang (Swinburne)
\vspace{0.1cm}

\normalsize
{\bf Abstract}

The Keck Wide-Field Imager (KWFI) is a proposed 1-degree diameter field of view UV-sensitive optical camera for Keck prime focus. KWFI will be the most powerful optical wide-field camera in the world and the only such 8m-class camera sensitive down to $\sim$3000\AA\ for the foreseeable future. Twenty science cases are described for KWFI compiled largely during 2019--2021, preceded by a brief discussion of the instrument, components, and capabilities for context.

\vspace{0.5cm}
{\bf The Keck Wide-Field Imager - Summary}

\noindent The Keck Wide-Field Imager (KWFI) is a proposed 1-degree diameter field of view UV-sensitive optical camera for Keck prime focus [1,2]. KWFI will be the most powerful wide-field camera in the world and the only such 8m-class camera sensitive down to $\sim$3000\AA\ for the foreseeable future.  

\vspace{0.1cm}
KWFI exploits the superior atmospheric UV transmission on Maunakea and the large aperture and fast focal ratio of Keck to enable extremely deep, 3000--10000\AA\ wide-field science. With a powerful combination of extreme sensitivity, including extreme blue sensitivity, and wide field of view, KWFI enables leading science that cannot be done by any other telescope, including JWST and future 30m-class telescopes. KWFI fills a crucial 3000--5000\AA\ imaging gap for upcoming wide-field optical/infrared NASA Roman and ESA Euclid space telescopes. In addition, KWFI will detect extremely faint, rare, and lensed targets for 30m-class telescopes and JWST and will acquire a high density of targets for upcoming or planned 10m-class wide-field spectrographs, such as Subaru Prime Focus Spectrograph, Keck FOBOS, and the Maunakea Spectroscopic Explorer.  The proposed addition of a deployable secondary mirror will enable KWFI to perform deep Target of Opportunity observations within minutes, localizing sources detected at all wavelengths and messengers. Furthermore, KWFI's fast data processing, will enable deep imaging of target fields and deep spectroscopy with Keck optical and infrared instruments during the same night.

\vspace{0.1cm}
KWFI will strengthen Keck's leadership of blue-sensitive science and help the Keck community resolve several long-standing problems, including areas of high-impact science discovery. Examples include game-changing blue depths over wide fields to detect escaping Lyman continuum radiation from high redshift galaxies to understand cosmic reionization; localize the bulk of electromagnetic counterparts to distant gravitational wave sources detected by next-generation facilities; discover and characterize the far-UV of high redshift supernovae to enable the study of the first generation of stars; map the faint diffuse circumgalactic medium, detect early and exotic transient events, explore the diffuse and faint emission Universe, perform metallicity selection and mapping of the Milky Way, local galaxies and globular clusters, and to help search Planet 9, NEOs and asteroids, and comet emission to better understand our solar system.  

\vspace{0.1cm}
Wide-field imagers are the highest demand instruments on their respective telescopes (cf. CTIO Blanco DECam [3], CFHT MegaCam [4], Subaru Hyper-SuprimeCam (HSC) [5]) and are a testament to their essential role in nearly every area of astronomy. The Keck telescopes were designed to accommodate prime focus instruments and a wide-field imager, with rapid instrument exchange enabled by modules similar to Cassegrain focus instruments. 

\begin{figure}[!h]
\begin{center}
\scalebox{0.36}[0.36]{\includegraphics{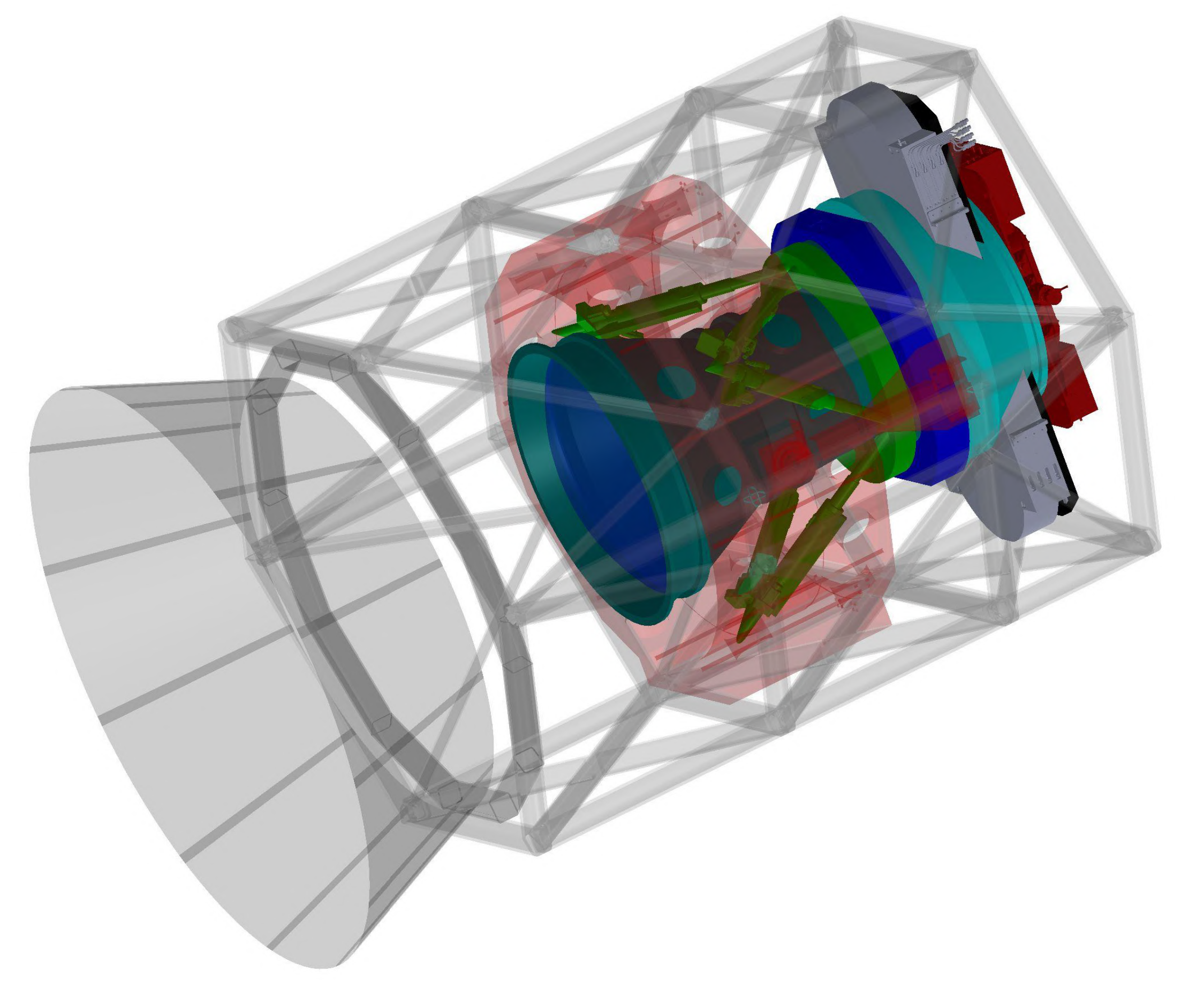}}
\hspace{0.2cm}
\scalebox{1.1}[1.1]{\includegraphics{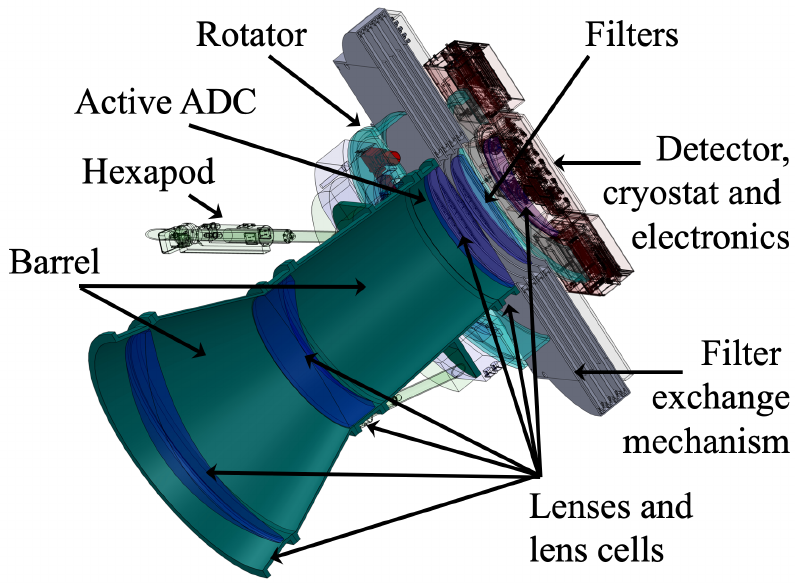}}
\caption{\small {\bf Left:} The Keck Wide-Field Imager (KWFI; colored components) inside the carrier module (translucent red) located in the Keck prime focus cage (translucent light gray) and extended telescope baffles (translucent dark gray).  {\bf Right:} Cross-section of KWFI with several components labeled. The hexapod, rotator, back-end enclosure are shown translucent and other components, such as electronics and cable wraps are not shown for clarity.}
\label{cross-section}
\end{center}
\end{figure}

Although KWFI is a straightforward (albeit extremely powerful) camera, the project includes a number of innovations and new technologies that help lower cost, increase performance, and offer advances for future instruments. The straightforward design of KWFI (Fig.~\ref{cross-section}) has low technical risk based on our four-element optical design, the construction of detector components for a similar system (the Zwicky Transient Facility [6]), a proven filter exchange mechanism design (DECam), use of an existing Keck Observatory top-end module, and its development and installation without the need for an observatory upgrade.  

\begin{figure}[!h]
\begin{center}
\vspace{-0.1cm}
\scalebox{0.395}[0.395]{\includegraphics{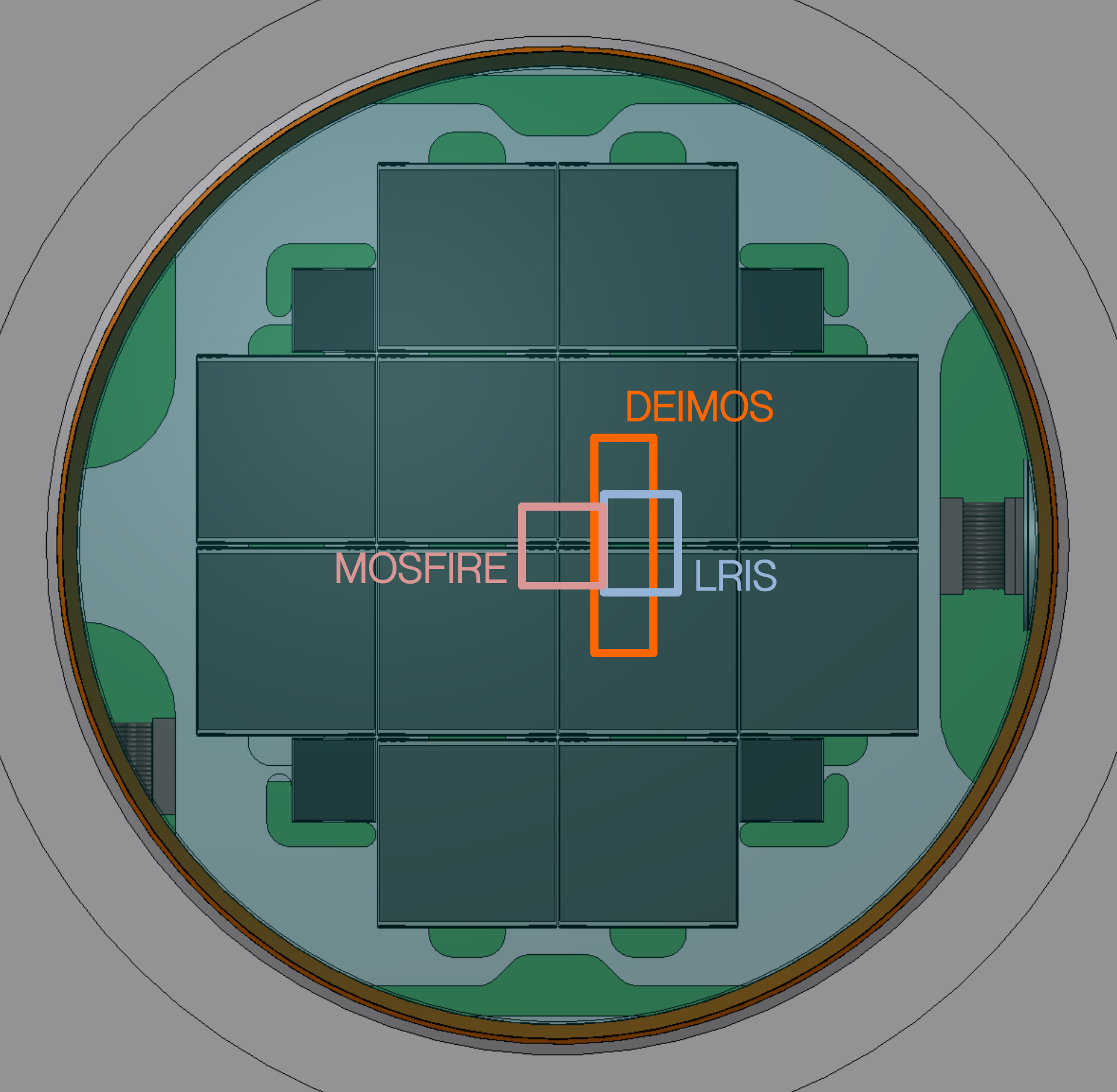}}
\hspace{0.1cm}
\scalebox{0.48}[0.45]{\includegraphics{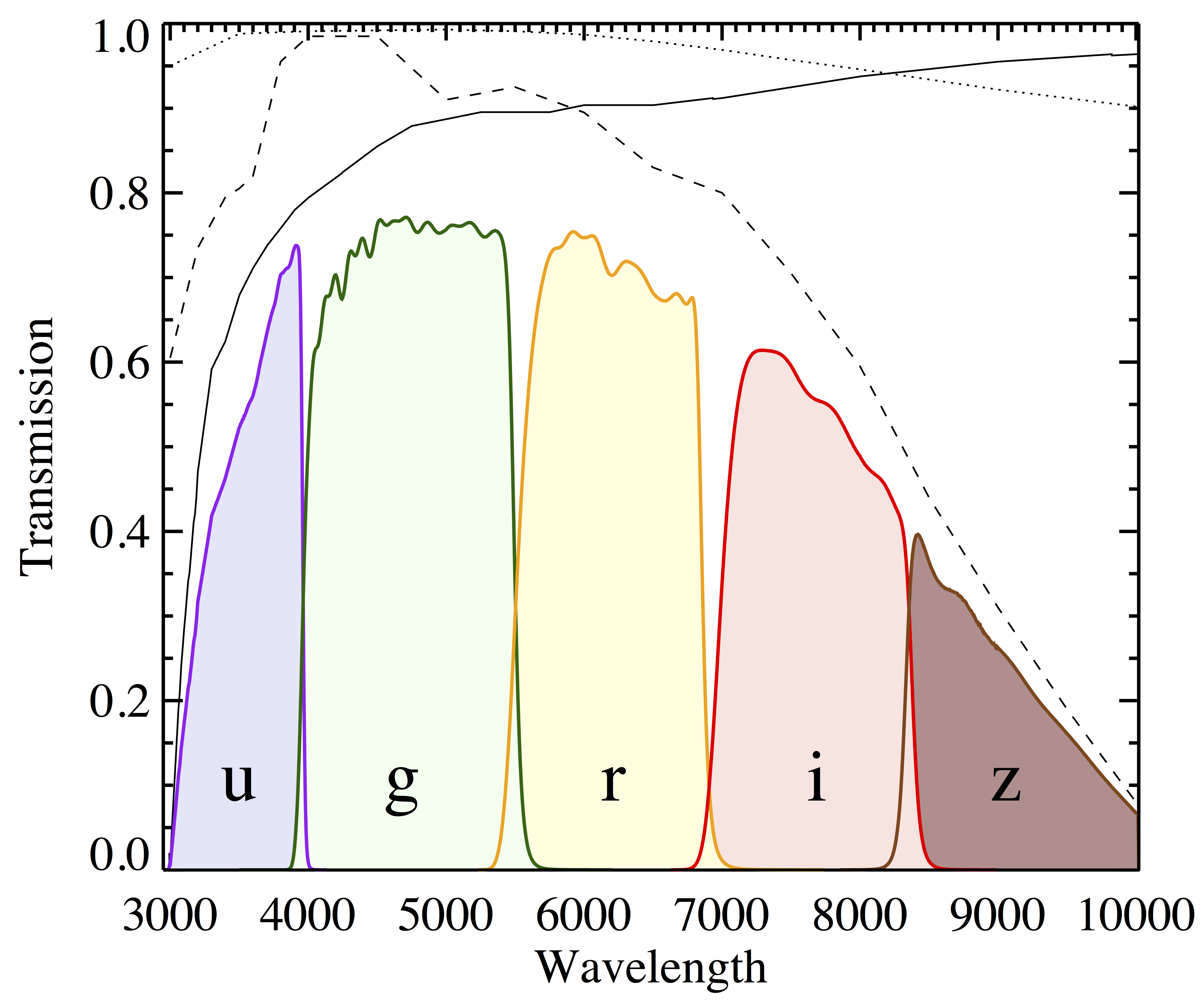}}
\caption{\small {\bf Left:} KWFI detector design. Twelve 6K$\times$6K CCDs (large squares) span 1 degree and image $\sim$0.8 deg$^2$ and corner 4K$\times$4K CMOS chips (small, dark green squares) are used for focus, guiding, and science. Overlaid are the fields of view of other Keck instruments (DEIMOS is not typically used for imaging). Image quality is excellent with little vignetting within a 1 degree diameter. {\bf Right:} KWFI projected throughput. Colored curves are KWFI $ugriz$ filter transmission (here, SDSS filters and a high-throughput u-band filter) convolved with the Maunakea atmospheric transmission (solid curve), optics throughput (dotted curve), and the currently explored CCD quantum efficiency (dashed curve). Thicker CCDs are being explored to increase red throughput without sacrificing blue throughput.}
\label{transmissions}
\end{center}
\end{figure}

Key components include: 
\begin{compactitem}
     \item {\bf Corrector optics:} A simple 4-element fused silica optical design, coated with sol-gel over MgF$_2$, provides excellent throughput from 3000--10000\AA\ ($\sim$98--90\%, see Fig.~\ref{transmissions}).  With four lenses, and only two mild aspheric surfaces, we have saved cost and increased throughput.
    \item {\bf Atmospheric Dispersion Correction (ADC) and instrument control:}  A hexapod compensates for flexure, focus, and alignment. An active ADC eliminates the cost of extra optical elements and UV transmission loss by using the hexapod to tilt the instrument with elevation, and a purely lateral shift (0--26 mm) of the third lens by an actuator.
    \item {\bf Filters:} Based on the DECam design, the filter exchange mechanism is fast ($\lesssim$ 10\,s) and highly reliable. It holds 8 exchangeable filters that can include broadband $ugriz$ and 3 others (e.g., narrowband).
    \item {\bf Focal plane:} Low cost-per-pixel 6K$\times$6K (or 4K$\times$4K) CCDs (see Fig.~\ref{transmissions}) built with high yielding processes, compact and lightweight detector cryostat, commercial electronics, and extensive CCD readout optimization with heritage from ZTF lowers technical risk and saves cost. 
    \item {\bf CMOS detectors:} We will explore Complementary Metal–Oxide–Semi-conductor (CMOS) detectors in the focal plane corners for guide/focus and deep fast-cadenced (down to millisecond) imaging.
    \item {\bf Deployable secondary mirror:} An optional deployable secondary mirror would accommodate flexible scheduling, rapid response time-sensitive observations, and explore new science by rapidly switching between KWFI and Keck optical and infrared spectrographs.
    \item {\bf Smart technology and digital twin:} The use of smart sensors and digital twin models of the instrument, observatory, and instrument/human interaction, enable more efficient design and operation, higher productivity, time and cost savings, safety, and predictive maintenance.
    \item {\bf Fast data reduction pipeline:} KWFI aims to have a `quick-look' (seconds), fast (minutes), and comprehensive (hours) data reduction pipeline to accommodate operation, fast science cases with rapid-response spectroscopic triggers, and archived data for all other science.
    \vspace{0.05cm}
 \end{compactitem}

The fast focal length of the Keck telescopes accommodates a large field of view (FOV) at prime focus. Practical optical designs working within the space and mass budget at prime focus can deliver a 1 deg diameter FOV with minimal vignetting (Figure~\ref{transmissions}). The main considerations in determining FOV are the science requirements and cost. 

KWFI is designed to be extremely blue sensitive and more sensitive over $\sim$3000-9000\AA\ than any other wide-field camera in the world. Expected sensitivities for a 2\,hr integration are: u(AB) = 28.2, g(AB) = 28.7, r(AB) = 28.0, i(AB) = 27.0, and z(AB) = 25.9, 5$\sigma$ {\it [0.8$^{\prime\prime}$ seeing FWHM, point source, 1.6$^{\prime\prime}$ aperture]}. KWFI can quickly reach m $\sim$ 27--28, the goal of current `Ultra Deep' surveys and will break new ground, reaching an unprecedented m $\sim$ 29, 5$\sigma$, from the ground in u g r filters in $\sim$8, $\sim$3.5, $\sim$12\,hrs, respectively, over its wide 1 degree diameter field. KWFI is capable of reaching m $\sim$ 30, 5$\sigma$, with a dedicated effort {\it (e.g., $\sim$20\,hr in g-band)}. Such extreme sensitivity over its wide field enables science cases unique to Keck.

\hd{References}

{\bf [1]} Gillingham et al. 2020, SPIE, 11203;
{\bf [2]} Radovan et al. 2022, SPIE, 12184;
{\bf [3]} Flaugher et al. 2015, AJ, 150, 150;
{\bf [4]} Boulade et al. 1998, SPIE, 3355, 614;
{\bf [5]} Miyazaki et al. 2018, PASJ, 70, S1;
{\bf [6]} Bellm et al. 2019, PASP, 131, 8002

\clearpage

\begin{center}
  {\bf \Large Table of Contents}
\end{center}

Twenty science cases are described for KWFI compiled largely during 2019--2021. The format includes the science case and optional input on science design requirements, including considerations of field of view, operational mode, filters, and a deployable secondary mirror capability. The science cases are listed roughly in order from the distant Universe to the solar system.
\bigskip

{\bf Science Case \hspace{14.8cm} Page}
Lyman Continuum Galaxies and Cosmic Reionization \dotfill 5\\
Roman and Euclid Science \dotfill 8\\
Lyman Break Galaxies \dotfill 10\\
High-Energy Neutrinos and VHE Gamma-Rays \dotfill 13\\
High-Redshift Supernovae and Superluminous Supernovae \dotfill 15\\
The Circumgalactic Medium \dotfill 19\\
Ly$\alpha$ Emitters in the Redshift Gap \dotfill 21\\
Fast Radio Burst Counterparts \dotfill 24\\
Early Transient Detection \dotfill 26\\
Gravitational Wave Science \dotfill 28\\
Untriggered Kilonova Searches \dotfill 33\\
Fast Transients \dotfill 35\\
Multiwavelength Very Fast Transients \dotfill 37\\
Low-Redshift and Local Galaxies \dotfill 40\\
The Low Surface Brightness Universe \dotfill 42\\
Globular Clusters, Compact Elliptical Galaxies and Ultra Diffuse Galaxies \dotfill 46\\
Supernova Remnants and Light Echos \dotfill 48\\
Milky Way Star Formation and Feedback \dotfill 51\\
Milky Way Stellar Populations and Metallicities \dotfill 53\\
Solar System Science \dotfill 55\\

\clearpage

\hd{KWFI Science Case}

\begin{center}
  {\bf \Large Lyman Continuum Galaxies and Cosmic Reionization}
\end{center}

\hd{Contributing authors}

Jeff Cooke (Swinburne)\\
Uro\v{s} Me\v{s}tri\'{c} (INAF)

\hd{Executive Summary}

Star forming galaxies are believed to have dominated the production of ionizing Lyman continuum (LyC; $<$ 912\AA) photons that reionized the Universe at z $\gtrsim$ 6 and set the stage for subsequent galaxy formation and evolution. Mapping the 2D and 3D distribution of ionizing sources enables a deep understanding of cosmic reionization and provides a direct complement to deep 21 cm {\sc Hi} mapping by radio facilities like the upcoming Square Kilometre Array (SKA) and collecting large samples of LyC emitting galaxies enables insight into their physics, morphology, and the mechanisms behind the escaping flux. However, LyC flux can {\bf only} be detected for z $\lesssim$ 5 galaxies, as the IGM opacity at higher redshifts makes the direct measurement of LyC flux impossible. The only way forward is to detect and study z $\sim$ 3--5 LyC emitting galaxies and calibrate their LyC flux with rest-frame $>$ 1216 \AA\ atomic transitions and rest-frame optical nebular emission lines that are observable at z $>$ 6. As a result, very deep (m $\sim$ 28--30) wide-field imaging at $\lesssim$ 5000\AA\ is required to detect LyC flux from the ground, as LyC flux falls exclusively in the u-band for z $\sim$ 2.5--5. 

\hd{Background}

\begin{figure}[!b]
\begin{center}
\vspace{-0.3cm}
\scalebox{0.61}[0.63]{\includegraphics{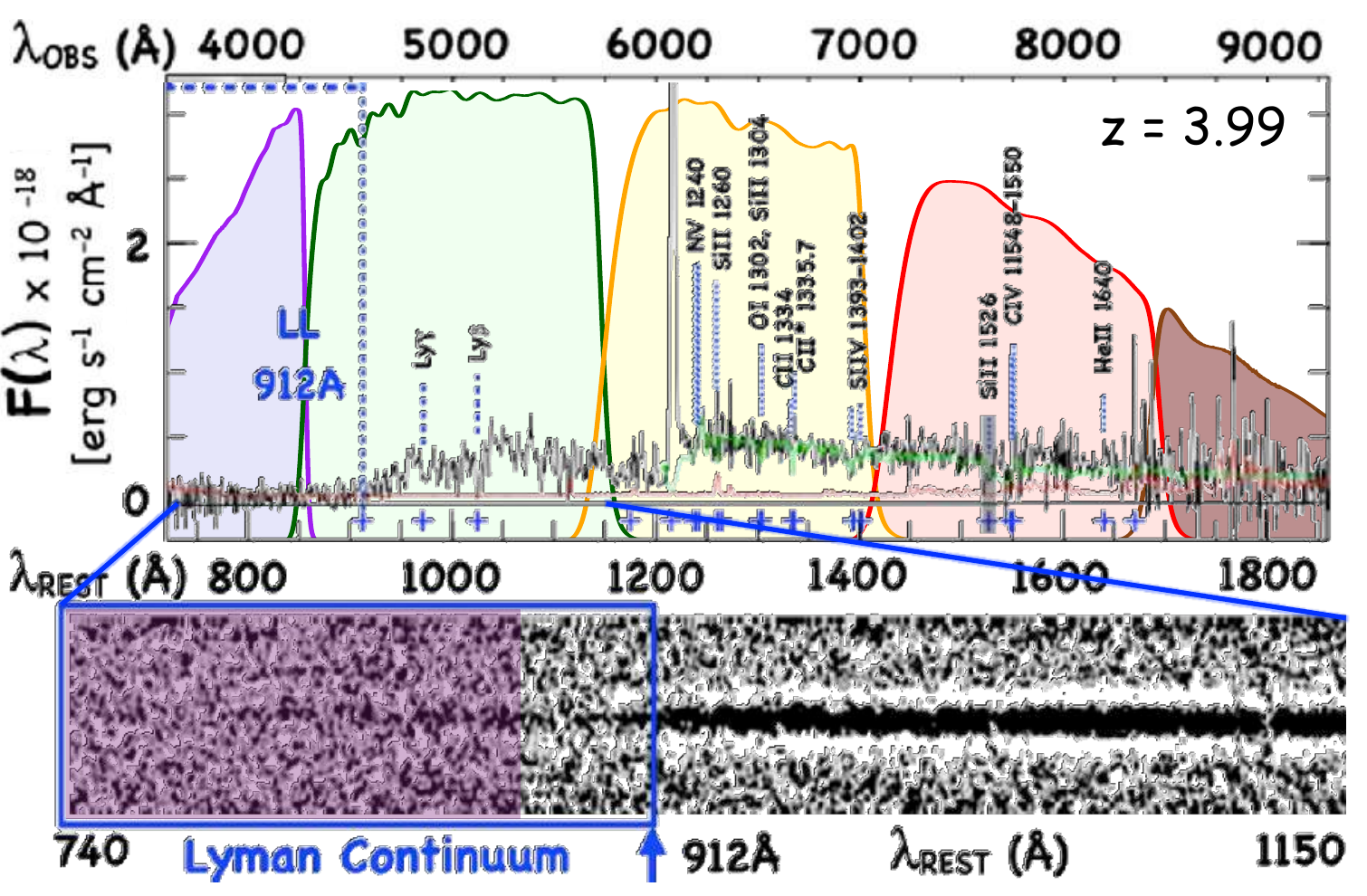}}
\scalebox{0.22}[0.22]{\includegraphics{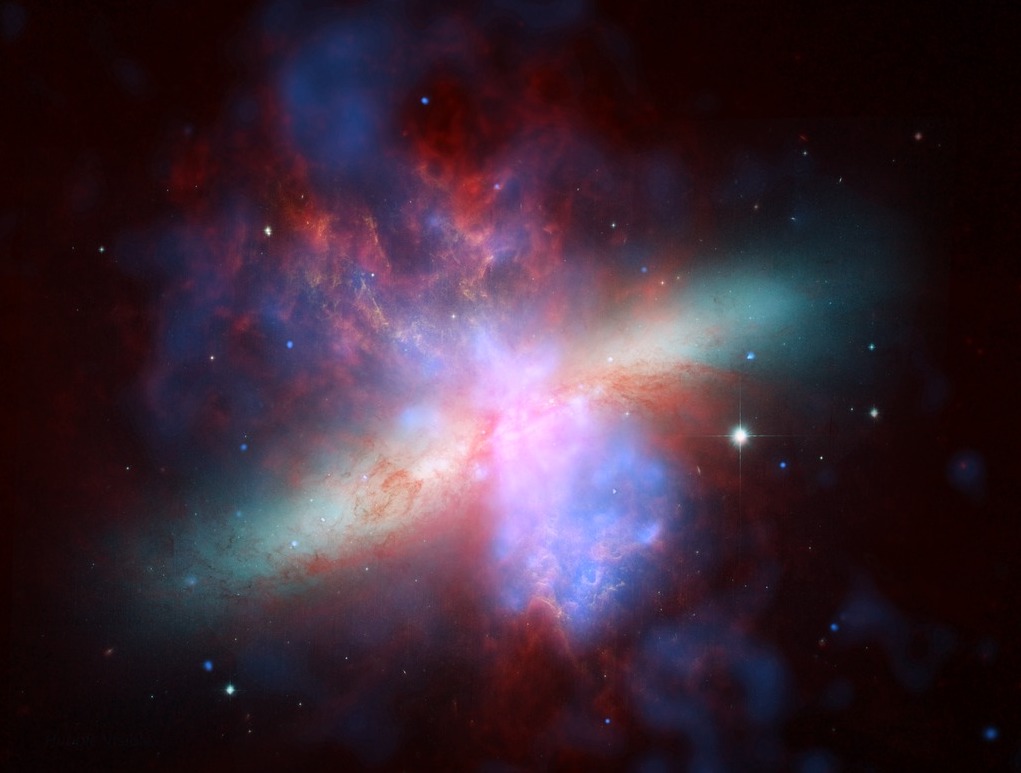}}
\vspace{-0.2cm}
\caption{\small {\bf Left:} A bright, weakly-lensed Lyman continuum emitting galaxy Ion3 at z = 3.99 [12].  The 1D spectrum (top) showing observed-frame (upper x-axis) and rest-frame (lower x-axis) with the KWFI filter transmissions underlaid. The 2D spectrum shown below denotes the LyC wavelength region (blue box) and the region probed by the u-band (shaded in purple). At z $\gtrsim$ 3.4, the u-band probes the LyC flux cleanly, whereas the g-band is contaminated by the \lya\ forest. The LyC flux is fully attenuated by the IGM at redshifts higher than z $\gtrsim$ 5 before the \lya\ is redshifted out of the g-band.  {\bf Right:} The galaxy M82 is an illustrative example of one means of radiation escape (blue), here in biconical flows along the minor axis. Given the orientations, morphologies, duty cycles, and varying sightline transparencies, galaxies with escaping Lyman continuum flux reaching Earth are rare. Thus, a wide-field imager is needed to gather large galaxy samples over large scales and to understand the physical mechanisms behind the escape.} 
\vspace{-0.0cm}
\label{LCGs}
\end{center}
\end{figure}

The origin and evolution of the UV ionizing background is fundamentally tied to many astrophysical processes involved in the formation and evolution of galaxies including the reionization of the intergalactic medium (IGM), the suppression of cloud collapse and star formation, and the distribution and availability of neutral gas in galaxy outskirts [1--4]. Because the space density of QSOs drops quickly toward higher redshift, it is believed that these objects did not make a significant contribution to the UV background beyond $z\gtrsim$ 3 (e.g., [5--7]). Instead, the radiation field from $z\sim3$, to well beyond the epoch of reionization, is believed to have been dominated by young, star forming galaxies. Thus, directly measuring the fraction of Lyman continuum (LyC; $<$ 912\AA) photons that escape high redshift galaxies has been a fundamental goal of observational cosmology for the last few decades  (Figure\,\ref{LCGs}). 

\begin{figure}[!b]
\begin{center}
\vspace{-0.6cm}
\scalebox{0.37}[0.36]{\includegraphics{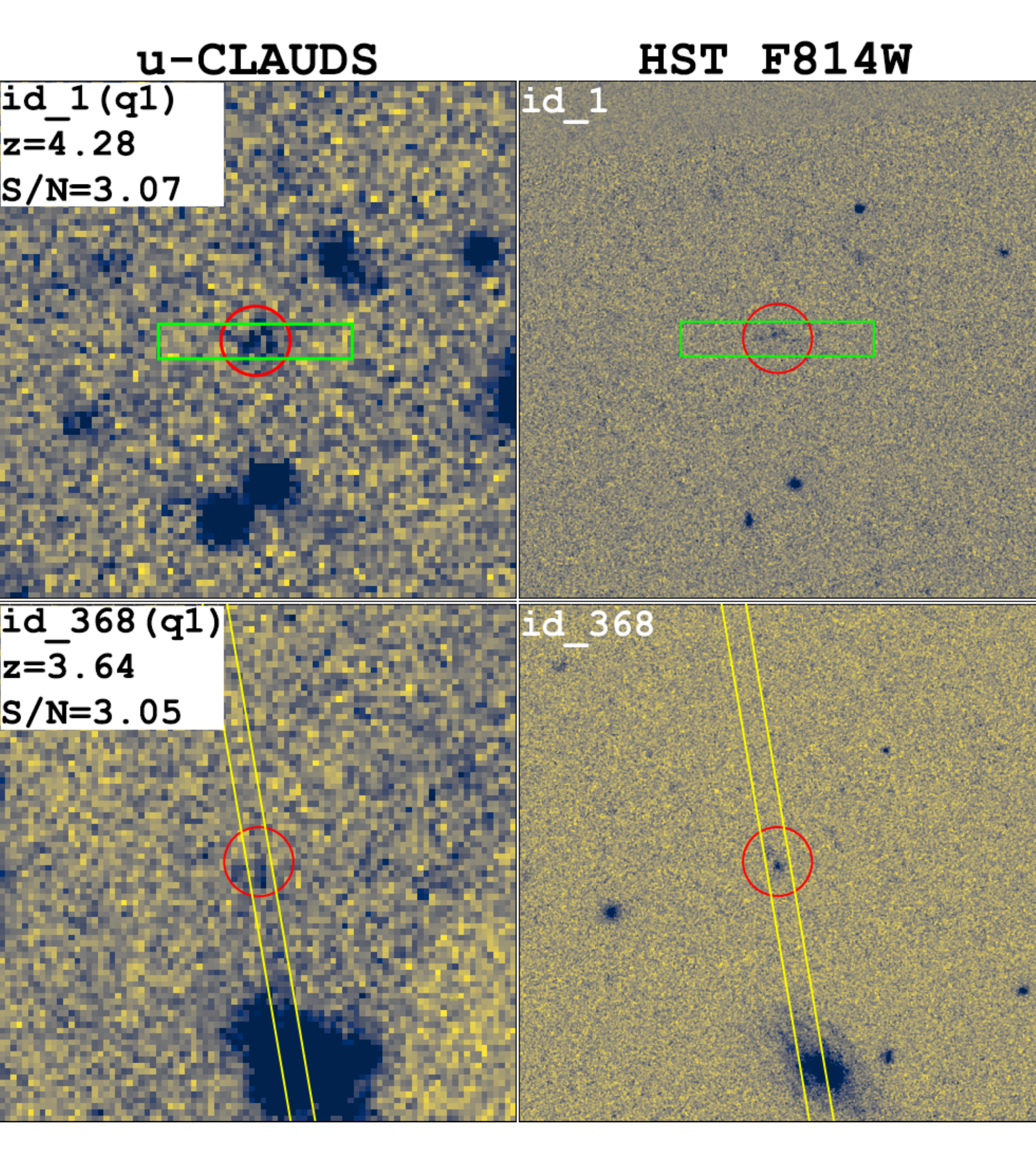}}
\hspace{0.2cm}
\scalebox{1.06}[0.90]{\includegraphics{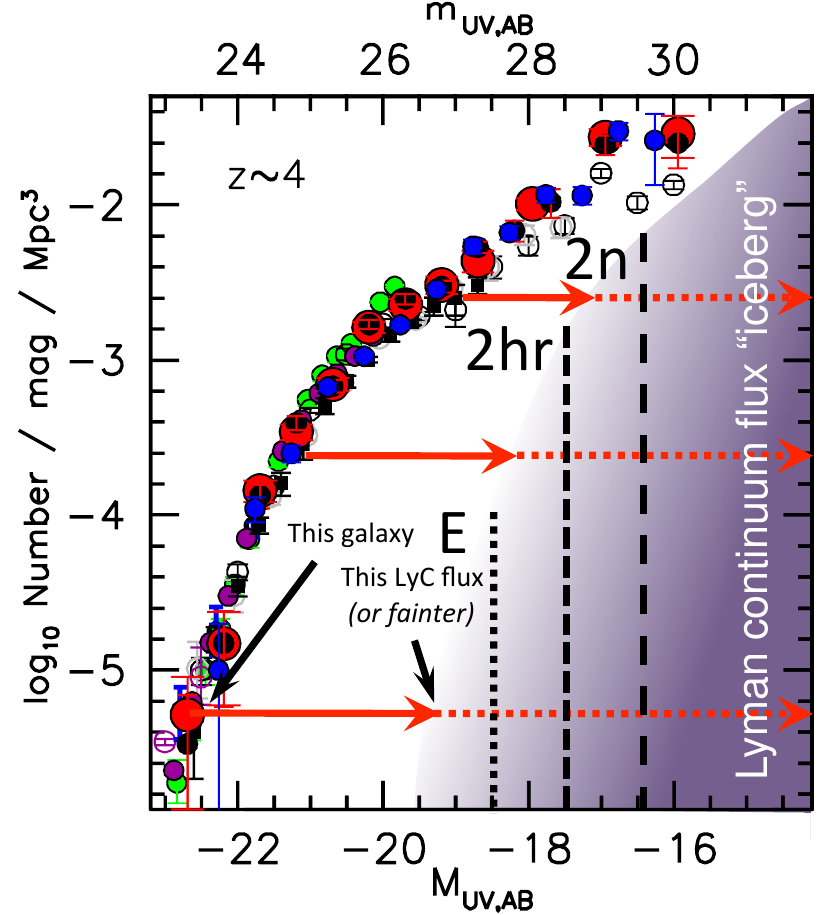}}
\vspace{-0.4cm}
\caption{\small {\bf Left:} Two Lyman continuum (LyC) emitting galaxies at z = 4.28 and z = 3.64 in the COSMOS field [21]. The u-CLAUDS u-band imaging {\it (left column)}, which is the deepest wide-field imaging to date (m $\sim$ 27.5) cleanly probes restframe wavelengths blueward of 912\AA\ and the $HST$ 814W imaging (right column) probes the non-ionizing continuum. Wide-field rest-frame optical space-based imaging (e.g., Roman and Euclid) will complement wide-field KWFI rest-frame far-UV imaging to ensure that the LyC emission is not contaminated by lower redshift sources near the line of sight. {\bf Right:} The Lyman break galaxy luminosity function at z $\sim$ 4 measured in deep $HST$ images over small areas (modified from [22]). The purple region is an expectation for LyC luminosity for galaxies horizontal in position {\it (see red arrows and text)}. For example, here a galaxy with log N mag$^{-1}$ Mpc$^{-3}$ of $-$3 has a magnitude of m $\sim$ 25.5 (UV luminosity M $\sim$ $-$20.5) would require a depth of m $\sim$ 28 or fainter. The vertical dotted line labeled `E' is the depth of the deepest existing wide-field u-band imaging to date [20] and finds $\sim$10 LyC emitting galaxies per deg$^2$ at z $\sim$ 4 [21]. A 2-hour u-band integration with KWFI will detect $\sim$2 per arcmin$^{-2}$ and a deep effort ($\sim$15 hr) will detect $\sim$10 arcmin$^{-2}$, and for much fainter galaxies that are expected to emit a larger fraction of LyC flux [10].}
\vspace{-0.0cm}
\label{LFs}
\end{center}
\end{figure}

The intergalactic medium (IGM) at z $\gtrsim$ 5 is opaque to LyC photons. The LyC for z $\lesssim$ 5 galaxies corresponds to $\lesssim$ 5000\AA\ in the observed frame (cf. Figure\,\ref{LCGs}). LyC flux is $\gtrsim$ 3 mags fainter than non-ionizing ($L^\star$: m $\sim$ 24.5--25.5) UV continua [8,9], with the largest contributions likely arising from galaxies at the faint-end of the luminosity function [10]. As a result, z $\sim$ 3--5 LyC flux can only be detected with extremely deep m $>$ 28 u-band imaging (Figure\,\ref{LFs}).  To date, the deepest searches have made a handful of LyC detections, and these detections are just at the u-band imaging threshold, or for rare bright or lensed events [9,11--13]. The deepest space-based efforts have reached m $\sim$ 29.1--29.3 [14--18], over very small areas ($\sim$2.7 $\times$ 2.7 arcmin each) requiring 23.6--24.5 hrs each and have detected $\sim$10 LyC emitting galaxy candidates over $>$10 fields. Equivalent depths can be achieved by KWFI in one-third the time over more than 100$\times$ the area.

\vspace{0.1cm}
The detection of escaping ionizing LyC flux and the physics behind the escape are affected by the (preferential) IGM line-of-sight transparency, HI covering fraction, galaxy orientation and morphology, duty cycle, and large-scale environment [9,19].  As a result, their low sky density requires wide-field, very deep imaging.  The CFHT Large Area U-band Deep Survey (CLAUDS [20]) is the largest wide-field u-band imaging program to date.  CLAUDS includes the `UltraDeep' u(AB) $\sim$ 27.5 over 1.3 deg$^2$, the deepest wide-field u-band of any program to date. Searches in CLAUDS for LyC emitting galaxies reveal that the depth just {\it `touches the tip of the LyC flux iceberg'} (see Fig.~\ref{LFs}), requiring 1--2 mags deeper to explore the population and to identify targets for JWST and 30m-class spectroscopy [19,21].  

{\bf LyC emission line detection -} Star forming galaxies are expected to produce emission lines shortward of 912\AA. Deep, wide-field $<$ 3500\AA\ KWFI narrowband imaging will be a powerful means, unique to Keck, to efficiently select galaxies by their expected LyC emission lines for deep follow up with wide-field Subaru Prime Focus Spectrograph, Maunakea Spectroscopic Explorer, JWST, and 30m-class spectroscopy. KWFI is the \underline{\it only} instrument in the world able to achieve this science and fill a critical gap in our understanding of the earliest epochs of the Universe, a key science driver for JWST and 30m-class telescopes.

\hd{The big questions}

{\it Big Questions 1}

Are star forming galaxies the main contributors to cosmic reionization?  What is their contribution over time and as a function of environment?

{\it Big Questions 2}

What physics is behind the escape of Lyman continuum radiation? How is Lyman continuum flux escape tied to other galaxy properties, in particular those observable at z $>$ 6?

{\it Big Questions 3}

How did reionization come about and evolve with time? Did it occur as is illustrated in cosmological simulations over large volumes? What is the tomography of ionization as mapped along with neutral hydrogen via the 21 cm line?

\hd{What KWFI can do for this science case}

KWFI will dominate reionization research, as wide-fields and m $\sim$ 28--30 optical depths are required. KWFI is the only wide-field instrument with the necessary extremely deep, blue sensitivity to detect LyC emitting galaxies and the capability to map the population over large cosmological volumes. A deep search over the $\sim$2 degree COSMOS field with existing relatively shallow u-band imaging has uncovered only a few  LyC emitting galaxy candidates (e.g., [21]). However, targeted deep $HST$ F336W (similar to the u-band) imaging to m $\sim$ 29 [GO 15100; PI Cooke] finds 3 spectroscopic and 32 photometric LyC emitting galaxy candidates [18] in two $\sim$2.7$\times$2.7 arcmin random fields in COSMOS using the method of [23]. Finally, the mechanisms behind LyC flux escape is not well understood and may depend on galaxy morphology, HI covering fraction, kinematics, interactions, environments on small and large scales, and other properties. As a result, wide fields and large numbers of relatively rare LyC emitting galaxies are needed. 

One efficient means to search for LyC galaxies would via the method in [23] that has been successful in detecting these galaxies in random searches [18,21] and provides a means for mapping galaxies in deep fields. For example, very deep u-band (m $\sim$ 29--30) of legacy fields (e.g., COSMOS, SXDS). Roman space-based wide-field infrared (rest-frame optical) images would provide a check for nearby galaxy contaminants to any detected u-band flux and also to help map the LyC flux localization contours to the star forming flux contours. In addition, deep KWFI u-band coverage of the Roman and Euclid deep fields would be the next logical step. To explore how reionization occurred, mapping volumes similar to those produced by cosmological simulations is needed. Typical simulation boxes are 100 $h^{-1}$ Mpc on a side and a 1 degree diameter field of view corresponds to $\sim$100 Mpc on a side at z $\sim$ 3--4.

The extreme and blue sensitivity of KWFI will enable a completely unique means to search for LyC emitting galaxies using narrowband filters, akin to the search for Lyman-$\alpha$ emitters (LAEs). LyC emission-lines are expected for star forming galaxies and KWFI will enable their detection using narrowband filters that are aligned with Lyman-$\alpha$ narrowband filters and broadband filters. The extreme blue sensitivity of KWFI will enable the detection of LyC emission deep into the emitting galaxy population, largely independent of galaxy continuum luminosity, and in a wholesale manner to map over wide-fields.

\hd{References}

{\bf [1]} Fan et al. 2002, AJ, 123, 1247;
{\bf [2]} Efstathiou 1992, MNRAS, 256, 43;
{\bf [3]} Bullock et al. 2000, ApJ, 539, 517;
{\bf [4]} Somerville 2002, ApJ, 572, 23;
{\bf [5]} Hopkins et al. 2007, ApJ, 654, 731;
{\bf [6]} Jiang et al. 2008, AJ, 135, 1057;
{\bf [7]} Fontanot, Cristiani, \& Vanzella 2012, MNRAS, 425, 1413;
{\bf [8]} Steidel et al. 2001, ApJ, 546, 665;
{\bf [9]} Steidel et al. 2018, ApJ, 869, 123;
{\bf [10]} Robertson et al. 2015, ApJ, 802, 19;
{\bf [11]} Bian et al. 2017, ApJL, 837, L12;
{\bf [12]} Vanzella et al. 2018, MNRAS, 476, L15;
{\bf [13]} Fletcher et al. 2019, ApJ, 878, 87;
{\bf [14]} Siana et al., 2015, ApJ, 804, 191;
{\bf [15]} Oesch et al., 2018, ApJS, 237, 12;
{\bf [16]} Smith et al., 2018, ApJ, 853, 191;
{\bf [17]} Smith et al., 2020, ApJ, 897, 41;
{\bf [18]} Prichard et al. 2022, ApJ, 924, 14;
{\bf [19]} Bassett et al. 2021, MNRAS, 502, 108;
{\bf [20]} Sawicki et al. 2019, MNRAS, 489, 5202;
{\bf [21]} Mestric et al., 2020, MNRAS, 494, 4986;
{\bf [22]} Bouwens et al. 2021, AJ, 162, 47;
{\bf [23]} Cooke et al. 2014, MNRAS, 441, 837

\clearpage
\hd{KWFI Science Case}

\begin{center}
  {\bf \Large Roman and Euclid Science}
\end{center}

\hd{Contributing authors}

Jason Rhodes (JPL/Caltech)\\
Jeff Cooke (Swinburne)

\hd{Executive Summary}

NASA's Nancy Grace Roman Space Telescope and ESA's Euclid Space Telescope are wide-field optical/infrared space missions to be launched in the next few years. These missions are driven to better understand the dark energy equation of state, dark matter, weak lensing and gravity, large-scale structure and cosmology, to discover and characterize exoplanets, and other areas of infrared astrophysics. Both facilities are sensitive from $\sim$5000--5500\AA\ to $\sim$2 $\mu$m and plan deep, high-resolution, wide-field imaging and grism spectroscopy surveys to m $\sim$ 28--30. Both telescopes rely on complementary ground-based imaging for a number of science cases.  For instance, deep u- and g-band wide-field imaging is essential for multiple science cases, including photometric redshifts and high redshift galaxy drop-out selection. KWFI will be the only facility able to acquire the missing extremely deep, wide-field $\sim$3000--5500\AA\ imaging (i.e., u-band and g-band filters). 

\hd{Background}

\begin{figure}[!h]
\begin{center} 
\scalebox{0.64}[0.64]{\includegraphics{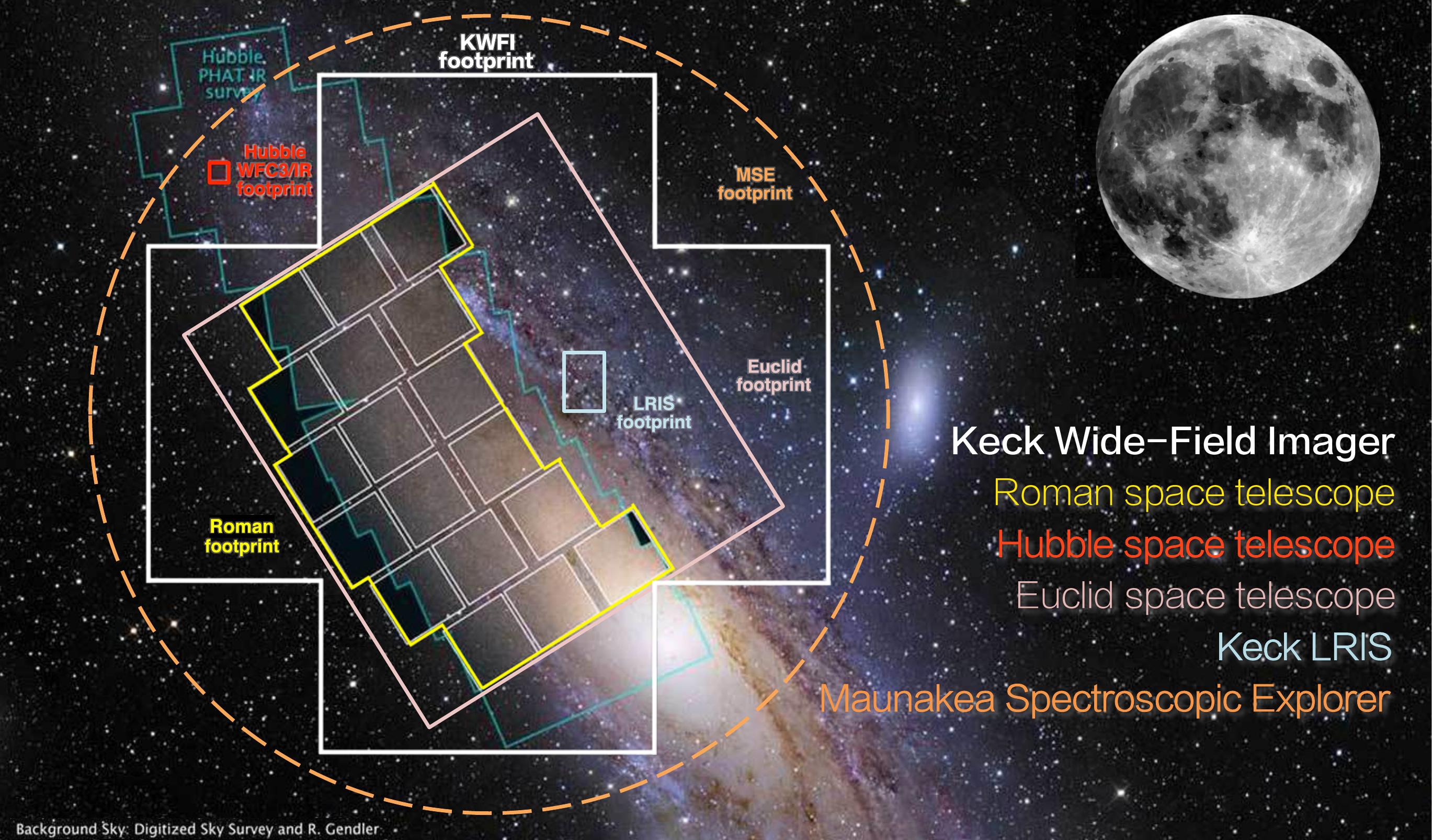}}
\vspace{-0.2cm}
\caption{\small KWFI 1 degree diameter CCD footprint (white) in comparison to Euclid (pink), Roman (yellow), and the Hubble Space Telescope (red) fields of view (FOV). Also shown are the FOVs of Keck LRIS, the only viable optical imager on Keck, and the Maunakea Spectroscopic Explorer wide-field fiber spectrograph. Imagers on 8m-class and 30m-class telescopes have FOVs similar to that of LRIS and do not have the blue sensitivity of KWFI.}
\label{fovs}
\end{center} 
\end{figure}

The Euclid Space Telescope (Euclid) is planned for launch from late-2022 to 2023 and the Nancy Grace Roman Space Telescope (Roman) is planned for launch before mid-2027. Both space telescopes will offer wide-field $\sim0.5-2\mu$m imaging and grism spectroscopy. Euclid plans a 40 deg$^2$ deep survey to m $\sim$ 28--29 and, similarly, Roman will plan deep fields to m $\sim$ 29--30. Wide-field UV satellites UVEX and ULTRASAT being proposed aim to cover $\sim$2100--2800\AA\ (albeit lower resolution and shallower). As a result, KWFI fills the $\sim$2800--5500\AA\ deep, wide-field imaging `wavelength gap' and has similar field of view (Figure~\ref{fovs}).

\hd{Roman and Euclid science synergy with KWFI}

Roman and Euclid space-based resolution infrared imaging and KWFI wide-field optical imaging can achieve many science cases.  Several are listed below in the areas of time domain, gravitational lensing, high redshift galaxies, and cosmic reionization. 

{\bf Time domain -} Roman's planned survey programs include an {\bf Galactic Plane Time Domain Survey} (driven by the primary science case of exoplanet microlensing) over 2 deg$^2$ and a {\bf High-Latitude Time Domain Survey} (driven by the primary science case of observing supernovae) over 5 deg$^2$. Both surveys are time domain surveys that benefit from fast, deep cadenced imaging. In addition, deep, wide-field blue imaging is required to identify type Ia supernovae from core-collapse supernovae and other events, as only a small fraction of these events can be followed up with deep spectroscopy. Finally, Roman and KWFI have powerful synergy for {\bf z $\sim$ 2--6 superluminous supernovae} (SLSNe) in the High-Latitude Time Domain Survey and other programs. The rare events are detected in deep, wide-field KWFI optical and Roman and Euclid infrared imaging, with redshifts from KWFI filter drop out selection. Euclid and Roman space-based imaging can determine the transient location in the host, study host properties, and KWFI-Roman will help connect rest-frame far-UV behavior to rest-frame optical, in which the events are classified, that is necessary to progress SLSNe detections to z $\sim$ 13 with Roman and Euclid.

{\bf Weak lensing -} A key science case for Euclid and Roman is weak lensing at z $\sim$ 1--2 where deep, wide-field 3000--5500\AA\ imaging, in particular u-band imaging, is crucial for {\bf photometric redshifts}. A KWFI legacy survey of the entire 40 deg$^2$ Euclid deep fields to equal depths over 3000--5500\AA\ would require $\sim$28\,hr in g-band and $\sim$55\,hr in u-band {\it (8--10 nights)} and would produce a rich dataset for deep 10m-class and JWST spectroscopy and upcoming 10m-class wide-field and 30m-class spectroscopy.   

{\bf High redshift galaxies -} Lyman break galaxies (LBGs) comprise the bulk of star forming galaxies at high redshift. LBGs have been used to study the properties of high redshift galaxies, to identify high-redshift clusters and gravitationally lensed galaxies, to trace star formation and large-scale structure, among other science. Selecting LBGs at z $\sim$ 2 and z $\sim$ 3 depends on their `drop out' or flux decrement in the u-band. As a result, {\bf z $\sim$ 2--3 LBG science cannot be done without deep u-band imaging}, in particular the faint-end of the luminosity distribution, which includes the bulk of the galaxies. Roman and Euclid space-based imaging helps to confirm and study lensed LBGs, confirm galaxy morphological predictions based on KWFI broadband color criteria predictions, confirm and investigate galaxies involved in interactions, and study the morphology-density relation at high redshift.

{\bf Cosmic reionization -} KWFI is the only instrument capable of detecting the extremely faint ionizing photons (Lyman continuum photons, $>$ 912\AA) over wide fields from galaxies that are believed to have been responsible for the reionization of the Universe. {\bf Deep, wide-field u-band imaging is the only means to detect the faint (m $\sim$ 28--30) Lyman continuum flux} from z $\sim$ 3--5 galaxies, as the flux falls in the u-band and the IGM becomes opaque to higher redshift galaxies. Wide fields are needed as the transmission out of the distant galaxy to Earth depends on transparent sightlines through the IGM, the galaxy morphology, their duty cycle, and other physical mechanisms. Roman and Euclid space-based wide-field imaging is essential to eliminate lower redshift contaminant galaxies very close in the line of sight and to provide morphological and interaction information.

\hd{What KWFI can do for this science case}

There are no existing imagers, or imagers planned in the foreseeable future, that can provide this necessary deep, wide-field 3000--4000\AA\ imaging. KWFI is the only instrument that can provide this extremely deep u-band imaging. As a result, the Keck community will lead in these research fields.

\hd{Requirements for KWFI}

The Roman and Euclid footprint and planned deep surveys require the widest field possible for KWFI.
\vspace{-0.2cm}

High sensitivity in the blue down to $\sim$3000\AA, or the atmospheric limit.
\vspace{-0.2cm}

Broadband filters, including u-band, narrowband filters.

\clearpage

\hd{KWFI Science Case}

\begin{center}
{\bf \Large Lyman Break Galaxies}
\end{center}

\hd{Contributing author}

Garry Foran (Swinburne)

\hd{Executive Summary}

Lyman break galaxies (LBGs) are a population of high redshift star forming galaxies that are efficiently identified in deep broadband images at desired redshifts intervals (e.g., [1,2]). Remarkably, recent work has shown that broadband imaging alone is sufficient to select subsets of LBGs with specific spectral properties [3,4] and up to 10 other internal and environmental properties, such as galaxy kinematics, morphology, local and large-scale environment, star formation rate, ISM absorption-line strengths, galaxy outflows, etc [e.g., 5--10]. In addition, LBGs can help measure the UV background and ionising budget of the Universe and have applications in cosmology [11]. As a result, broadband imaging of LBGs is an extremely powerful means to understand large-scale behavior down to internal structure of high redshift galaxies, connecting observed galaxies to cosmological simulation halos, identify the precursor to the morphology-density relation observed at low redshift, and better understand galaxy formation and evolution.

\bigskip

\hd{Background}

One of the most important and well-studied populations of primordial galaxies are the Lyman break galaxies (LBGs) or so-called `dropout' galaxies. LBGs can be efficiently selected in large numbers, on large scales across a wide range of redshift pathlengths using as few as three broadband optical filters sensitive to the rest-frame UV [e.g., 1,2]. LBGs at redshifts $3 \lesssim z \lesssim 5$ are identified by the break in their UV flux blueward of the Lyman limit (912~\AA), and at $z\gtrsim5$ by the strong decrement in flux in the Lyman-$\alpha$ forest blueward of Lyman-$\alpha$ (\lya, 1216~\AA). Modified color-color selection criteria can extend the LBG selection window to $1.5 \lesssim z \lesssim 2.5$ at which redshifts the Lyman continuum break is not observable from the ground [2,12].  Accordingly, ground-based LBG surveys provide an unrivaled view of the rest-frame UV properties of galaxies from the epoch of reionization through the peak in cosmic star-formation rate density, and a means by which consistently-selected and physically-related populations of star-forming galaxies can be efficiently assembled and compared over a wide redshift range and across hundreds to thousands of Mpc.  Moreover, the observational advantages that pertain to the study of rest-frame-UV selected samples and the convenience and efficiency with which such galaxies can be collected homogeneously and in high numbers makes them an important target population in the current generation of large-area and all-sky photometric campaigns.

\begin{figure}[!h]
\scalebox{0.9}[0.9]{\includegraphics{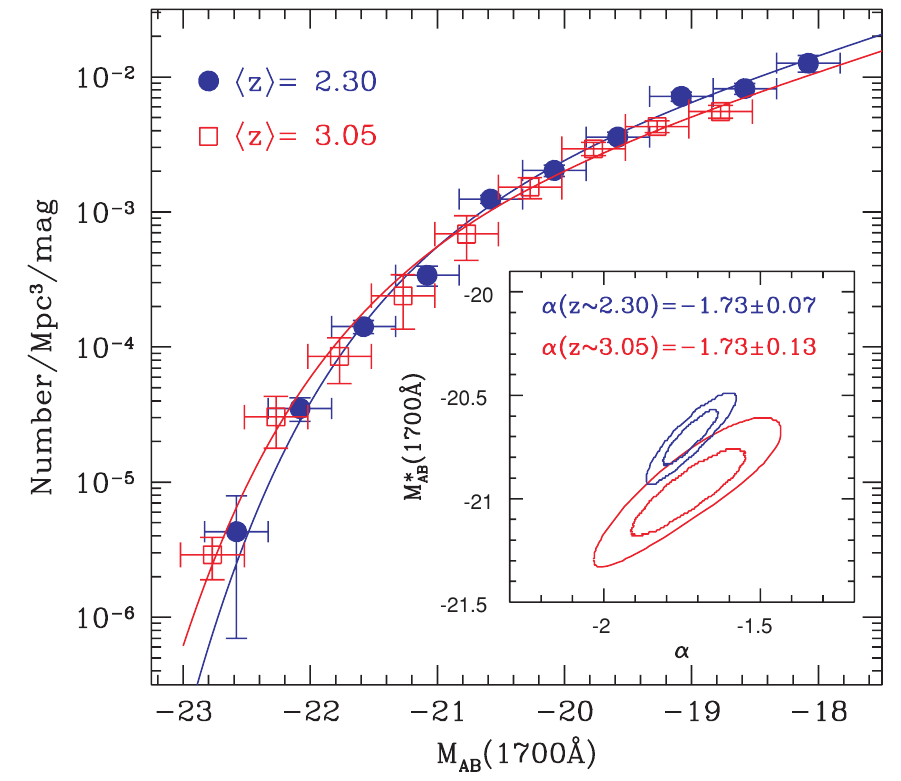}}
\scalebox{1.2}[1.24]{\includegraphics{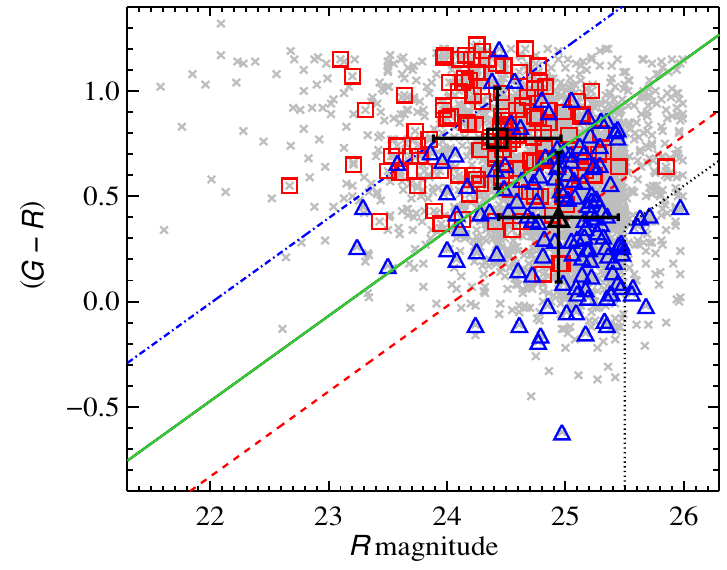}}
\vspace{-0.3cm}
\caption{\small {\bf Left:} Luminosity functions for $z\sim2$ and $z\sim3$ Lyman break galaxies [15]. To date, the galaxy luminosity function on any semi-cosmological scale is probed down to M $\sim$ $-$18 for $z\sim2$ and M $\sim$ $-$18.5 for $z\sim3$ in fields other than very small ($\sim$3$\times$3 arcmin) $HST$ deep fields. These depths equate to apparent magnitudes of m $\sim$ 27. The form of the faint end of the luninosity function and the point that it turns over is still unclear.  {\bf Right:} An example of the segregation of LBGs by their \lya\ spectral feature using broadband imaging alone [3]. LBGs with dominant \lya\ in absorption are indicated by red squares and those with dominant \lya\ in emission by blue triangles. LBGs selected using deep imaging separate on the color-magnitude diagram (here, ($G-R$) vs.\ $R$ for $z\sim3$ galaxies, deep u-band needed for $z\sim2$) such that simple color criteria (e.g., the dotted and dot-dashed lines) can identify galaxies with \lya\ in emission and those with \lya\ in absorption for individual galaxies in large-area photometric surveys using only the broadband information alone. Moreover, each of these LBG subsets have other galaxy properties found to be associated with them (e.g., kinematics [8] and morphology[7]). Thus, KWFI imaging can provide sufficient information to understand many properties of the millions of galaxies detected in the deep images. 
}
\label{reddy}
\end{figure}

{\bf UV luminosity function:}  Measurement of the UV luminosity function is critical to our understanding of the ionizing flux budget of the Universe and its evolution across cosmic time  (Figure~\ref{reddy}). Data from existing surveys to typical depths of m $\lesssim$ 27, suggest that LBGs comprise the majority of star-forming galaxies and dominate the star formation rate density at $z\sim2-3$ [13,14].  It follows that LBGs measured to even greater depths will be required to probe the faint end slope and extent of the UV luminosity function. Given that the broadband filter that is sensitive to the Lyman break (u-band at $z\sim3$) needs to probe $\gtrsim$1 mag deeper than the redder detection band, this field can only be progressed with an instrument like KWFI that can achieve u- and other band depths of m $\sim28-29$ on large scales and in a reasonable amount of time.

We note that Subaru HSC has no u-band sensitivity and cannot select LBGs at redshifts $\lesssim4$. In addition, the estimated depths of Ruben/LSST 10-year stacks are m = 26.1 for the u-band, m = 27.4 (g-band), and m = 27.5 (r-band) -- depths achievable with 2--45 minutes per pointing with KWFI. The LSST four `deep drilling fields' may probe to approximately 1 mag deeper than the full LSST. These depths are quickly reached with KWFI and, unlike Rubin, can target any specific field for science.

{\bf Spectroscopic properties and environment from only broadband imaging:} Broadband imaging alone can segregate pure samples of $z\sim2$ and $z\sim3$ LBGs based on their \lya\ spectral type, displaying either dominant \lya\ in absorption (aLBGs) or dominant \lya\ in emission (eLBGs) [3,4] (Fig.~\ref{reddy}). There are $\sim$10 galaxy internal and environmental properties that are related to the \lya\ feature emission or absorption strength (e.g., [5--10]). These properties including galaxy kinematics, morphology, outflows, large-scale environment (i.e., residing within groups, clusters, cluster outskirts or in the field), interactions, luminosity, halos mass, UV continuum reddening, ISM absorption-line strengths, and escaping ionizing radiation.

Broadband imaging selection has already been utilized to study the clustering properties of $z\sim3$ aLBGs and eLBGs by correlation function analysis [9], and as a tool to predict the nebular emission-line kinematic type of \lya-absorbing and \lya-emitting $z\sim2-3$ LBGs [8], among other properties. These results suggest a framework within which a wide range of properties known to be associated with \lya\ can be studied in large samples and over large scales in data from current and future large-area photometric campaigns.

At $z\sim2$ where the Lyman limit lies beyond the range of ground-based instruments, deep u-band photometry of the \lya forest is essential for LBG selection.  The photometric requirements are particularly demanding for the important, but fainter, \lya-emitting fraction that at typical survey depths is numerically swamped by the brighter more abundant \lya-absorbing population making statistically reliable photometric segregation more challenging.  The ability to probe to u-band depths of m$\sim28$ and beyond would greatly increase the number of net \lya\ emitters (and indeed LAEs) in the LBG selection, and thereby  enhance the statistical reliability and practical utility of the \lya\ spectral-type segregation approach at $z\sim2$. Similarly, the importance of KWFI for the application of the \lya\ spectral type method is highlighted by the lack of u-band capability on any comparable wide-field imaging instrument.

Thus, the u-band sensitivity and high throughput of KWFI means that it will stand alone for its ability to assemble large samples of LBGs, LAEs, and \lya\ absorbing galaxies on large scales and, in practical terms, unrivaled depth at $z\sim2-3$, the epoch encompassing the cosmic peak in star-formation and black-hole accretion activity, and during which the morphology--density relation was taking shape.

{\bf Cosmology using dropouts:} LBGs -- along with the related \lya\ emitters (LAEs) -- are important targets for the study of cosmic large-scale structure. LBGs and LAEs have been posited as critical populations that will meet the demanding requirements of cosmological studies in the era of large-area photometric surveys [11], particularly at redshifts $z\gtrsim2$, where (1) only methods based on \lya\ emission or Lyman break detection can be applied in large numbers and over large scales, (2) beyond the sampling of H-$\alpha$ emitters enabled by the Euclid and Roman space telescopes, and (3) where other tracers of cosmic density fluctuation such as luminous red galaxies and QSOs rapidly decline in number density, and become more difficult to efficiently select with low contamination rates.

For all the above science cases, the efficient sampling of large cosmic volumes is imperative in order to progress the field. Indeed, for $z\sim2-3$ LBGs, there is no instrument, current or planned, to pursue the science objectives above. While current and proposed large-area photometric surveys go some way toward meeting the requirements of these studies, KWFI is the only instrument that combines the necessary wide area, dpeth, and u-band sensitivity to deliver the required number of $z\sim2-3$ LBGs in the volumes necessary -- and in a reasonable timeframe. Moreover, the additional science that will be enabled by an optimised design in terms of area and efficiency should not be underestimated. With telescope resources under increasing demand, the highest return on investment will derive from the largest and most capable design that is optimised for throughput especially for partner institutions with small time allocations, for whom many projects would be not viable with a less efficient instrument.

Accordingly, representatives of LBG science are strongly in favor of the full 1 degree diameter FOV design for KWFI.  We note that not only is the smaller FOV design less capable due to its smaller area (less than half), it is less efficient by an even larger factor due to the tiling and overlap requirements that are attendant on any instrument configuration, but which are exacerbated as the FOV is reduced.

\hd{The big questions}

{\it Big Questions 1}

How can we better connect observed galaxies to halos in cosmological simulations to better understand galaxy formation and evolution on a galaxy-by-galaxy basis? How do high redshift galaxies connect to their low-redshift counterparts? How do galaxy spectroscopic properties differ on large scales in different environments? At what point does the galaxy luminosity function turn over at high redshift?

{\it Big Questions 2}

When did the framework for the morphology-density relation observed at low redshift begin? How did massive elliptical galaxies form and when? When did spiral galaxies form their bulges and their discs? How does environment affect the ultimate morphology of a galaxy?

\hd{What KWFI can do for this science case}

Broadband filters, the sensitivity, and wide field of KWFI are all that is needed to achieve this science. 

Deep u-band is absolutely essential. Work at z $\sim$ 2 and z $\sim$ 3 cannot be done without it.

Imaging over many square degrees is necessary, thus, the widest field possible for KWFI is preferred. For the smallest, focused programs, $\sim$1 deg$^2$ is needed to accommodate investigations of the observations with cosmological simulations.

KWFI is the only wide-field imager able to image 1--2 magnitudes deeper in u-band than other filters (necessary to measure the drop-out feature), to fainter magnitudes than what has been achieve to date, and to reach the faint-end of the luminosity functions over large areas to m $\sim$ 28--29 in reasonable timeframes.

\hd{References}

{\bf [1]} Steidel et al. 1998, ApJ, 492, 428;
{\bf [2]} Steidel et al. 2003, ApJ, 592, 728;
{\bf [3]} Cooke 2008, ApJL, 704, 62;
{\bf [4]} Foran et al. 2021, {\it submitted};
{\bf [5]} Shapley et al. 2003, ApJ 588, 65;
{\bf [6]} Law et al. 2007, ApJ, 656, 1;
{\bf [7]} Law et al. 2012, ApJ, 745, 85;
{\bf [8]} Foran et al. 2021, {\it submitted};
{\bf [9]} Cooke et al. 2013, MNRAS, 433, 2122;
{\bf [10]} Cooke et al. 2014, MNRAS, 441, 837;
{\bf [11]} Wilson \& White 2019, JCAP, 10,15;
{\bf [12]} Steidel et al. 2004, ApJ, 604, 534;
{\bf [13]} Reddy et al. 2005, 633, 748;
{\bf [14]} Marchesini et al. 2007, 656, 42;
{\bf [15]} Reddy et al. 2009, ApJ, 692, 778

\clearpage

\hd{KWFI Science Cases}

\begin{center}
{\bf \Large High-Energy Neutrinos and VHE Gamma-Rays}
\end{center}

\hd{Contributing author}

Fabian Sch\"ussler (IRFU, CEA Paris-Saclay)\\

\hd{Executive Summary}

KWFI will be a game changer in the field of high-energy neutrino, gamma ray, and UHECR source searches. Its unique capabilities to rapidly provide deep observations that cover the sizeable localization uncertainties of high-energy neutrinos detected by current and future observatories with a single pointing will revolutionize the field. Reactions to detections of very-high-energy gamma-ray transient phenomena will similarly allow emission region pinpointing and provide important and complementary information.

The localization and characterization of counterparts will enable global multi-wavelength and multi-messenger campaigns with unprecedented rapidity and efficiency. KWFI will thus allow for powerful synergies with next-generation neutrino telescopes like IceCube-Gen2 and KM3NeT as well as with the Cherenkov Telescope Array (CTA), the global, next-generation gamma-ray observatory.

\bigskip

\hd{Background}

Ultra-high energy cosmic rays (UHECRs), particles with energies up to several $10^{20}~\mathrm{eV}$, have puzzled mankind since their detection over 100 years ago. Tremendous instrumental efforts have allowed us to obtain increasingly high-resolution observations of the average UHECR properties (e.g., energy spectrum, mass composition, large scale anisotropies, etc.). Nevertheless, the sources and acceleration sites that are able to achieve the phenomenal observed energies, orders of magnitude above the largest man-made accelerators, remain elusive. 
 
Over recent years it has become more and more obvious that multiple messengers and novel techniques are needed to achieve this task. The detection of high-energy neutrinos is now opening a new window into the high-energy universe and may provide the long-sought breakthroughs. Combined with the unique capabilities of deep observations by KWFI, crucial and very complementary information will be obtained about the most violent phenomena in the universe. High-energy neutrinos are directly linked to hadronic interactions of UHECRs in or around their acceleration sites and therefore provide vital information. As weakly interacting particles, neutrinos can escape the densest astrophysical sources and travel long distances through the various radiation fields. On the other hand their detection is difficult and, to date, only a small number of individual high-energy neutrino events are tentatively associated with multi-wavelength counterparts and astrophysical objects. 

The most prominent is the blazar TXS 0506+056 that could be linked at the 3-sigma level to the high-energy neutrino IceCube-170922A during a several week long flaring period in 2017 [1]. Correlation studies like these still leave many open questions. For example, none of the proposed modeling attempts of TXS 0506+056 are able to reproduce the full multi-wavelength and multi-messenger emission [2,3]. Additional observations like the increased neutrino emission observed by IceCube in 2014/2015 from the same direction, that was not accompanied by an equally strong flare in other wavelengths, only increases the mystery [4]. Other high-energy neutrino events have been claimed to be associated with tidal disruption events [5,6]. Further observations and increased rapidity and efficiency in the detection and classification of potential multi-wavelength counterparts to detected neutrino events are clearly necessary.

Similarly, ever more sensitive and optimized observations of gamma-rays in the GeV and TeV energy range have provided significant insights into the underlying phenomena. Space-borne instruments like Fermi-LAT detect and monitor an increasing list of Galactic and extragalactic high-energy sources. Ground-based imaging air Cherenkov telescopes like H.E.S.S./MAGIC/VERITAS and the future Cherenkov Telescope Array (CTA) are providing complementary, high-resolution observations at the highest photon energies. Major and very recent breakthroughs in the study of transient very-high-energy phenomena include the detection of TeV emission from gamma-ray bursts [7--9] and the identification of recurrent novae as TeV emitters [10]. With the continued observations of the current instruments and especially the start of CTA, new surprises and detections of (transient) sources are to be expected. While the angular resolution of CTA may approach a few arcmin at the highest energies and for the strongest sources, the larger localization uncertainties of new, weak or rapidly fading sources do not typically allow time for crucial spectroscopic observations. 

\hd{The big questions}

{\it Big Question 1 -}
What are the sources of UHECRs? Exploiting the direct link between the acceleration sites of UHECRs and the production of high-energy neutrinos: what are the astrophysical sources of high-energy neutrinos?

{\it Big Question 2 -}
What processes are able to accelerate particles to energies orders of magnitude above man-made accelerators? How are the multi-messenger and multi-wavelength emissions (up to TeV energies) linked to each other? 

\hd{What KWFI can do for this science case}

KWFI will overlap with the operation of the next generation neutrino telescopes IceCube-Gen2 and KM3NeT and thus create exciting synergies for joint multi-messenger studies. Similar synergies are expected with CTA, the next generation VHE gamma-ray observatory.

Deep and rapid observations of the neutrino localization regions is a very promising avenue to resolve the UHECR/high-energy neutrino puzzle: the large FoV of KWFI often covers the entirety of the localization uncertainty region of high-energy neutrino events in a single observation, eliminating the need for tiling and enabling increased depth and faster identification. In addition, the fast readout and filter exchange enables fast-cadenced, deep observations with color information to record the event evolution for classification. The rapid detection of (possibly transient) sources will enable rapid triggered observations and will optimize extended, global monitoring campaigns that study source characteristics like flux and spectral evolution, redshifts, etc. Moreover, with a deployable secondary mirror and fast KWFI data processing, rapid localization and rapid deep spectroscopic follow up is possible with Keck instruments, providing a next-generation capability of speed, depth, and spectral information.

Detection of optical counterparts, afterglows, etc. and identification and characterization of their host galaxies (if applicable) is also crucial for our understanding and characterization of sources detected in VHE gamma-rays. The rapid and precise localization of the source will enable an increase in the monitoring and spectral coverage of the detected phenomena and, therefore, play a fundamental role in their interpretation. 

The deployable secondary mirror will allow, in addition to the fundamental determination of redshifts, the capability to classify the detected transients with unprecedented rapidity and depth, thus enabling a paradigm shift in the search for high-energy neutrinos sources and UHECRs as well as the study of VHE phenomena.

\hd{References}

{\bf [1]} IceCube collaboration, 2018, Science, 361, 1378;
{\bf [2]} Keivani, A., et al. 2018, ApJ, 864, 84;
{\bf [3]} Cerruti, M., et al. 2019, MNRAS, 483, 12;
{\bf [4]} IceCube collaboration, 2018, Science, 361, 147;
{\bf [5]} Stein, R., et al., 2021, Nature Astronomy, 5, 510;
{\bf [6]} Reusch, S,, et al., 2022, PRL, 128, 221101;
{\bf [7]} Abdalla, H., et al. 2019, Nature, 575, 464;
{\bf [8]} MAGIC collaboration, 2019, Nature, 575, 455;
{\bf [9]} H.E.S.S. collaboration, 2021, Science, 372, 1081;
{\bf [10]} H.E.S.S. collaboration, 2022, Science, 376, 77

\clearpage

\hd{KWFI Science Case}

\begin{center}
{\bf \Large High-Redshift Supernovae and Superluminous Supernovae}
\end{center}

\hd{Contributing authors}

Jeff Cooke (Swinburne)\\
Anais M\"{o}ller (Swinburne) \\
Charlotte Angus (NBI)\\ 

\hd{Executive Summary}

Core-collapse supernovae (CCSNe) and superluminous supernovae (SLSNe) have been discovered at $z\sim$ 2--4 in increasing numbers and SLSNe can be detected to $z\sim$ 13--20 with upcoming facilities (e.g., the Roman and Euclid space telescopes, JWST, and Tokyo Atacama Observatory (TAO) equipped with SWIMS [1--4]). High-redshift time dilation and their intrinsic moderate to very slow evolution, permits classically scheduled, i.e., non-Target of Opportunity, Keck, 30m-class, and JWST spectroscopy. Their detection, rest-frame UV light curves, and follow-up spectra are key to understanding models of their explosion physics, pre-expelled circumstellar material, their progenitors, rates, and evolution with redshift. SLSNe in particular have great utility. They are brighter than their host galaxies and are one of our only means to probe the $z\gtrsim10$ Universe, providing temporary beacons through their faint host proto-galaxies and IGM. SLSNe trace star formation to very high redshift, place direct constraints on the high-mass end of the stellar initial mass function, and provide our best means to detect pair-instability supernovae and Population III star deaths. Finally, SLSNe show promise as UV and optical standardizable candles from  $z\sim$ 0--20 [5--7]. Deep, wide-field optical imaging is necessary to search for and study $z\gtrsim$ 2--7 SLSNe. Understanding the UV behavior of $z\sim$ 2--7 supernovae is essential to identify events at higher redshifts, where only the UV is accessible, even with JWST mid-IR. Due to their extreme distance, comparable rarity, and time dilation, CCSNe and SLSNe require very deep (m $\gtrsim$ 27) imaging over wide fields once per 1--2 months for several years. Only KWFI is sufficiently sensitive to make such a program feasible and to achieve this science.

\hd{Background and detection}

\begin{figure}[!h]
\scalebox{0.98}[0.98]{\includegraphics{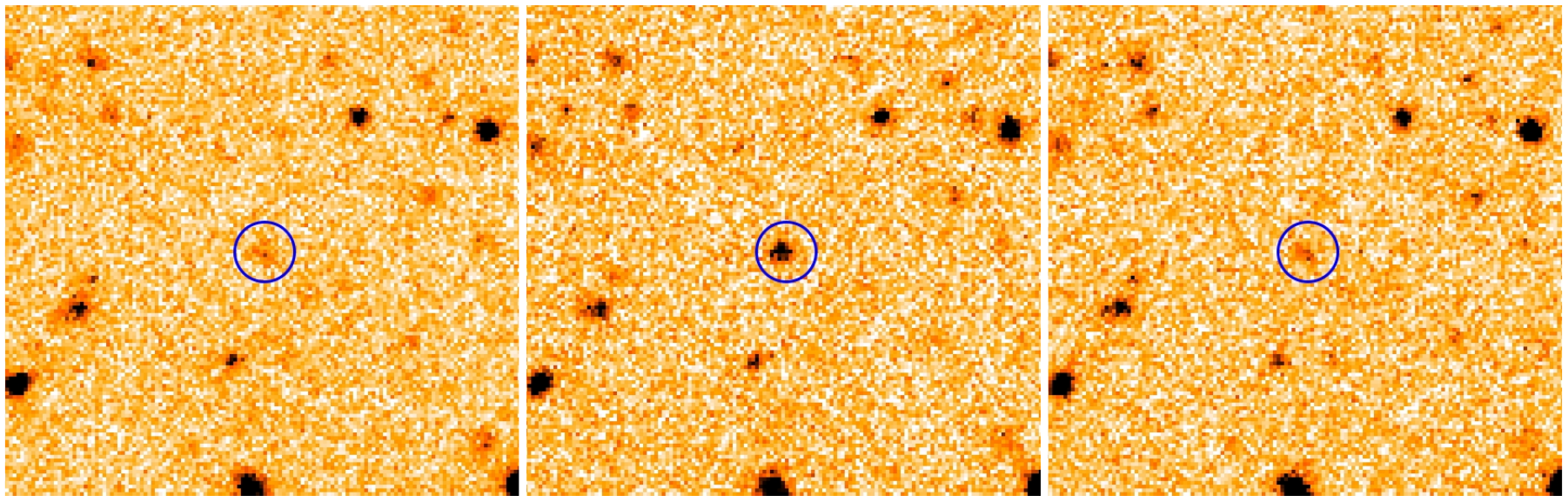}}
\caption{\small Supernova 235017 at z $\sim$ 2 detected using the host galaxy monitoring method [1]. Images (20" $\times$ 20") are centered on the host galaxy denoted by the blue circle. Each image is a `seasonal stack' made by stacking $\sim$25--30 nights of CFHT (3.6m) nightly images taken over the 6-month visibility of the field for 3 consecutive years. The supernova is seen in the center image (second year) and is brighter than the faint, low luminosity host galaxy.}
\label{SNimages}
\end{figure}

\begin{figure}[!h]
\scalebox{1.0}[1.0]{\includegraphics{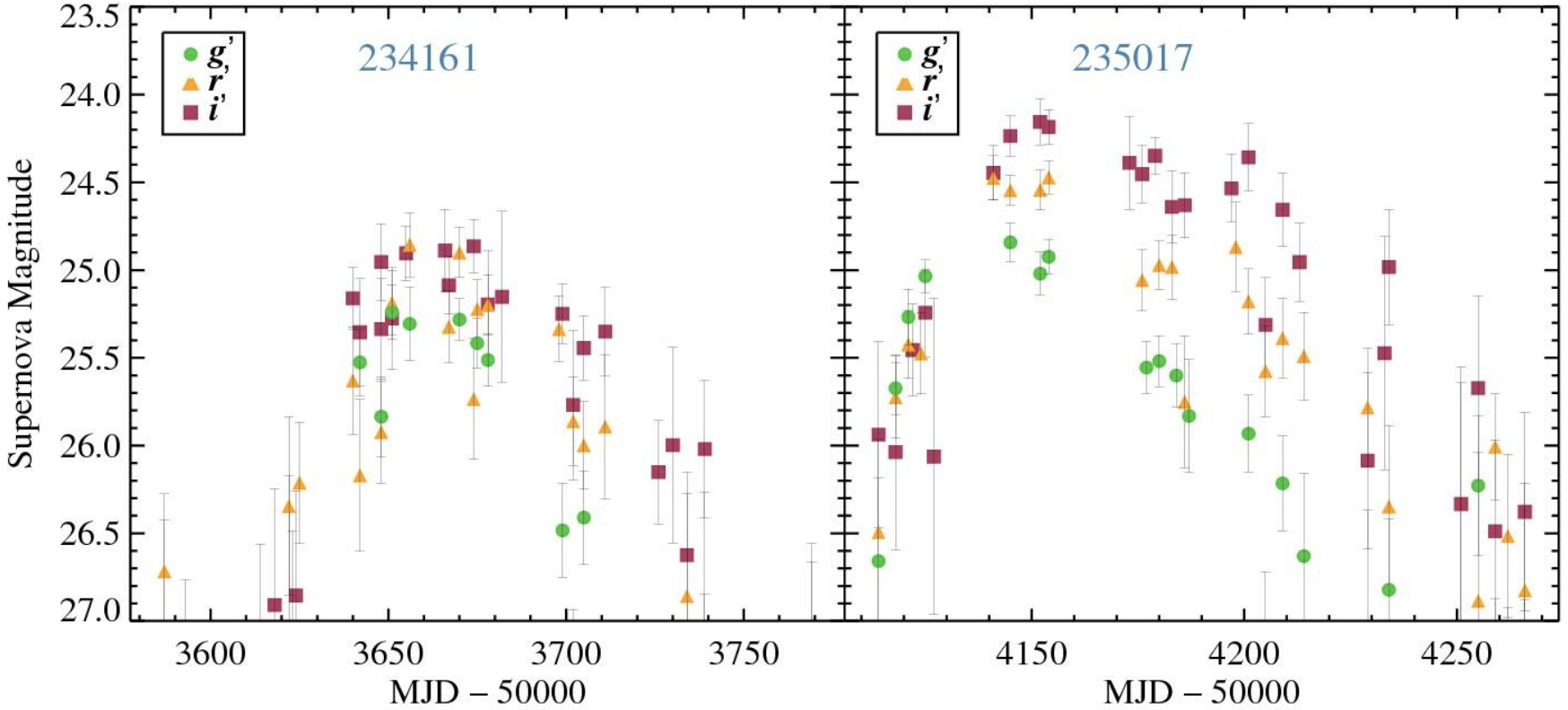}}
\caption{\small Core-collapse supernovae (CCSNe) at z $\sim$ 2 [1]. Shown are the rise and fade as observed in the g, r, and i filters that probe roughly 1600\AA, 2100\AA, and 2500\AA, restframe. These data are reliable to m $\sim$ 26. CCSNe can be quite luminous in the UV, as evidenced here, with peak magnitudes of M$_{UV}$ $\sim$ $-$19.8 and $-$20.5 at 2500\AA\ (i-band).  These events probe the tip of the supernova luminosity distributions that range from $\sim$ $-$15 to $-$20, peaking near $\sim$ $-$17 to $-$18.5, depending on supernova type. Per-epoch imaging to m $\sim$ 28 will reach to M $\sim$ $-$16.5 to get the bulk of supernovae at z $\sim$ 2 and to M $\sim$ $-$18.5 at z $\sim$ 6.}
\label{CCSNe}
\end{figure}

\begin{figure}[!h]
\scalebox{0.91}[0.94]{\includegraphics{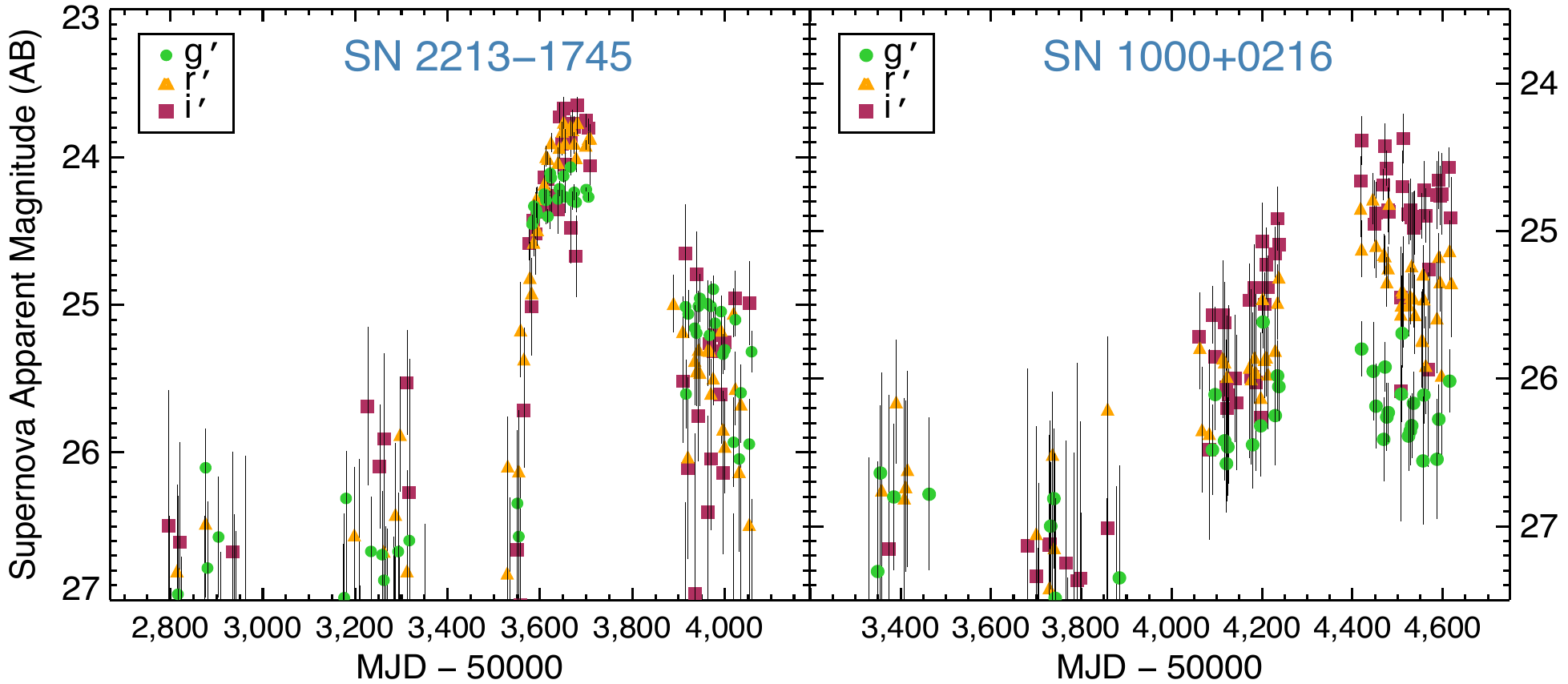}}
\caption{\small Superluminous supernovae (SLSNe) at z $\sim$ 2 (left) and z $\sim$ 4 (right). SLSNe evolve very slowly, and with time dilation, extend over multiple years of observation. Plotted here are four years of data taken over four 6-month periods when the field was visible. For both cases, the events were not detected in the first two years (limiting magnitude m $\sim$ 25.9, 26.4, left to right), and the burst occurred at the start (or prior to the start) of the third year's observations. The z = 2.05 SLSN light curve in the left panel spans two years (observed-frame) and the gri filters correspond to $\sim$1600\AA, 2100\AA, and 2500\AA\ restframe, respectively. The z = 3.90 SLSN light curve in the right panel likely spans beyond 2 years (observed-frame) and the gri filters correspond to $\sim$1000\AA, 1300\AA, and 1600\AA\ restframe, respectively. At z $\sim$ 4, the g-band probes the Lyman-$\alpha$ forest as is reflected by the decrement in flux in the g-band compared to the other filters. This SLSN an example of a class that exhibits evidence for a pre-peak `hump' [10--12], as mentioned in the text, with detections for this SLSN occurring near MJD 54100.}
\label{SLSNe}
\end{figure}

Most core-collapse supernovae (CCSNe), such as Types II-P, Ic, and IIn supernovae, and all superluminous supernovae (SLSNe) are UV luminous. SLSNe are 10--100$\times$ more luminous than `normal' supernovae and have peak absolute magnitudes brighter than M $\sim$ $-$21, with some recent classes extending down to M $\sim$ $-$20.5 [8,9]. Deep ugriz imaging is needed for detection at z $ \sim$ 2--6 and to perform efficient host galaxy monitoring [1,2] (Figure~\ref{SNimages}). Due to their comparative rarity, wide-fields are essential. In the UV, high-redshift CCSNe rise $\sim$2 mags in $\sim$10--20\,d in their restframe and fade in $\sim$20--30\,d (Figure~\ref{CCSNe}). SLSNe evolve much slower, rising in $\sim$50\,d in their restframe and fading in $\sim$150\,d in the UV (Figure~\ref{SLSNe}), including the interesting enigmatic class with pre-peak `humps' [2,10--12].  With time dilation, observations for high redshift supernovae need only occur roughly once per month, depending on redshift probed. 

Previous searches to m $\sim$ 25--26 per epoch using the CFHT Legacy Survey Deep fields and the Hyper-SuprimeCam Subaru Strategic Program [1--4] only probe the tip of the CCSN luminosity functions and only partially into the SLSN luminosity function (Figures~\ref{CCSNe} \&~\ref{SLSNe}). As a result, imaging  to m $\sim$ 27--28 per epoch (0.5--2\,hr with KWFI) is needed over $>$ 2 deg$^2$ for 4--5\,yr to reach sufficiently into the SLSN luminosity functions at z $>$ 2 to sample the different populations, as probing deeper into these luminosity functions rapidly increases the number of detectable events. Such surveys will detect statistically significant numbers and enable SLSN detections to z $\sim$ 6 and these data will also detect z $\sim$ 2--4 CCSNe. This observational strategy equates to 2 nights per observing epoch (roughly a month) over each 6-month/year field visibility using the 1 degree diameter KWFI design and 4 nights per epoch using the 0.707 degree diameter design for 4--5\,yr. As these observations require dark/gray nights for deep u- and g-band imaging, the 0.707 degree diameter design becomes difficult to regularly schedule. 

{\bf Extending to higher redshifts -} Historically, CCSNe and SLSNe are classified using their restframe optical spectra. Little is known about their UV behavior and spectral features. KWFI detections with JWST and 30m-class telescope follow-up spectroscopy will connect the restframe UV behavior of each SN type with their rest-frame optical classifications. This connection is vital to identify and understand higher redshift events where only the far-UV is accessible with Roman and JWST (e.g., 1800\AA\ is $\sim$3 $\mu$m at z $\sim$ 15). Some SLSNe can be classified by their photometry alone. These events have pre-peak humps [e.g., 2,10--12] that may enable classification, once confirmed via spectroscopy. Deep wide-field optical imaging with KWFI is the only path to enable Roman, JWST, and 30-class telescopes to identify and understand SLSNe to z $\sim$ 7--20 and exploit their utility.

\hd{The big questions}

{\it Big Questions 1}

How do the rates of various types of CCSNe and SLSNe evolve over cosmic time? Is there an evolution of core-collapse diversity and properties throughout cosmic time?

{\it Big Questions 2}

Does the high-mass end of the stellar initial mass function (IMF) evolve over cosmic time? Can counting the number of SLSNe in galaxies with measured star formation rates selected in well-defined volumes, as is done with the host galaxy monitoring method, answer this question?

{\it Big Questions 3}

What is the metallicity evolution of supernova host galaxies at z $\sim$ 2--20? What does the CGM and IGM in the line-of-sight to galaxies at z $>$ 10 look like?

{\it Big Questions 4}

Can SLSNe be used as standardizable candles and probe dark energy and the expansion rate of the Universe to z $\sim$ 20 through deceleration? Do pair-instability supernovae exist? Are they found in large numbers at very high redshifts as predicted?

\hd{What KWFI can do for this science case}

CCSNe and SLSNe can be detected conventionally by subtracting the most current science image with an image of the field taken in the past to search for flux changes indicating the presence of a new source. This method is efficient for relatively faster evolving events, such as supernovae in the local Universe, However, for high redshift supernovae, an efficient method is to pre-select high redshift galaxies, e.g., Lyman break galaxies that can be color-selected at well-defined redshifts, and monitor them using `seasonal stacked' images. These stacks are images combined from 6-months of data for maximum depth to search for the very faint flux from high-redshift supernovae (they evolve slowly over the season), and to help eliminate the `fog' of lower redshift transient in the deep images necessary to detect the high-z events. Supernova candidates are then investigated in the nightly data for their behavior. As the events have slow evolution and time dilation, they can still have spectra taken months after burst, observed-frame.

Using the high redshift LBG host galaxy selection and monitoring method, deep, wide-field u-band imaging ($\sim$1--2 mags deeper than griz) is absolutely essential to select z $\sim$ 1.5--3.5 to detect the Lyman limit break, or `drop out', and to eliminate lower redshift galaxies with similar spectral energy distributions. Similarly, z $\sim$ 3.5--4.5 hosts require deeper (by $\sim$1--2 mag) g-band compared to riz; z $\sim$ 4.5--5.5 hosts require deeper r-band compared to iz; and z $\sim$ 5.5--6.5 require deeper i-band compered to z. Each of these targeted redshifts requires deep shorter wavelength broadband imaging shortward of the Lyman limit (for z $\sim$ 1.5--4.5 and Ly-$\alpha$ (1216\AA\ for for z $\gtrsim$ 4.5) to confirm the `drop-out' nature of the targeted redshift galaxies in which supernovae will be searched. 

Given the depths required, wide-field searches are needed to detect the relative rare SLSN and CCSNe (a few to a few dozen, respectively, per square degree per year to m $\sim$ 27) and to sample the different supernova types and their behaviors. The UV photometirc behavior and spectra and then be tied to the restframe optical classification for future z $>$ 7 supernova detections. 

Finally, although CCSNe and SLSNe evolve relatively slowly, their identification and rapid follow up spectroscopy yields crucial information on their progenitors, explosion physics, cicumstellar and ejected material, and for classification [13]. In addition, Type II-P supernovae rise very quickly (as little as hours), in particular in the rest-frame UV. As such, the deployable secondary mirror and the fast KWFI data processing would provide a game-changing capability to obtain these deep spectra within minutes using Keck instruments.

\hd{References}

{\bf [1]} Cooke et al. 2009, Nature, 460, 237;
{\bf [2]} Cooke et al. 2012, Nature, 491, 229;
{\bf [3]} Moriya et al. 2019, ApJS, 241, 16;
{\bf [4]} Curtin et al. 2019, ApJS, 241, 17;
{\bf [5]} Inserra et al. 2014, ApJ, 796, 87;
{\bf [6]} Inserra et al. 2021; MNRAS, 504, 2535;
{\bf [7]} Khetan et al. 2021, MNRAS, {\it submitted};
{\bf [8]} Gal-Yam 2012, Science, 337, 927;
{\bf [9]} Quimby et al. 2018, ApJ, 855, 2;
{\bf [10]} Leloudas et al. 2012, A\&A, 541, 129;
{\bf [11]} Nicholl et al. 2015, ApJL, 807, L18;
{\bf [12]} Angus et al. 2019, MNRAS, 487, 2215;
{\bf [13]} Gal-Yam et al. 2014, Nature, 509, 471

\clearpage

\hd{KWFI Science Case}

\begin{center}
  {\bf \Large The Circumgalactic Medium}
\end{center}

\hd{Contributing authors}

Joseph N. Burchett (New Mexico State University)\\
Glenn G. Kacprzak (Swinburne University of Technology)

\hd{Executive Summary}

Galaxy formation and evolution is primarily driven by their gas reservoirs and their environment. Quasar absorption lines have been able to detect the presence and evolution of the circumgalactic gas (CGM), yet how this gas directly relates to their galaxies remains unknown. This is due in large part to the fact that quasar absorption spectroscopy affords detections in only `pencil beam' apertures, and we must infer trends across galaxy populations in a statistical sense by acquiring many QSO-galaxy pairs.  The only way to accurately quantify the relation between gas and their galaxies is to directly image and map the gas.

\hd{Background}

\begin{figure}[!h]
\begin{center} 
 \scalebox{1.27}[1.27]{\includegraphics{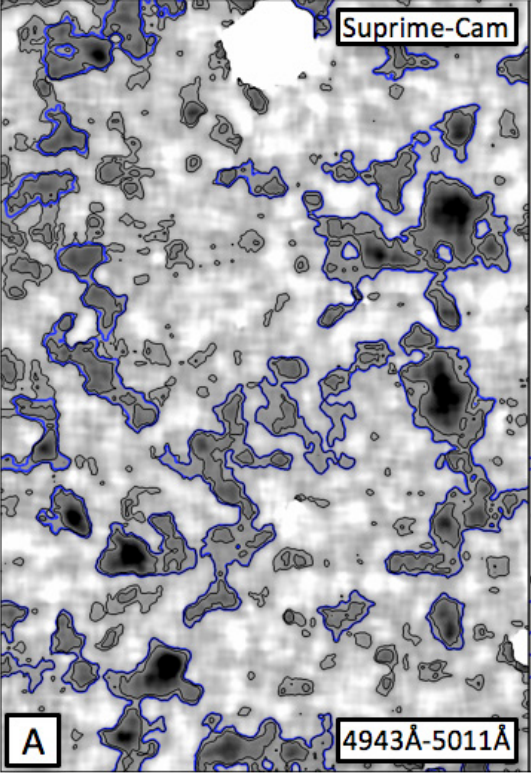}}  
\caption{\small Cosmic web filaments detected in Ly$\alpha$ at $z=3.09$ [1] with the Subaru Suprime-Cam. These structures were followed up by MUSE to constrain gas kinematics in addition to the large scale distribution. Note the wavelength bandpass (4943 \AA--5011 \AA) indicated on the image.  KWFI could image such structures at $z=1.5$, where the Subaru Hyper-SuprimeCam has no sensitivity, opening a new window in the diffuse universe at cosmic noon and enjoying a factor of 6.5$\times$ boost due to cosmological surface brightness dimming alone. }  
\label{umehata}
\end{center} 
\end{figure}

KWFI will enable unprecedented capabilities for studying the circumgalactic medium (CGM).  Deep, wide-field narrowband imaging can map diffuse emission from the CGM itself (Figure~\ref{umehata}). This emission can be mapped in \lya\ as well as metal-line diagnostics of $10^{4-6}$~K gas such as MgII and OVI in redshift regimes z $=$ 1.5--3.0 for \lya, z $=$ 0.1--1.0 for MgII (accessible in H\textsc{i} with radio) and z $=$ 2.0--3.5 for OVI, inaccessible with any other instrument, and can {\it explore galaxy environments over scales of tens of Mpc for each pointing}. Seeing limited observations of z $=$ 2 galaxies will also enable the mapping of CGM around individual galaxies at a resolution of a few kpc. The same filters that target UV lines at $z>2$ can be used to observe [OII] in the Milky Way and local Universe galaxies.  The deep KWFI imaging enables complementary DEIMOS/LRIS/FOBOS spectroscopy of the CGM emitting galaxies and followup kinematic studies of the CGM and galaxies using KCWI or FOBOS IFU bundles.

\hd{The big questions}

{\it Big Question 1}

How does the CGM map along cosmic structures and how does CGM abundance vary as a function of galaxy populations at cosmic noon?

{\it Big Question 2}

At the peak of cosmic star formation, how does the circumgalactic environment react to stellar/AGN feedback and regulate star formation?

{\it Big Question 3}

What is the prevalence of large-scale outflows? 

\hd{What KWFI can do for this science case?}

\begin{figure}[!h]
\begin{center} 
 \scalebox{0.65}[0.65]{\includegraphics{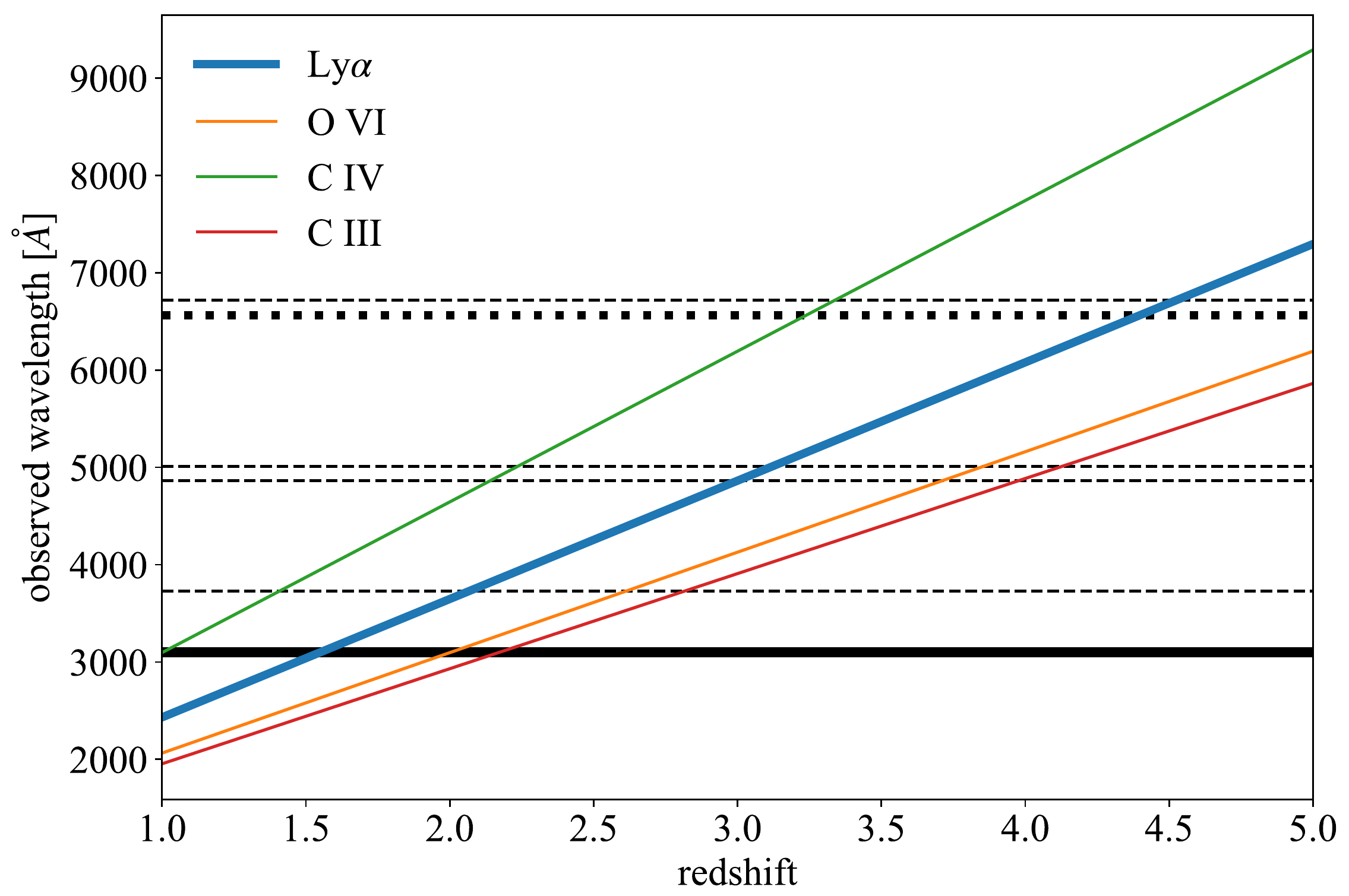}}  
\caption{\small Redshift coverage of key rest-frame UV diffuse gas diagnostics (see legend).  The bold horizontal line indicates the nominal blue limit of KWFI.  The dashed lines indicate standard emission line: such [O II] 3727, H$\beta$ 4862, [O III] 5007, and [S II] 6718.  The dotted line represents H$\alpha$.  This demonstrates the power of a even a standard narrowband filter suite to map these gas tracers across a a wide span of cosmic time. }
\label{redshiftCoverage}
\end{center} 
\end{figure}

The primary requirements for CGM science is narrow-band filters. These filters could be designed in tandem with narrow-band filters planned to map nebular emission lines and low redshift. Figure~\ref{redshiftCoverage} shows redshift as a function of observed wavelength for four common CGM emission emission lines. The horizontal dashed lines indicate the wavelengths of [O II], H$\beta$, [O III], H$\alpha$ and [S II] $z=0$ emission lines. By using co-designed filters, KWFI will map the CGM over a range of CGM emission lines, while covering the peak epoch of star formation at cosmic noon. Additional narrow-band filters could be designed to target specific redshifts or known comic over-densities. The FWHM width of these filters should be narrow and could range from 100~{\AA} to around 10~{\AA}.

\hd{References}

{\bf [1]} Umehata, H., et al. 2019, Science, 366, 97

\clearpage

\hd{KWFI Science Case}

\begin{center}
  {\bf \Large \lya\ Emitters in the Redshift Gap}
\end{center}

\hd{Contributing author}

Isak Wold (NASA/GSFC)

\hd{Executive Summary}

Lyman alpha (\lya) emission is a primary tool for studying the high-redshift universe and one of the very few constraints on the reionization period. However, to use samples of \lya\ emitters (LAEs) as a probe of the high-redshift Universe, we must understand the intrinsic properties that facilitate the escape of \lya\ emission and how LAEs are drawn from the overall galaxy population. It is difficult to address these questions at high redshift due to the opacity of the intergalactic medium (IGM), cosmological dimming, and the cost of spectroscopic followup. To make progress, we can conduct detailed studies of lower-redshift samples and examine how they evolve with redshift. However, there is currently a 7 Gyr gap between low-redshift LAEs at $z=0.3$ and high-redshift LAEs at $z>2$ due to instrumentation and atmospheric constraints. KWFI’s sensitivity below 3700\AA\ offers a unique opportunity to bridge the gap from low to high-redshift LAE studies by observing $z\sim$ 1.5--2 LAEs using narrowband imaging.
\bigskip

\hd{Background}

\lya\ luminosity functions (LFs) and their evolution can be used to constrain the timing of reionization because \lya\ emission is resonantly scattered by any neutral hydrogen that it encounters making it very sensitive to the ionization state of the IGM. Beyond a redshift of $z\sim6$, many studies have found that \lya\ LFs decline (e.g., Konno et al.\ 2014; Inoue et al.\ 2018) at a rate exceeding the decline expected from the parent population of star-forming galaxies  (e.g., Finkelstein et al.\ 2015; Bouwens et al.\ 2015). This decline may arise from the increasing opacity of the $z\sim7$ IGM and the onset of the reionization epoch. However, theoretical studies have suggested that the evolving properties of LAEs may also play an important role (e.g., Mesinger et al.\ 2015). To isolate the effects of the IGM, we would like to determine which physical properties govern \lya\ emission and how this evolves with redshift. 

Given the difficulties of studying LAEs at the highest redshifts, many local LAE studies have been conducted.  While these $z\sim0$ LAE studies have provided insight into the physical conditions that facilitate strong \lya\ emission (e.g., Hayes et al.\ 2013; Ostlin et al.\ 2014; Rivera-Thorsen et al.\ 2015; Alexandroff et al.\ 2015; Henry et al.\ 2015; Izotov et al.\ 2016; Runnholm et al.\ 2020; Kim et al.\ 2021), it is very difficult to make statistical comparisons to high-redshift LAE populations because – unlike the high-redshift samples – the $z\sim0$ studies have not been selected based solely on their \lya\ emission. There is not currently a survey instrument capable of observing a large number of $z\sim0$ LAEs. Thus, local LAEs are typically pre-selected from identified high equivalent width H$\alpha$ emitters, compact [OIII] emitters, or ultraviolet-luminous galaxies and subsequently observed with the {\it Hubble Space Telescope} ({\it HST}) to investigate the existence of \lya\ emission.

\begin{figure}[!h]
\begin{center} 
\scalebox{0.55}[0.55]{\includegraphics{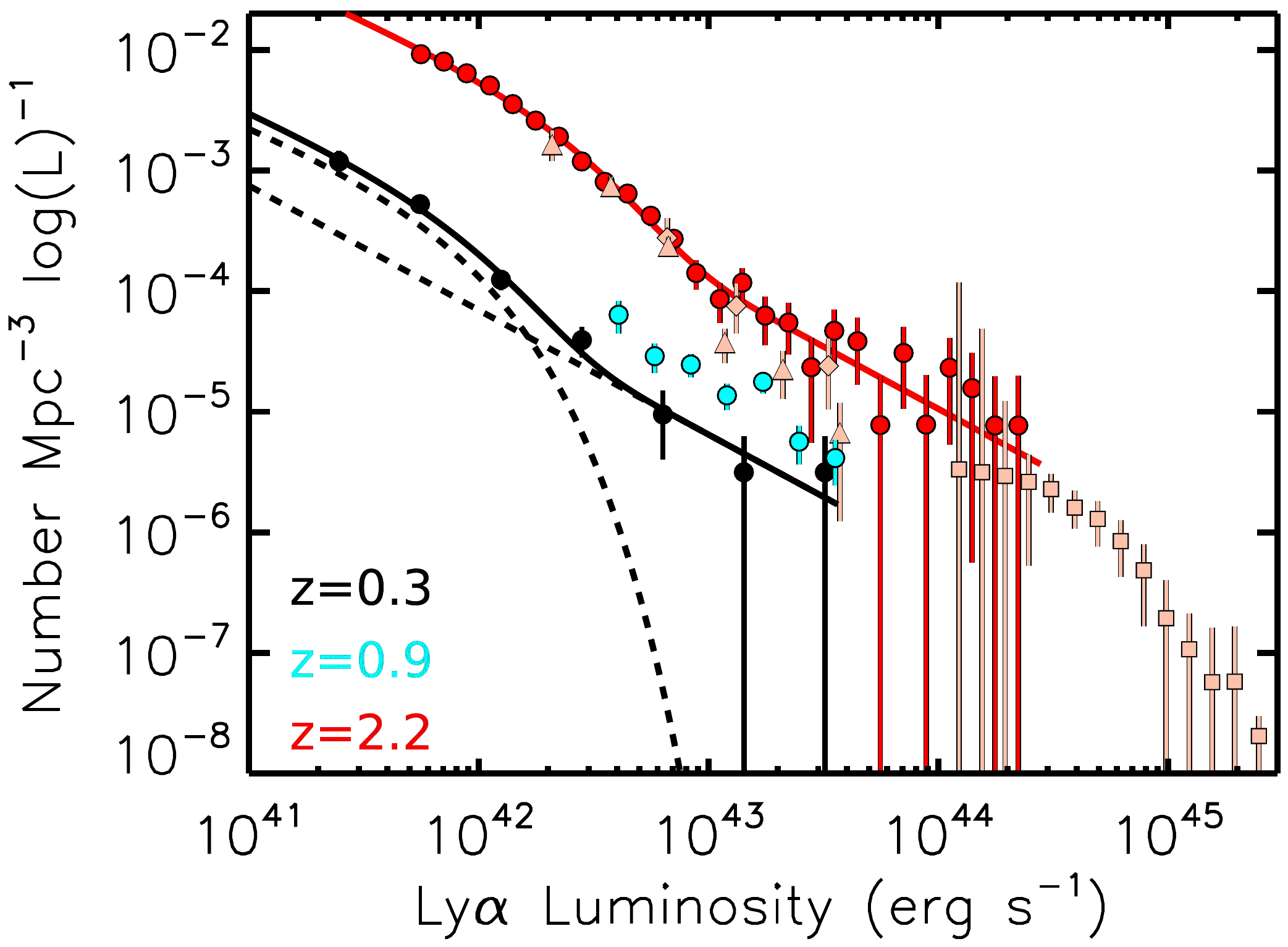}}
\vspace{-0.3cm}
\caption{\small Evolution of the Ly$\alpha$ luminosity functions (LFs) from $z=0.3$ to $z=2.2$ with best-fit Schechter function + AGN power law adopted from Wold et al.\ (2017). The black and cyan points show $z=0.3$ and $z=0.9$ LF data from Wold et al.\ (2014, 2017). The red points show the $z=2.2$ LF data from Konno et al.\ (2016; circles), Sobral et al.\ (2017; triangles), Matthee et al.\ (2017; diamonds), and Spinoso et al.\ (2020; squares). A modest KWFI narrowband survey could detect $z\sim1.5$ LAEs with Ly$\alpha$ luminosities greater than 10$^{41}$ erg s$^{-1}$ and easily probe a volume comparable to the $z=0.3$ and 2.2 surveys, offering a unique opportunity to study the LAE evolution over a 7 Gyr gap in our samples.}
\label{LAE_LF}
\end{center} 
\end{figure}

The now decommissioned {\it Galaxy Evolution Explorer} ({\it GALEX}) space telescope and its FUV/NUV grism capability have given us the only direct LAE surveys at the low redshifts (Barger et al.\ 2012; Wold et al.\ 2014, 2017). While at $z>2.2$, \lya\ emission is redshifted into optical bandpasses allowing for many ground-based surveys. In Figure~\ref{LAE_LF}, we show the current state-of-the-art $z\leq2.2$ LAE LFs.  Recent studies have shown that LAE populations transition from a Schechter function distribution dominated by star-forming galaxies (SFGs) to a power-law  distribution (with exponential cut-off observed at $z=2.2$) dominated by AGN as \lya\ luminosity increases. The current $z=0.9$ LAE survey lacks the sensitivity to probe the SFG regime, hampering our ability to study the evolution of LAEs.

\begin{figure}[!h]
\begin{center} 
\scalebox{0.65}[0.65]{\includegraphics{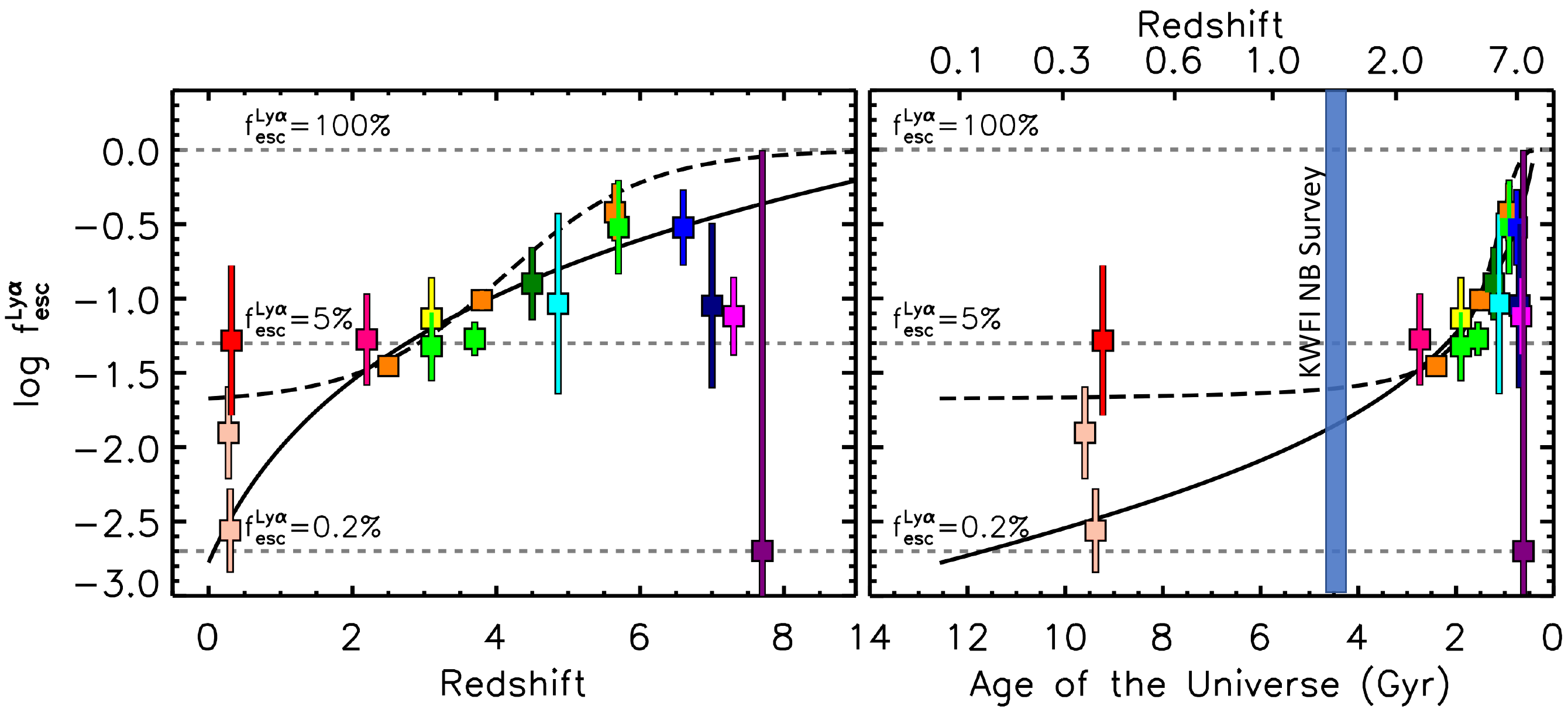}}
\caption{\small {\bf Left:} Evolution of the sample-averaged Ly$\alpha$ escape fraction adopted from Hayes et al.\ (2011) but with the latest $z=0.3$ constraint from Wold et al.\ (2017; red symbol). The two light red $z=0.3$ constraints were used by Hayes et al.\ (2011; solid curve) and Blanc et al.\ (2011; dashed curve) to help infer evolutionary trends, but are based on a continuum-selected {\it GALEX} sample (see Wold et al.\ for details). The decline in the Ly$\alpha$ escape fraction seen at the highest redshifts has been attributed to the increasing opacity of the IGM and the onset of the reionization epoch. However, the evolving properties of LAEs may also play an important role in the $z\sim7$ decline (Mesinger et al.\ 2015). To help understand the physical properties that govern \lya\ emission -- free from IGM effects -- many different groups have studied the $z<6$ evolution (e.g., Hayes et al.\ 2011; Blanc et al.\ 2011; Konno et al.\ 2016; Wold et al.\ 2017).  {\bf Right:} Same as the left but the x-axis is now cosmic time to emphasize the 7 Gyr gap between LAE samples. A KWFI $z\sim$ 1.5 LAE survey can help fill in this gap and can be used to determine whether the \lya\ escape fraction declines rapidly (consistent with the solid curve proposed by Hayes et al.\ 2011) or levels off (roughly consistent with the dashed curve proposed by Blanc et al.\ 2011) as the universe ages.}
\label{LAE_escape}
\end{center} 
\end{figure}

Regardless, multiple groups have used the evolution of LAE LFs to help identify the physical properties that regulate \lya\ emission.  Specifically, the evolution of the LAE LFs relative to the encompassing SFG LFs are used to measure the sample-averaged \lya\ escape fraction as a function of redshift (see Figure~\ref{LAE_escape}).  Some studies have suggested that the evolving dust extinction in the average SFG can explain the observed evolution of the \lya\ escape fraction (e.g., Hayes et al.\ 2011; Blanc et al.\ 2011), while other studies have emphasized the importance of the HI column density in suppressing \lya\ escape (e.g., Konno et al.\ 2016).  A major shortcoming of all these studies is a lack of \lya\ data for half the history of the universe, from $z=0.3$ to $z=2$.

KWFI’s ability to explore below 3700\AA\ offers a unique opportunity to help fill in this 7 Gyr gap by observing $z\sim$ 1.5--2 LAEs using narrowband imaging.  This redshift regime is where the star-forming properties of galaxies change very rapidly and where the star formation is near its peak. It is a key area for connecting local studies to the high-redshift Universe. Furthermore, a KWFI LAE survey will be extremely efficient given the instrument’s sensitivity, large field-of-view, and the lack of strong foreground emitters that plague high-redshift LAE surveys.  
\bigskip

\hd{The big questions}

How does the LAE luminosity function evolve over the current 7 Gyr gap in our LAE samples? What physical properties regulate \lya\ emission and how do they evolve with redshift?

A KWFI \lya\ survey can be used to directly address these questions.
\bigskip


\hd{What KWFI can do for this science case}

A KWFI narrowband survey with a very modest exposure time (5 hours of narrowband imaging + 2 hours of broadband imaging) could detect $z\sim1.5$ LAEs with luminosities greater than 10$^{41}$ erg s$^{-1}$.  A single $\sim$1 degree KWFI pointing would probe a survey volume comparable to the $z=0.3$ and $z=2.2$ LAE surveys ($\sim10^{6}$ cMpc$^{3}$).  Thus, we will have the depth to study the regime dominated by star-forming galaxies and the volume needed to study the SFG / AGN transition.  We estimate that this nominal single-pointing narrowband survey will detect $\sim$500 LAEs giving us the statistical leverage to perform clustering studies.

We note that there are existing H$\alpha$ surveys and LFs at $z=1.47$ (Sobral et al.\ 2013).  This favors a $z\sim1.5$ LAE survey because a robust comparison sample is available that will facilitate studies of the sample-averaged \lya\ escape fraction.

\clearpage

\hd{KWFI Science Case}

\begin{center}
  {\bf \Large Fast Radio Burst Counterparts}
\end{center}

\hd{Contributing author}

Jeff Cooke (Swinburne)

\hd{Executive Summary}

Fast radio bursts (FRBs) are extragalactic millisecond bursts in the radio (e.g., [1.2]) and, although several host galaxies have been localized [3,4], the cause of the bursts is unknown. Some FRBs have been observed to repeat and there may be several classes based on the observed properties. FRBs have been found to occur in star forming galaxies, passive galaxies, spiral arms, outside of galaxies and even in globular clusters (e.g., [5]), further obscuring their nature. Several theories of their origin and burst physics include mechanisms that produce optical emission, such as inverse Compton scattering processes, magnetar bursts, and blitzars. Only fast-cadenced images on a very large aperture telescope can detect these faint signals and test these models. Fast (down to millisecond) exposure and readout CMOS chips installed in the corners of KWFI can be used to monitor repeating or localized FRBs for optical signals to either confirm or rule out these theories and help uncover the nature of FRBs.


\hd{Background}

\begin{figure}[!h]
\vspace{-0.2cm}
\begin{center} 
\scalebox{0.71}[0.71]{\includegraphics{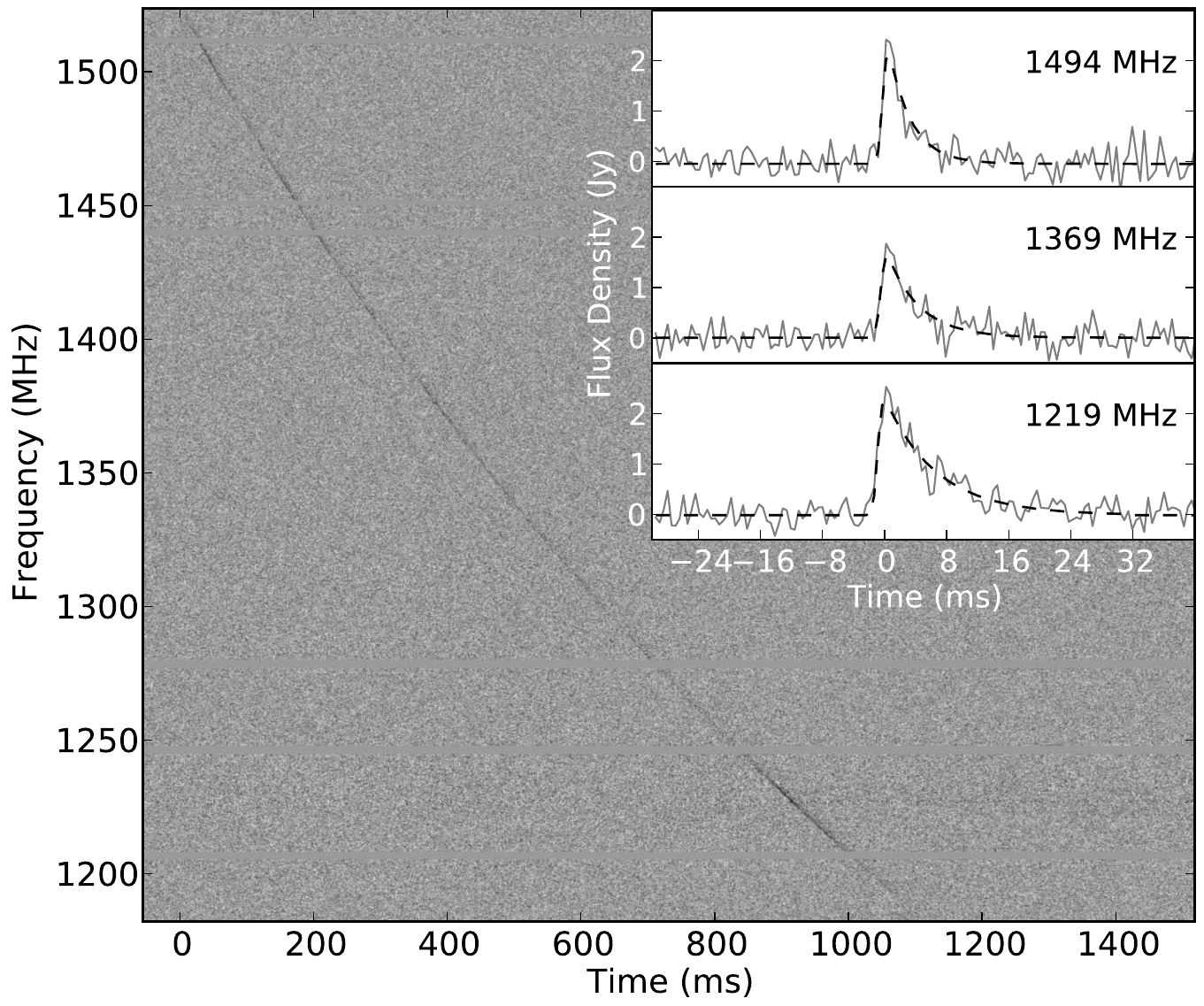}}
\vspace{-0.3cm}
\caption{\small A radio spectrum of FRB 110220 showing the frequency-dependent signal delay [1]. Higher frequencies (here, optical) arrive to Earth before the radio. The insets show dedispersed sub-bands illustrating the pulse profile.}
\label{frb}
\end{center} 
\end{figure}

Fast radio bursts (FRBs) were first discovered in 2007 [6] and since that time, over 1000 have been detected with facilities such as the Parkes and ASKAP radio telescopes in Australia and the CHIME radio telescope in Canada [e.g., [1,2,7]. A fraction of FRBs have been observed to repeat [8,9], with some having predictable periods. FRBs have been detected over a range of radio frequencies, but not at any other wavelength. However, the Galactic magnetar SGR193+2154 exhibited FRB-like radio bursts along with high-energy emission. The radio burst was a few orders of magnitude weaker than extragalactic FRBs, yet some FRBs may be magnetars. Given the diversity in FRB extragalacitc locations (e.g., in star forming galaxies, passive galaxies, spiral arms, outside of galaxies and even in globular clusters), it is likely that many FRBs are not magnetars. Finally, longer wavelength frequencies from FRBs arrive later than shorter frequencies as a result of traveling through ionized baryons on the way to Earth (Figure~\ref{frb}). This dispersion measure is used as our only direct means to measure the ionized baryon content of the Universe.

Several theories indicate optical emission from FRBs [e.g, 10--13]. If found to be correct, the nature of some FRBs would be identified. Depending on the duration of the optical bursts, rapid spectroscopic follow up could better understand the physics behind FRBs. If the optical emission is coherent with the radio burst, it will arrive to Earth seconds {\bf before} the radio emission. Thus the optical telescope needs to be on target before the FRB is detected in the radio (Figure~\ref{frb}). 

{\bf Millisecond optical bursts -} For very fast optical bursts, on the order of milliseconds, very fast imaging is required. The faster the imaging, the shallower the images as a result of short exposure times. To date, only very shallow efforts have been possible to search for very fast FRB optical emission.  The guiding/focus CMOS chips on KWFI can be used for science. By placing localized FRBs on a CMOS chip, the large aperture of Keck will enable sufficiently deep exposures to test the predictions of several models.

{\bf Slower optical bursts -}  For cases where the optical emission from FRBs may last seconds to hours, or for the detection of new FRBs with previously unknown locations, the wide field of KWFI can be used with programs like the Deeper, Wider, Faster (DWF) program [e.g., 14,15]. DWF coordinates  wide-field telescopes in the radio, mm/sub-mm, infrared, optical, UV, X-ray, and gamma-ray, along with particle detectors, to search for fast transients  (millisecond-to-hours duration). Wide fields are needed as single-burst FRBs are rare and DWF uses wide-field radio search telescopes capable of identifying and localizing FRBs in real time. Coordinating all the telescopes on the same field at the same time enables shorter wavelength detection of FRB counterparts before the radio detection.  

For wide-field optical coverage DWF currently coordinates DECam on the 4m Blanco telescope in the Southern hemisphere performing a continuous stream of 20\,s images that reach m $\sim$ 22--23. However, some predictions for optical FRB emission requires depths down to m $\sim$ 25--26. Such depths are only possible with KWFI (e.g., m(g) = 25.8, 5$\sigma$ in 30\,s). DWF also performs real-time data processing and analysis, with fast transient identification in minutes after the light hits the telescopes. With a deployable secondary mirror, KWFI could trigger minutes-later deep spectroscopy to uncover the nature of FRBs.

\hd{The big questions}

{\it Big Questions 1}

What is the nature of fast radio bursts? Is there more than one population (i.e., single burst and repeaters, long and short FRBs)? Do they emit radiation in the optical?

{\it Big Questions 2}

What is the physics behind fast radio bursts and their emission mechanism(s)? Deep optical emission and spectroscopy would help understand their physics and nature.

{\it Big Questions 3}

The dispersion measure of FRBs currently provides our only direct measure the ionized baryon content in the line of sight to the FRB host galaxy. How can optical emission and optical spectroscopy of FRBs help understand the line-of-sight intergalactic medium?

\hd{What KWFI can do for this science case}

KWFI can help progress this field in three ways. Firstly, the CMOS corner chips and the 10m aperture of Keck can enable fast (millisecond), deep detection of FRB optical bursts. Secondly, the high throughput, wide field of view, and large aperture will enable the depth required to test seconds-to-hours duration optical emission models with fast-cadenced (e.g., $\sim$30\,s exposures) when coordinated with a wide-field multi-wavelength program like DWF. Finally, fast data processing and analysis and a deployable secondary mirror would enable spectroscopy of longer-duration emission and help uncover the nature of FRBs.

\hd{References}

{\bf [1]} Thornton et al.\ 2013, Science, 341, 53;
{\bf [2]} Petroff et al.\ 2015, MNRAS, 447, 246;
{\bf [3]} Chatterjee et al.\ 2017, Nature, 541, 58; 
{\bf [4]} Bannister et al.\ 2019, Science, 365, 565;
{\bf [5]} Bhandari et al. 2021, arXiv210801282;
{\bf [6]} Lorimer et al.\ 2007, Science, 318, 777;
{\bf [7]} Fonseca et al.\ 2020, 891, 6;
{\bf [8]} Spitler et al.\ 2016, Nature, 531, 202; 
{\bf [9]} Kumar et al.\ 2019, ApJ, 887, 30;
{\bf [10]} Yang et al.\ 2019, ApJ, 878, 89; 
{\bf [11]} Metzger et al. MNRAS, 2019, 485, 4091; 
{\bf [12]} Yi et al. 2014, ApJL, 792, 21;
{\bf [13]} Lyutikov \& Lorimer 2016, ApJL, 824, 18;
{\bf [14]} Andreoni \& Cooke, 2019, IAUS, 339, 136;
{\bf [15]} Cooke et al., {\it in prep.}

\clearpage

\hd{KWFI Science Case}

\begin{center}
  {\bf \Large Early Transient Detection}
\end{center}

\hd{Contributing authors}

Katie Auchettl (UniMelb/UCSC)\\
Ryosuke Hirai (Monash)\\
Charlotte Angus (NBI/DARK)

\hd{Executive Summary}

The progenitors and explosion physics of core-collapse and Type Ia supernovae, as well as a number of other recently discovered fast transients are uncertain. To progress the field, early detections of these explosions and high-cadence observations are needed to catch shock breakouts from core-collapse supernovae, the interaction of supernova ejecta with their companion stars, and the fast rise of events such as the Fast Blue Optical Transients. These outbursts are short-lived, rare, and highly energetic. As a result, deep, wide-field, blue imaging is needed to probe sufficient cosmological volumes to enable detections and to understand these fast, hot bursts and the physics and nature of a variety of transient phenomena.

\bigskip

\hd{Background}

One of the most uncertain aspects of stellar evolution is the link between the nature of supernovae (SNe) and their progenitor stars. The type of SN explosion is thought to depend strongly on the mass, metallicity and rotation of the progenitor, as well the history of interactions in a binary (or higher-order multiple) system. As a consequence, it can be difficult to confidently predict which stars undergo core collapse and produce a SN of a given type, and which of those produce neutron star or black hole remnants. Depending upon the wavelength observed and the phase of the explosion, multi-wavelength and time-domain observations of these events, including very early detections, allow us to understand the properties of both the progenitor and the explosive engine, and the late stages of mass loss by the progenitor star.

Type Ia SNe play a central role as standard candles in the cosmological distance ladder, yet the nature of their explosion mechanism and the role of a binary companion remain unclear. Do white dwarfs explode due to mergers with other white dwarfs, accretion from a non-degenerate companion, or both? Without a better understanding of these events, our ability to use them as standard candles across cosmic history ($z\sim$ 2) is limited.  However, to probe the progenitor system(s) of these events one can probe their very early evolution by using rapid photometry and spectroscopy. Rapid observations will allow us to study the systems through emission, absorption, reflection, and ionization signatures that can illuminate key features. Obtaining a full set of multi-wavelength and time domain observations of Type Ia SNe will allow us to fully capture the amount of $^{56}$Ni and the presence of different elements to constrain the details of the explosion dynamics.

\begin{figure}[!h]
\begin{center} 
\scalebox{1.2}[1.2]{\includegraphics{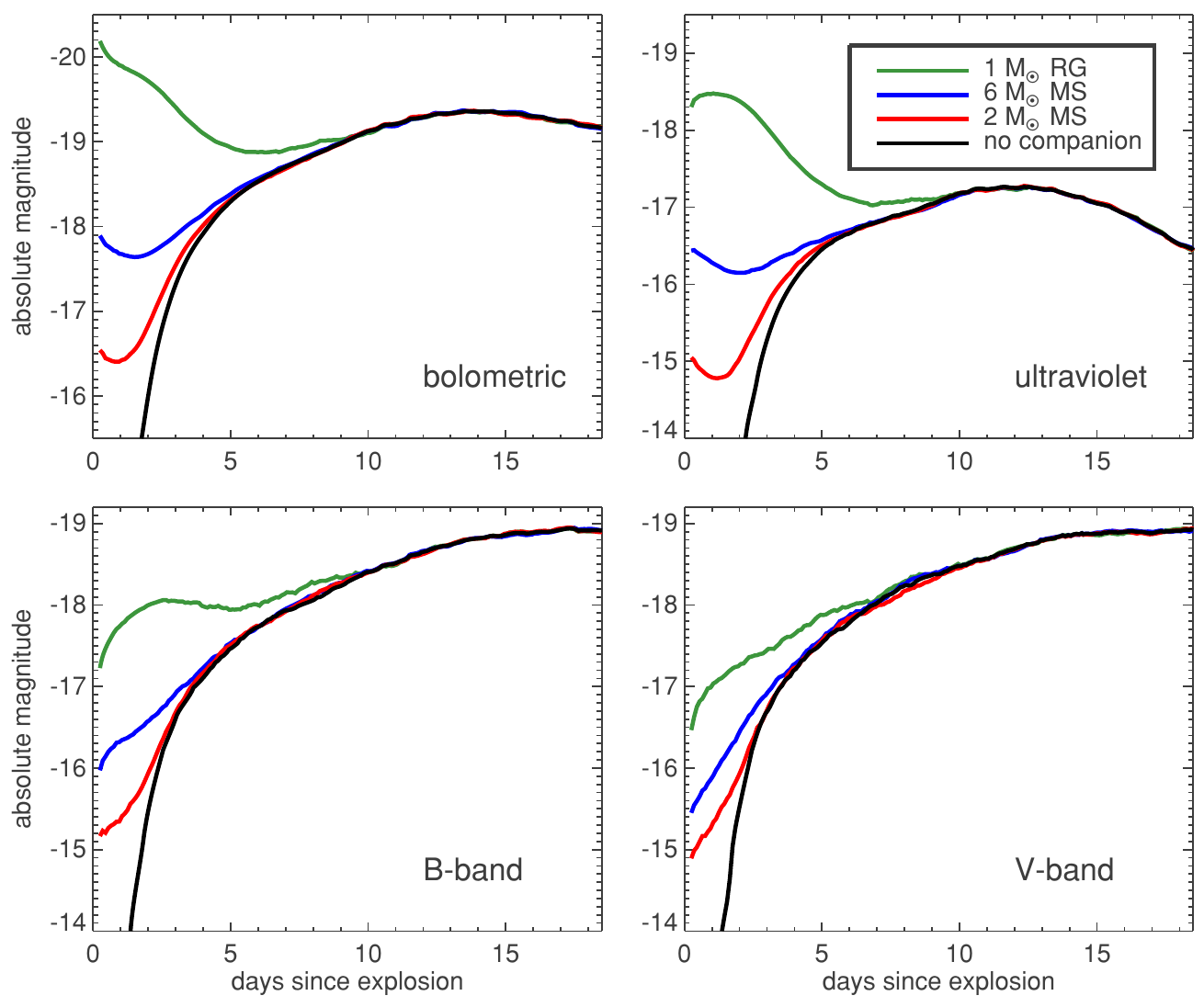}}
\vspace{-0.3cm}
\caption{\small Light curve signatures of Type Ia supernova interaction with companion stars [3]. Three different progenitor scenarios are modeled (a red giant and two different mass main sequence companion stars). The fast signature (accessible over $\sim$1 observing night) can be detected with blue imaging (here, B- and V-band), increasing strongly with shorter wavelength. The blue/UV sensitivity of KWFI will enable observations that can detect these fast bursts and its wide field is essential to survey large areas for these fleeting bursts in relatively rare supernovae.}
\label{SN-Ia-collision}
\end{center} 
\end{figure}

\hd{The big questions}

{\it Big Question 1}\\
{\it What are the progenitors of core-collapse and Type Ia supernovae?}\\
The very early phases of core-collapse supernova explosions contain rich information about the details of the progenitor. The initial shock breakout phase can directly inform us the progenitor size and/or the circumstellar matter distribution [1,2]. Monitoring the supernova throughout the rise and fall of the light curve is critical for accurately estimating the ejecta mass and explosion energy. Many core-collapse stripped-envelope and type Ia supernova progenitors are expected to have close binary companions, and the interaction between the ejecta and the companion can exhibit some noticeable signatures in the early parts of the light curve [3,4] that can help resolve their binary system origin and the nature of the companion (Figure~\ref{SN-Ia-collision}). Most of these early signatures are expected to be very short and hot, and therefore require high cadence surveys in the ultraviolet wavelengths. The wide field and u-band sensitivity of KWFI will be ideal for such searches. Combined with the supernova light curve and late-time follow-up of further circumstellar matter interactions or detection of surviving companions [5], these early signatures will provide unprecedented insight into the last stages of massive star evolution, explosion mechanisms, formation of compact objects and binary evolution.  

\hd{What KWFI can do for this science case}

To reach sufficient depths to probe large cosmological volumes necessary to detect the rare events described above, large aperture telescopes equipped with sensitive wide-field imagers are required. In addition, as these events are short-lived energetic bursts, they are predicted, and some have been observed, to emit in the UV/X-ray and blue optical wavelengths. KWFI on Keck offers the highest sensitivity of any wide-field imager in the world as a result of its high throughput and 10m aperture. In addition, a focus for KWFI for sensitivity in very blue wavelengths is needed for this science. 

As some of these events are short-lived (minutes to hours duration), having the capability of a deployable secondary mirror and fast data reduction will make it possible to detect an event, such as a supernova shock breakout, and obtain very rapid spectroscopic follow up, that will provide invaluable insight into the process.

\hd{References}

{\bf [1]} F\"{o}rster, F., et al.\ 2018, Nature Astronomy, 2, 808;
{\bf [2]} Waxman, E., \& Katz, B.\ 2017, Shock Breakout Theory, p967;
{\bf [3]} Kasen 2010, ApJ, 708, 1025;
{\bf [4]} Moriya, T. J., et al.\ 2015, MNRAS, 450, 3264;
{\bf [5]} Hirai, R., et al.\ 2018, ApJ, 864, 119

\clearpage

\hd{KWFI Science Case}

\begin{center}
  {\bf \Large Gravitational Wave Science}
\end{center}

\hd{Contributing authors}

Ryan Foley (UCSC)\\
Raffaella Margutti (UC Berkeley)\\
Jeff Cooke (Swinburne)

\hd{Executive Summary}

There is currently no large ($>$2m-class) telescope in the Northern Hemisphere with wide-field Target of Opportunity (ToO) search capability for electromagnetic counterparts to gravitational wave (GW) events and no such facility is planned for the foreseeable future. In the Southern Hemisphere, the Dark Energy camera on the 4m Blanco telescope is the only large (4m-class) facility capable of wide-field follow up. The upcoming Vera C. Rubin Observatory, also in the Southern Hemisphere will focus on its 10-year all-sky LSST survey and there is no formal ToO program and capability description for GW follow up and a program will likely comprise $\sim$1-2\% of the time (thus, $\sim$3--4 nights per year). As a result, $\gtrsim$90\% of all kilonovae produced by both binary neutron star and neutron star--black hole mergers will be missed, as 1m-class facilities can reach less than 1/200th the volume needed. Upcoming GW detectors will continue to become more sensitive, detecting events farther and will also have better localization. KWFI, with its extreme sensitivity, wide-field of view, and Target of Opportunity capability, will fill this much needed role and enable the detection of nearly all kilonovae in the Northern Hemisphere and $\sim$2/3rds of the Southern Hemisphere to understand the kilonova population, heavy element creation, measure the expansion rate of the Universe, provide insight into neutron star equation of state, among other physics.

\bigskip

\hd{Background}

The LIGO/Virgo gravitational wave (GW) detection of the nearby binary neutron star merger GW170817 that produced a kilonova [1,2] has helped give rise to multi-messenger astronomy and a new era of detecting and studying the physics of compact objects. GW events that produce kilonovae include binary neutron star (BNS) and neutron star--black hole (NS--BH) mergers, the latter of which can be detected further and are expected to produce fainter and redder kilonovae. The recent LIGO/Virgo O3 run detected $\sim$15 of these GW events, however, no convincing kilonova was identified as a result of (1) the very large localization regions for some events, making search, detection, and spectroscopic confirmation difficult before they quickly faded, (2) the sensitivity of GW detectors that now reach beyond 1m-class aperture optical search telescope sensitivities that perform the bulk of the searches, and (3) the expectation that most kilonovae are potentially fainter at peak than GW170817. Wide-field imagers on larger telescopes with higher sensitivities are needed for current and future gravitational wave detectors sensitivities to detect kilonovae, discover NS--BH kilonovae for the first time, and to understand their populations and physics.

There is only a single confirmed kilonova, but its properties combined with theory provide expectations for the population. The kilonova rises to peak in $\lesssim$ 12 hours (shorter in bluer bands) with the exact rise time and early colors depending primarily on the dominant emission source. The early, blue emission in GW170817 has been modeled as a shock breakout, radioactive decay, and heating from a cocoon around the jet. Each source has characteristics that can be distinguished by their early-time, blue observations. Later, the emission is comprised of at least two distinct components, a fast-evolving blue component from material composed of lanthanide-poor material and a slower-evolving red component from lanthanide-rich material. The composition is dictated by the electron fraction of the material with material from tidal tails having a low electron fraction and high lanthanide fraction while material irradiated by neutrinos originating from the merger remnant NS (which may be short lived) and hot accretion disk having higher electron fraction and low lanthanide fraction. Observations over the first several days after merger can precisely measure the lanthanide fraction and mass of each component.\\

\hd{Science and need for KWFI}

{\bf The utility of kilonovae -} Kilonovae are sites where the heaviest elements are created. They are an accurate independent tool to measure the Hubble parameter and the expansion rate of the Universe and they provide enormous information into the equation of state of neutron stars and GRB jet physics. However, with only one kilonova to date, GW science and understanding the kilonova population is severely limited. The rarity of events within $\sim$200 Mpc will only increase our knowledge incrementally over the next few decades.

There are currently no $>$ 2m-class Northern Hemisphere wide-field GW electromagnetic counterpart search telescopes and there are no plans for any $>$ 2m-class facilities in the future. A `wide field search telescope' is defined here as a telescope equipped with a $>$ 0.5 deg$^2$ field of view imager and the capability to perform large-area, multi-filter kilonova searches. Such searches require quick response (facilities with ToO programs and capability), deep, e.g., m $\gtrsim$ 24 imaging in $\sim$60--90s per exposure, fast (within seconds) filter exchange, and the ability to cover the search area in two or more filters, over two or more epochs, to understand the color and evolution of the fast-evolving transient sources to separate them from many other transients in the field -- all before the source fades.

The two existing $>$2 m-class wide-field telescopes in the Northern Hemisphere are the Canada-France-Hawaii Telescope (CFHT, 3.6m) equipped with the Megacam 1 deg$^2$ imager and the Subaru 8.2m telescope equipped with the Hyper-SuprimeCam imager (HSC; 1.8 deg$^2$). These facilities are not suited for GW search and Target of Opportunity (ToO) observations, e.g., 1 day top-end telescope installation, slow filter change prohibiting the necessary two+ filter, two+ epoch coverage, comparatively low or no blue sensitivity to catch the blue rise of kilonovae [3], and decommissioning/reduced availability (CFHT to make way for the Maunakea Spectroscopic Explorer and HSC optics will also be used for the Prime Focus Spectrograph).
As a result, there is a large gap in capability and kilonova detection in the Northern Hemisphere, as no wide-field imager is planned for 8m-class telescopes for the foreseeable future.

\begin{figure}[!b]
\begin{center}
\scalebox{0.44}[0.44]{\includegraphics{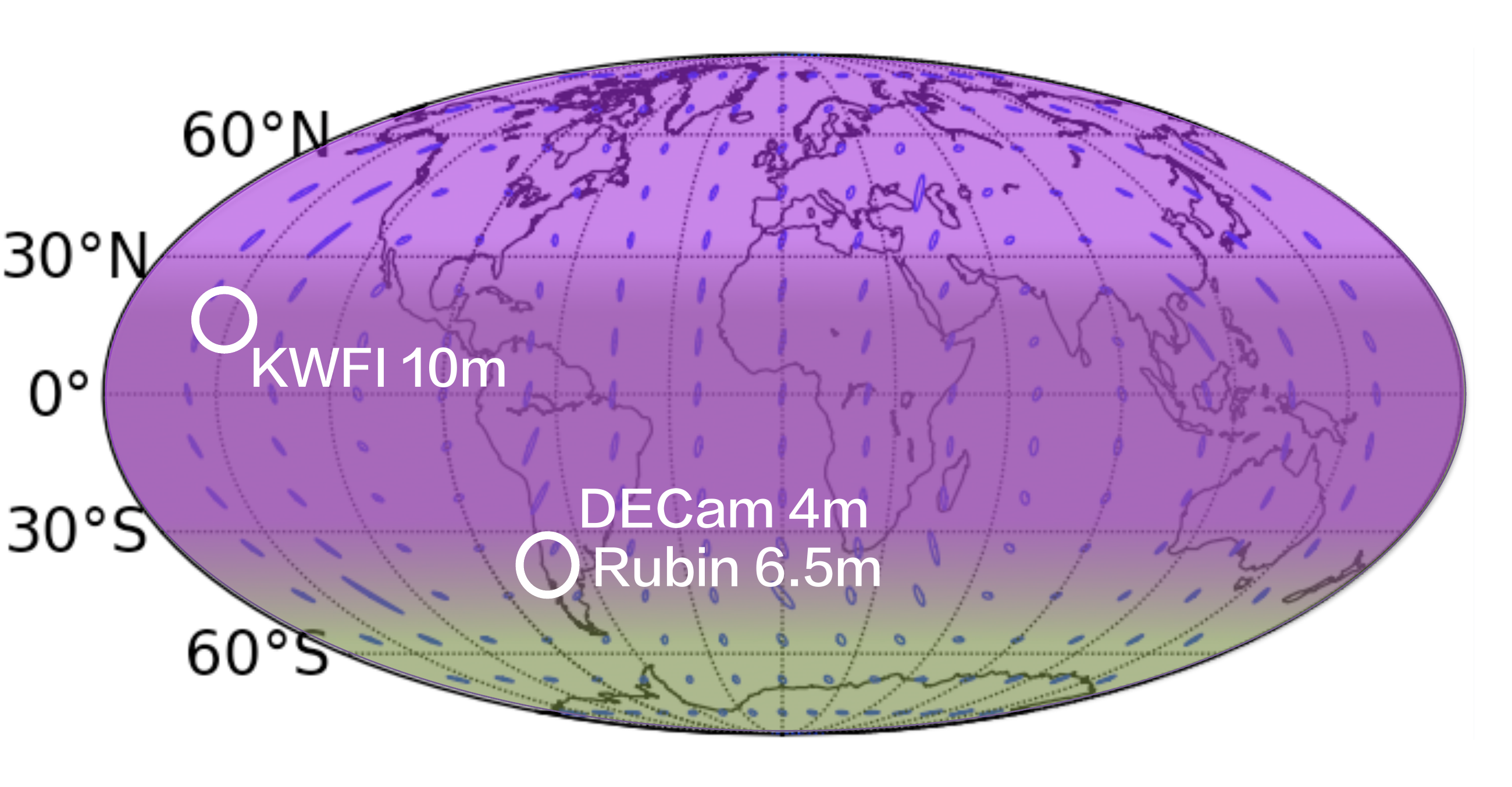}}
\scalebox{0.40}[0.40]{\includegraphics{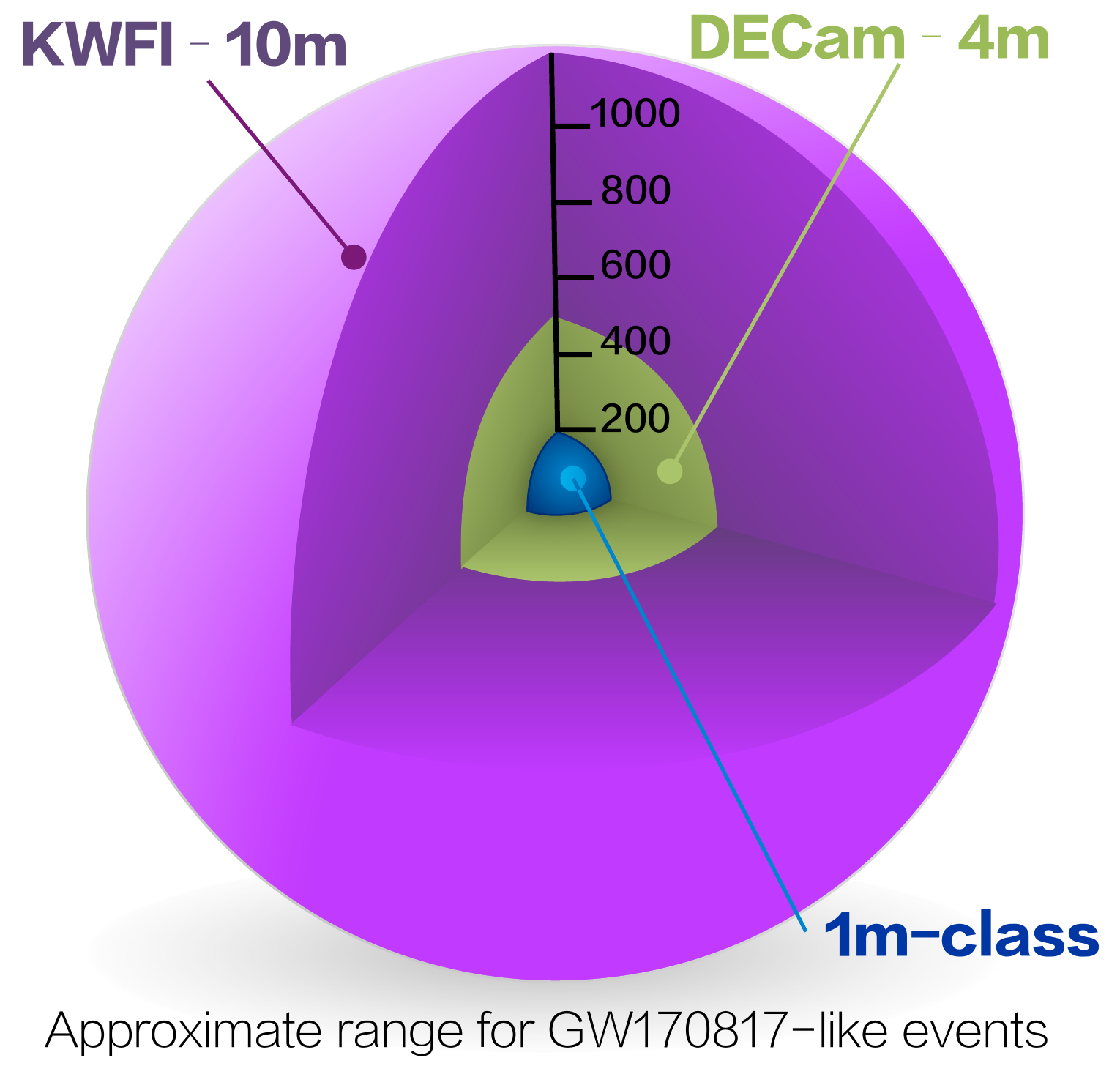}}
\vspace{-0.2cm}
\caption{\small {\bf Left:} Earth projection with gravitational wave (GW) sky localization capability (underlying blue ellipses) for LIGO/Virgo/KAGRA O5 and beyond [4]. Only two instruments will be capable of rapid, deep ToO GW searches through the foreseeable future {\it (see text)}: DECam (4m: sky coverage green) and the proposed KWFI (10m: sky coverage purple). {\bf Right:} Wide-field imager GW volume sensitivities by aperture (for 1 mag detection). Current GW detectors are sensitive to $>$ 400 Mpc and future runs will detect events to $\sim$500--1000 Mpc and will increase with time.}
\label{GWs}
\end{center}
\end{figure}

\begin{figure}[!h]
\vspace{-0.1cm}
\scalebox{0.232}[0.28]{\includegraphics{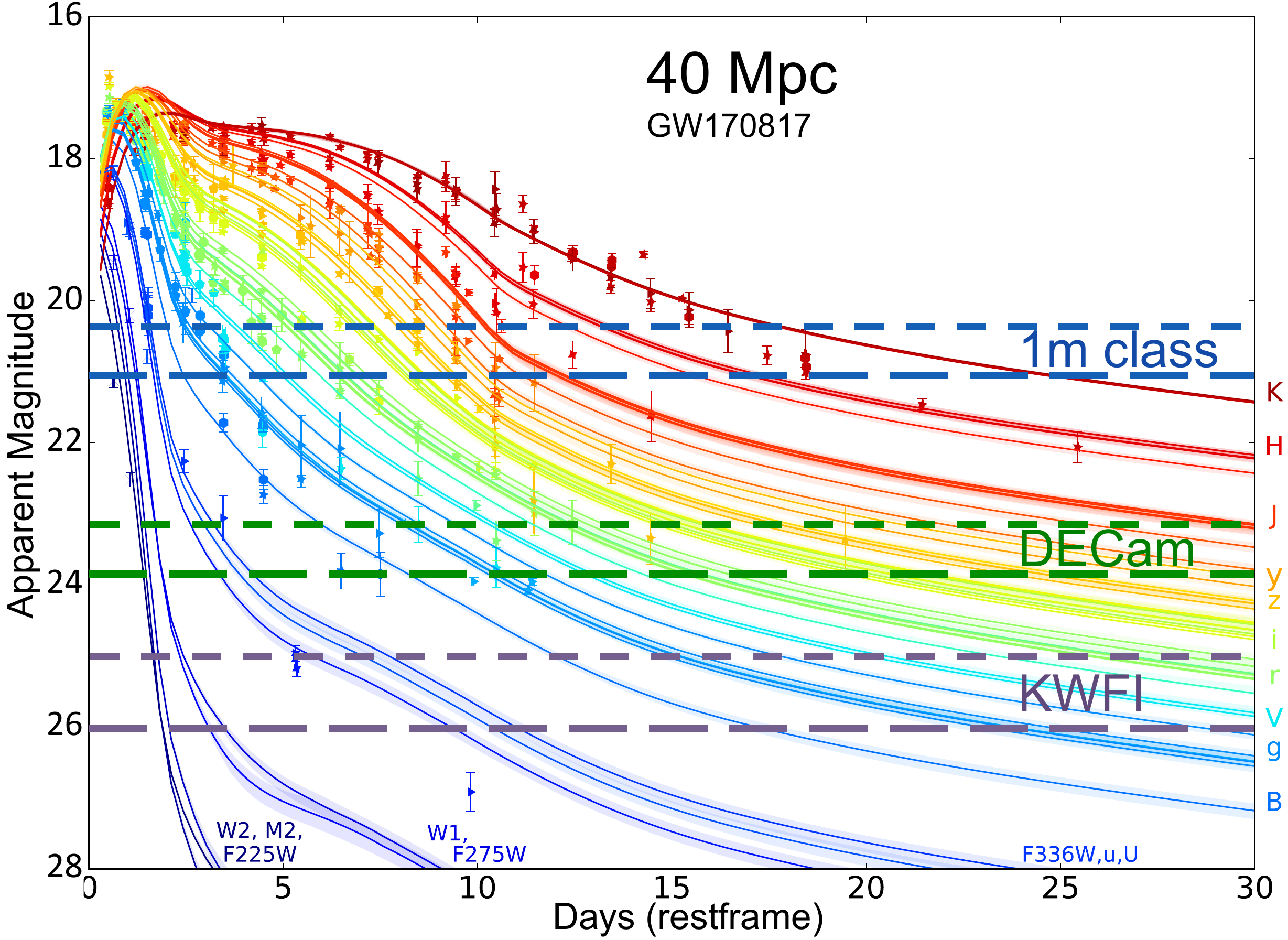}}
\scalebox{0.232}[0.28]{\includegraphics{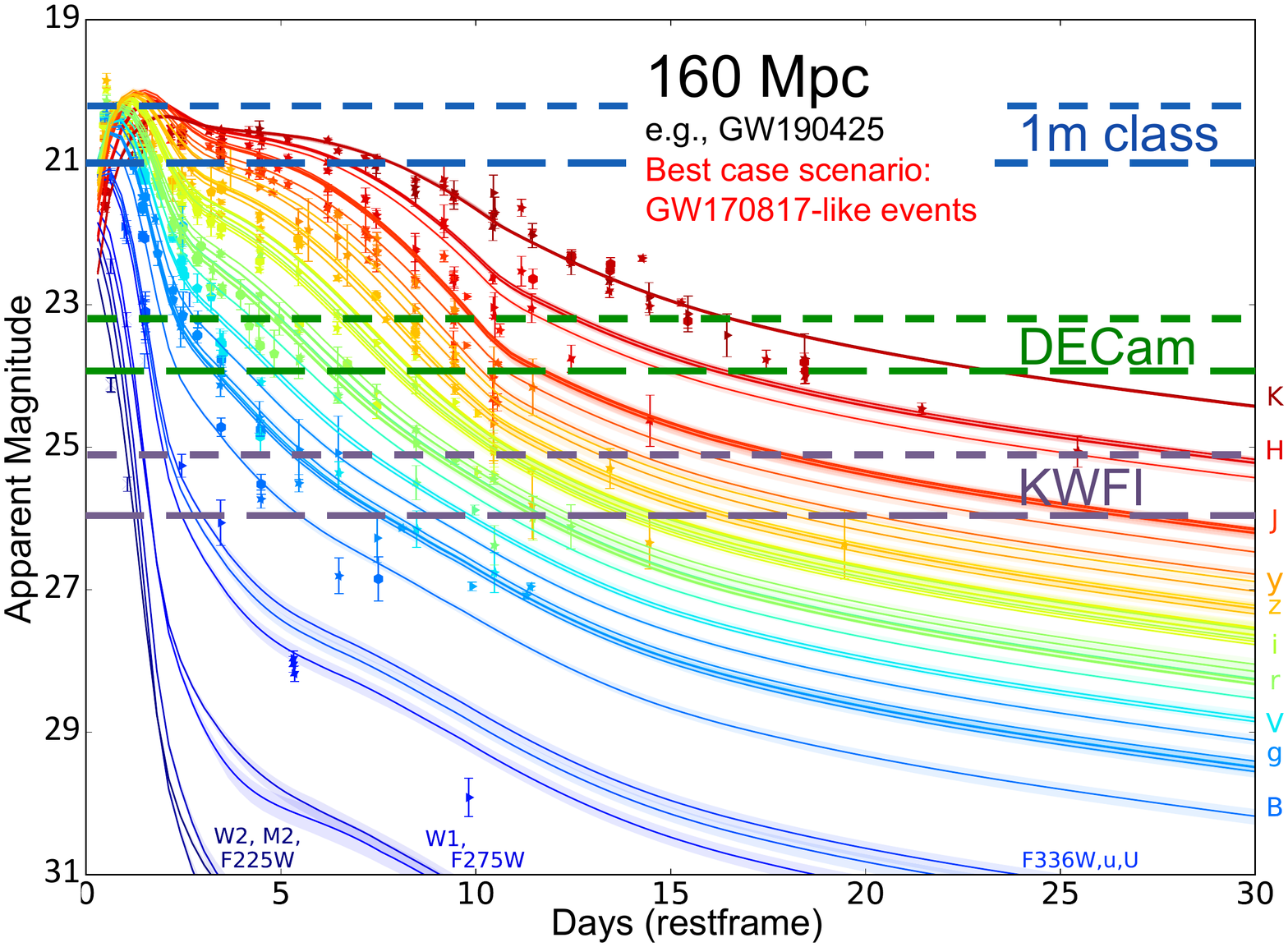}}
\scalebox{0.232}[0.28]{\includegraphics{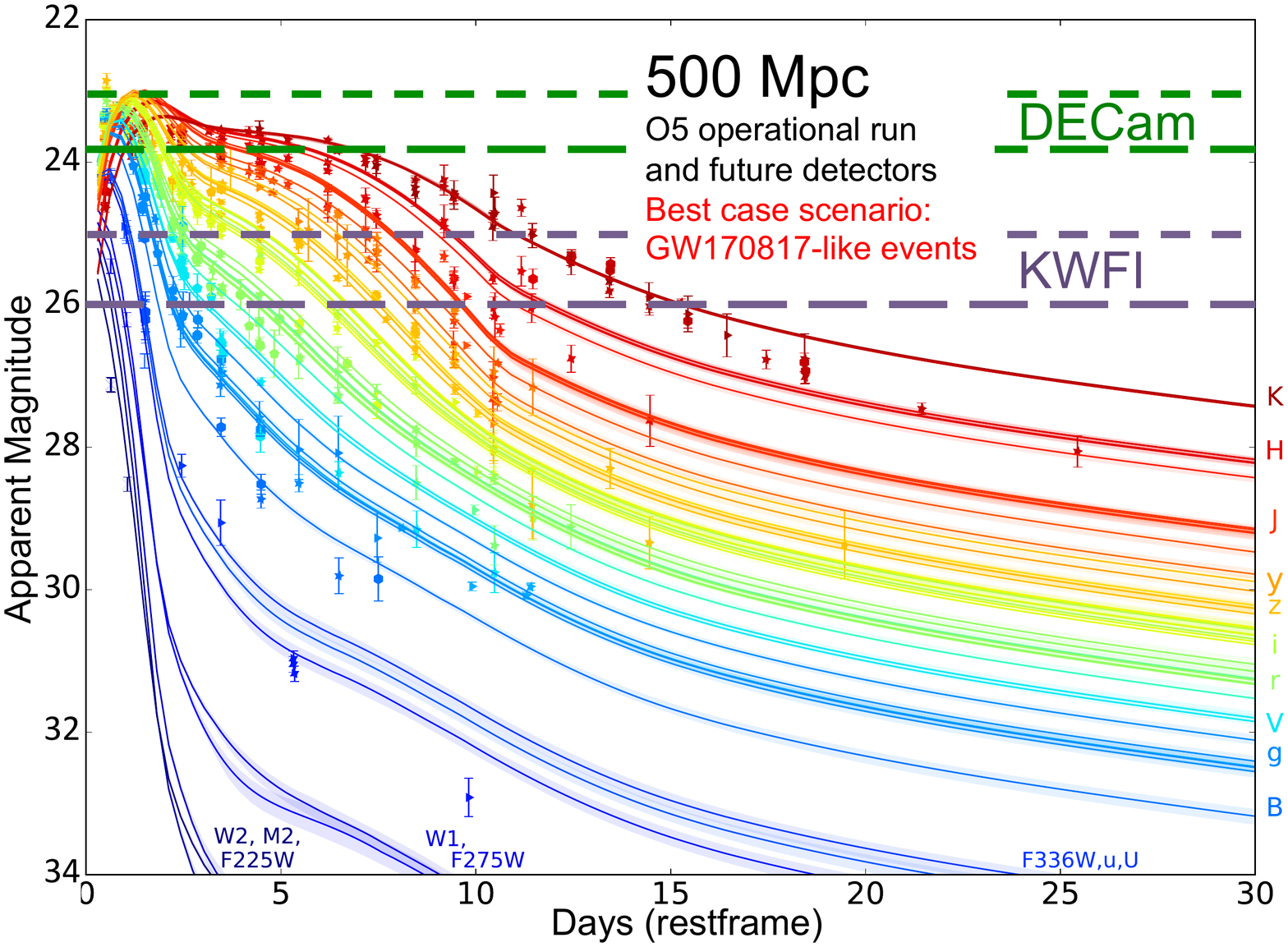}}
\vspace{0.3cm}
\caption{\small Light curves for binary neutron star (BNS) kilonovae at different distances using the single known GW170817  event [6] as a template. Note: Most BNS kilonovae are likely fainter than GW170817 shown on these plots, based on LIGO-Virgo O3 results, and neutron star--black hole (NS--BH) merger kilonovae are expected to be redder and $\gtrsim$1.5 mags fainter than BNS kilonovae [e.g., 6]. Sensitivity thresholds are shown for realistic (e.g., 90\,s exposure) search images for 1m-class (blue lines), DECam (4-m class, green lines) and KWFI (purple lines). Overlaid are depths for g-band (long-dashed lines) and i-band (short-dashed lines) imaging (i.e., the minimum filters for detection). {\bf Left:} The light curve for GW170817 in NGC4993 at 40 Mpc. This very close (rare/small volume) event was detectable by 1m-class telescopes. {\bf Center:} The expectations of GW190425 (LIGO/Virgo O3; 156 Mpc) if it were as bright as GW170817. Such events hit the threshold for 1m-class telescopes for g- and i-band imaging needed to discern event type. All $\gtrsim$ 200 Mpc events (and essentially all NS--BH events) in the Northern Hemisphere are out of reach. {\bf Right:} For the LIGO/Virgo O5 run ($\sim$2025) and beyond, BNS mergers are expected to be detectable to $>$ 330 Mpc and NS--BH mergers to $>$ 600 Mpc. At 500 Mpc, BNS events as bright as GW170817 will hit the threshold of DECam the 2-filter detection and fainter BNS and all NS-BH events will be out of reach. KWFI can reach all these events, as well as fainter and farther events, and can also get the fast, blue rise for BNS kilonovae.}
\vspace{-0.3cm}
\label{villar}
\end{figure}

{\bf The need for depth -} Imagers on 1m-class telescopes can detect GW170817-like BNS merger kilonovae out to $\sim$200 Mpc and fainter NS--BH merger kilonovae to $\sim$ 80 Mpc. In upcoming operational runs, LIGO/Virgo/Kagra GW detectors will detect BNS and NS--BH kilonovae to 330 and 590 Mpc, respectively and, in practice, to much further distances, as these ranges are orientation-averaged. In addition, future GW detector sensitivity improvements will only increase detection distances. As a result, the bulk of all Northern Hemisphere GW events will be missed, as 1m-class telescopes can only probe less than 1/200th the volume needed for BNS events alone, and most Southern Hemisphere events will be missed, as DECam (4m) probes less than 1/10th the necessary volume for BNS events (Fig.\,\ref{GWs}). The Vera C. Rubin Observatory LSST will randomly find relatively close untriggered events (i.e., events that may, or may not be GW events) with poor, very sparse cadence (two 15\,s images per day) during the 10-year LSST survey [5]. Importantly, the volume these telescopes probe is less than stated above if, as suspected from recent O3 operational run results, that the BNS kilonova population is fainter on average than GW170817. As a result, the missed fraction of kilonovae by existing facilities would be even larger, in particular for NS-BH kilnovae (Figure~\ref{villar}). Finally, future next-generation 40-km GW detectors, such as the Einstein Telescope and the Cosmic Explorer, along with the space-based LISA, will reach across the Universe, making the demand for a wide-field imager like KWFI even greater. 

{\bf The impact of neutron star--black hole kilonovae -} Detecting NS--BH kilonovae will result in high-impact science, as none have been discovered to date. They will inform us of heavy element production, the internal make-up of neutron stars, and fates for NS--BH systems. Moreover, both BNS and NS--BH mergers can be used to measure the expansion rate of the Universe. As NS--BH mergers can be detected to much greater distances than BNS mergers, due to their higher mass, these ``dark sirens" can be used to measure cosmic expansion and dark energy much farther back in time. However, their greater distances and fainter kilonovae require the sensitivity of KWFI.

The depth and speed of KWFI 1-degree diameter field of view will detect the elusive blue rise of KNe that encodes important physics, geometry, and nucleosynthetic information [3] and, reach deeper than any other imager through to z-band to enable the first detections of NS--BH kilonovae. Equally powerful, and unique to Keck, is the capability of KWFI to detect untriggered kilonovae to z $\sim$ 1 in \underline{classically-scheduled} programs, beyond the GW detector horizon (and also during months/years when LIGO/Virgo are offline) to amass kilonovae and understand the different populations and their redshift evolution. 

\hd{The big questions}

{\it Big Question(s) 1}

GW170817 was a kilonova with properties and evolution almost identical to certain theoretical models. Are all kilonovae similar to GW170817? Experience from the LIGO-Virgo O3 run seem to suggest otherwise. Do NS--BH mergers produce kilonovae as modeled? How can we use the observations and models to improve search techniques? 

{\it Big Question(s) 2}

What do BNS and NS--BH kilonovae tell us about heavy element production and the mysterious interior physics of neutron stars? Do some form blitzars? Are fast radio bursts associated with GW mergers?

{\it Big Question(s) 3}

What are the effects from, and on, the kilonova population, diversity, and sub-types from their environment? Do kilonova properties evolve with redshift? What are the true kilonova rates and do they evolve with redshift?

{\it Big Question(s) 4}

What is the expansion rate of the Universe and can measurements from kilonovae solve the ``Hubble tension"? 

\hd{What KWFI can do for this science case}

{\bf Sensitivity, blue sensitivity, and timeline for first light -} The current aim is for KWFI to come online in $\sim$2027. With LIGO/Virgo/KAGRA sensitivity improvements next year, and LIGO-India coming online by 2026--2027, GW events will be detected at larger distances and will have more accurate localizations ($\sim$9--30 deg$^2$ [4,8]). The latter will help resolve the localization problem, however, the larger volume means that the bulk of the detections will require 8m-class wide-field imagers (see Fig.~\ref{GWs}). 

At current LIGO/Virgo distance sensitivities, BNS kilonovae are expected to peak at m $\sim$ 22 and fainter, and NS--BH kilonovae peaking at m $\sim$ 23--24 and fainter, rising $\sim$3 mags to that peak mag in just a few hours and fading in $\sim$1--6 nights depending on wavelength (Fig.\,\ref{villar}). Searches for these events (and further and/or fainter events that will be detected) require depths several magnitudes fainter than their peak magnitudes, i.e., to m $\sim$ 24--27, and the capability to reach those depths very fast to cover the full search area at least twice in (at minimum) two filters to search for evolution and color evolution for identification amid the large number of transients. Moreover, often due to the location of the GW event, the visibility window for the field is small and can be only 1--2\,hrs.

To achieve the science here and lead the world in gravitational wave electromagnetic counterparts, KWFI will need the largest field of view possible. In addition, it will need fast filter change and ToO capability. KWFI will be the only instrument to reach m $\gtrsim$26 in $\sim$1 minute to quickly map the localization regions for discovery (Fig.\,\ref{villar}). In addition, the depth achieved in the u-band can detect the elusive rapid UV/blue rise of BNS kilonovae.

\vspace{0.1cm}
{\bf Deployable secondary mirror -} The current design has a deployable secondary mirror installed at prime focus with KWFI that moves in and out of the optical light path in minutes. With a deployable secondary mirror, KWFI can be installed permanently at prime focus and ready for rapid ToO imaging. Moreover, with fast data processing, kilonova candidate sources found in the KWFI images can be followed up in minutes with deep Keck spectroscopy by deploying the secondary mirror. Keck will be the only facility in the world able to trigger and search for kilonova events, have the sensitivity to detect the bulk of them, and enable rapid (minutes later) spectroscopy with Keck optical or infrared instruments. With KWFI, the Keck community will dominate GW science for the foreseeable future.  

\hd{References}

{\bf [1]} Abbott et al. 2017, ApJ, 858, 12;
{\bf [2]} Coulter et al. 2017, Science, 358, 1556;
{\bf [3]} Arcavi et al. 2018, ApJL, 855, 23;
{\bf [4]} Fairhurst 2014, PCS, 484, 012007;
{\bf [5]} LSST collaboration et al. 2009, arXiv:0912.0201;
{\bf [6]} Villar et al. 2017, ApJL, 871, 21;
{\bf [7]} Zhu et al. 2021, ApJ, 906, 94;
{\bf [8]} Abbott et al. 2020, LRR, 23, 3

\clearpage

\hd{KWFI Science Case}

\begin{center}
  {\bf \Large Untriggered Kilonova Searches}
\end{center}

\hd{Contributing author}

Jielai Zhang (Swinburne)

\hd{Executive Summary}

The origins of more than half the elements heavier than iron are unaccounted for. The main candidate sites for producing these elements are elusive explosions known as kilonovae (KNe) [1]. KNe are the optical thermal emission of the coalescence of a neutron star with another neutron star or a black hole. These events are a laboratory that reveals the equation of state of neutron stars and hence dense matter, a probe of short gamma ray burst (GRB) jet physics, and they offer an independent measure of the currently contested expansion rate of the Universe [e.g., 2]. However, only one event has been confirmed to date: GW170817/AT2017gfo [1,3]. Since the confirmation of AT2017gfo as a KN in 2017, over 50+ Nature/Science papers have been published on the topic. Several others GW events have been followed up since that time, but no other KN has been confirmed, as a result of the large search areas required and/or the faintness of the events. KWFI, with its superior sensitivity, fast filter exchange and wide field of view, can detect up to 1 KN in an untriggered search in $\sim$12 nights compared to 1 KN detected so far in triggered searches over last 5+ years. 

\hd{Background}

\textbf{Why untriggered searches} - The one confirmed KN, AT2017gfo, was discovered via targeted searches triggered by a gravitational wave event, GW170817 [4--7]. Despite major concerted efforts to date, no other KNe have been confirmed after gravitational wave or short gamma ray burst triggers [8--10]. An alternative KN search strategy is via “blind” or “untriggered” visible-wavelength searches. Visible-wavelength searches are sensitive to KNe at larger distances (up to $\sim$4 gigaparsec co-moving or z $\sim$ 0.3 with KWFI shallow/wide search) compared to gravitational wave searches (10 times less at $\sim$0.4 gigaparsec co-moving anticipated by the end of 2022 [11]). In addition, whereas KNe emit light isotropically, gravitational waves searches are only sensitive to merger systems with certain orientations [12]. The advantages of visible wavelength searches make them the most promising approach to identify a large population of KNe quickly. 

\begin{figure}[!h]
\begin{center} 
\scalebox{0.4}[0.4]{\includegraphics{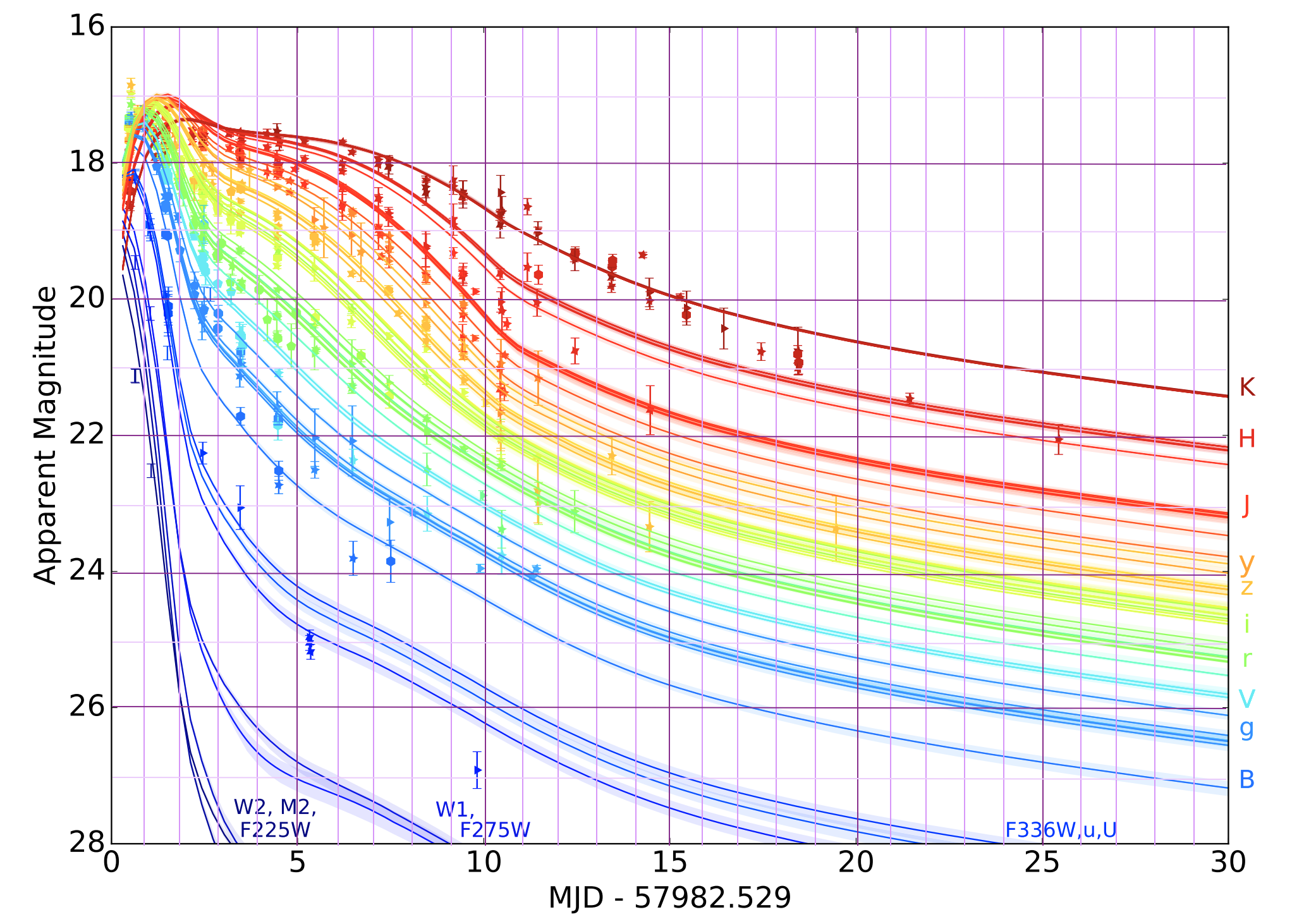}}
\caption{\small The brightness evolution of the only spectroscopically confirmed KN, AT2017gfo in UV, optical and infrared wavelength bands [7]. The x-axis is in days since the binary neutron star coalescence event, GW170817. Lines indicate the best fit KN model. Although this is the only confirmed KN, theoretical KN models fit the light curve of this event remarkably well. Note the rapid evolution in brightness. In optical light (B, g, v , r, and i curves), KN are near their peak brightness for about a day. AT2017gfo is localized to a host galaxy at a distance of 40 Mpc, the peak absolute magnitude at this distance is about M = $-$16 mag. A key outstanding question is whether AT2017gfo is typical for KNe.}
\label{fig:lc}
\end{center} 
\end{figure}

\textbf{Why existing general untriggered searches have yielded no confirmed KNe} - KNe evolve quickly, rising in $<$ 1 day and fade in $<$ 1 week, see Figure~\ref{fig:lc}. The fast evolution of KNe is one of the key reasons existing observations have not revealed a population of these elusive, but highly sought after, explosions. Time-domain surveys typically have cadences of 2--4 days or longer [13]. This cadence can capture at most a single instance of the KN before it fades and cannot differentiate a KN from the tens of thousands of other candidate events in the deep, wide fields and so will be missed. The rate of up to 1 KN per $\sim$12\,n is calculated such that KNe are discovered at least 1 mag brighter than the limiting magnitude searched, which can identify KNe with a few data points in multi-filter light curves. Imaging in \textit{g} and \textit{i}-band filters are one option to maximize the color lever arm without compromising on the image depth by choosing redder filters. A few surveys have reached day cadence, but are comparatively shallow ($\sim$18-20 mag), limiting them only to very rare nearby KN [14]. 

\hd{The big questions}

{\it Big Question 1:} \textbf{Kilonova Diversity}

What are the rates of KNe? What are the properties of kilonovae as a population and does it evolve with redshift? With only one confirmed KN example to date, these questions cannot be answered. 

{\it Big Question 2:} \textbf{Heavy Element Production}

Are kilonovae a dominant site producing elements heavier than iron? More than half of the elements heavier than iron require extreme neutron-rich environments where a nuclear process called rapid neutron capture (r-process). Astronomers have been on the hunt for this elusive environment since the 1950s [15]. The forerunner candidate for this elusive environment is the KN and its neutron rich ejecta. 

{\it Big Question 3:} \textbf{Dense Matter Equation of State}

What is the equation of state for neutron stars? Physical theories of dense matter are currently highly uncertain. The composition and structure of neutron stars can reveal the nature of dense matter, summed up in what is known as the equation of state. 

{\it Big Question 4:} \textbf{Hubble Tension}

What is the expansion rate of the Universe? The expansion rate relies on distance measurements or models of cosmology that are currently uncertain. KNe offer an alternative that can provide an independent measurement, helping resolve the current ``Hubble tension''.

{\it Big Question 5:} \textbf{Gamma Ray Burst Jet Physics}

Short gamma ray bursts (sGRBs) are released during the coalescence of neutron stars, and result from an energetic jet released at the poles of the colliding system. The jet physics of sGRBs is not well understood and a long predicted but not yet observed phenomena is an off-axis GRB where the jet is not pointed at us.

\hd{What KWFI can do for this science case}

The three defining features of KNe are (1) how quickly they evolve in brightness, (2) in color, and (3) how rare KN are [3,7,8,12,16]. An untriggered search for KNe requires sensitive, day cadence, two (or more) filter observations over large areas. This science case requires KWFI to have a fast filter exchange system and the largest field of view (FOV) possible. KNe needs to be discovered at least 1 mag brighter than the sensitivity limits of observations in order to differentiate them via their light and color curves from contaminants such as supernova (which are $\sim$300 times more common). KWFI can achieve $\sim$25 mag in g and i bands in 8\,s and 14,s respectively (3 days from new moon, 0.8$''$ seeing, 5 sigma). With KWFI's $1^{\circ}$ diameter FOV design (0.8 square degrees FOV), the instrument can discover up to 1 KN in $\sim$12 nights. This day-cadence, two-filter data is not only useful for the KN science case, but also other transient and non-transient science by coordinating fields. The KWFI economy design with a FOV of 0.375 square degrees would increase this to over 30 nights per expected KN.

\textbf{Without KWFI}, it would take the current best-suited telescope, CTIO DECam, over 72 nights. The only other competitive existing or upcoming instruments are the Subaru Hyper-SuprimeCam (HSC) and the upcoming Vera C. Rubin Observatory (Rubin). Rubin will be occupied with the Legacy Survey of Time and Space (LSST) for 10 years immediately after completion. Many cadences have been proposed for LSST, and it is likely the fastest cadence will be $\sim$3 days, and only for a portion of the surveyed area [17,18]. A three day cadence is too slow for KN discovery. The slow, 30-minute filter change system for HSC prohibits the program described here, as well as its availability only a few weeks per semester [19].

\hd{References}

\textbf{[1]} Metzger+ 2017, Living Reviews in Relativity, 23:1; 
\textbf{[2]} Dietrich+ 2020, Science, 370:1450; 
\textbf{[3]} Abbott+ 2021, PR, 11:021053;
\textbf{[4]} Abbott+ 2017, ApJL, 848:L12;
\textbf{[5]} Coulter+ 2017, Sci, 358:6370;
\textbf{[6]} Soares-Santos+ 2017, ApJL, 848:L16;
\textbf{[7]} Villar+ 2017, ApJL, 851:L21;
\textbf{[8]} Kasliwal+ 2020, ApJ, 905:145; 
\textbf{[9]} Coughlin+ 2019, ApJL, 885:L19; 
\textbf{[10]} Tanvir+ 2013, Natur, 500, 7464:547; 
\textbf{[11]} Abbott+ 2020, Living Review in Relativity, 23:3;
\textbf{[12]} Abbott+ 2017, PRL, 119:161101; 
\textbf{[13]} Astier+ 2006, A$\&$A, 447:31; 
\textbf{[14]} Bellm+ 2019, PASP 131:018002;
\textbf{[15]} Burbidge+ 1957, Rev. Mod. Phys. 29:547.; Cameron+ 1957, Atomic Energy of Canada, Ltd., CRL-41;
\textbf{[16]} Andreoni+ 2019, PASP, 131:068004;
\textbf{[17]} Lochner+ 2018 DESC White Paper Drafts on LSST WFD; 
\textbf{[18]} Scolnic+ DESC White Paper Drafts on LSST DDF;
\textbf{[19]} Subaru Telescope Schedule, \url{https://www.naoj.org/cgi-bin/opecenter/schedule.cgi}

\clearpage

\hd{KWFI Science Case}

\begin{center}
  {\bf \Large Fast Transients}
\end{center}

\hd{Contributing authors}

Katie Auchettl (UniMelb/UCSC)\\
Charlotte Angus (DARK/NBI)

\hd{Executive Summary}

As wide-format CCD arrays and technology progresses, more and more transient classes and the extent of their luminosity and evolution timescale distributions are uncovered. The largely uncharted region involves the fastest-evolving transients. KWFI will provide ground-breaking depth, blue and optical sensitivity, and speed to discover these transients and new phenomena, including `missed' events, help to understand their high-energies, explosion mechanisms, physics, and their connection to other classes.

\bigskip

\hd{Background}

\begin{figure}[!b]
\begin{center} 
\vspace{-0.3cm}
\scalebox{0.7}[0.7]{\includegraphics{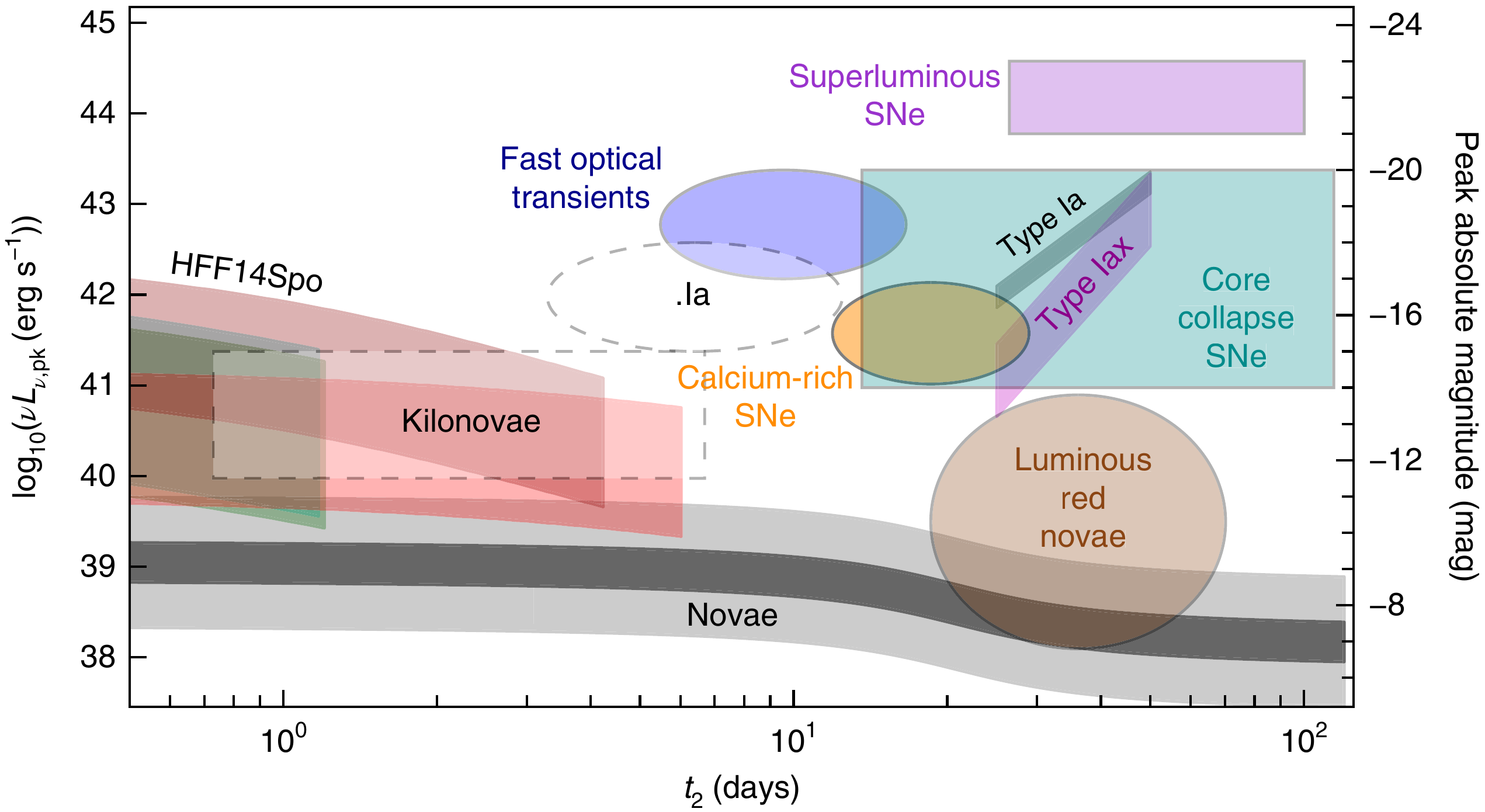}}
\caption{\small The luminosity-variability diagram, here in terms of peak luminosity vs. decline rate. The diagram highlights the locations of newly identified fast optical transients; HFF14Spo-like events, kilonovae and FBOTS [3].}
\label{rodney}
\end{center} 
\end{figure}

New wide-field, high-cadence optical surveys have discovered several novel classes of exploding transients. These surveys are discovering an exponentially increasing number of transients each year and this trend is expected to continue with the Vera C.\ Rubin Observatory coming online in late 2023. Recent advances in the observing strategies employed in these surveys have allowed us to identify multiple transients within hours of their original outburst and have uncovered many new classes of events; from the rapid-and-faint to the bright-and-enduring (Fig.\,\ref{rodney}). However, while we are getting better at discovering the full range of transients in the Universe, there is a large part of this transient parameter space that has remained relatively uncharted. 

In particular, the rapid-timescale regions of this parameter space, in which evolution can be measured on timescales of days to even hours from explosion, has only recently started being populated. These region can include infant supernovae (SNe), which when observed just hours from explosion can provide insightful information about their natal environments or their progenitors from variations in their light curves or evidence of pre-explosion/pre-cursor emission [1,2]. But it also includes exotic `fast' transients, whose entire evolution spans merely days. These transients are challenging to describe physically, and it is unclear how these events are connected (if at all) and how many sub-classes of rapid transients exist. 

We have observed rapid transients across a wide variety of luminosity scales; from the blue and bright, to the faint and red. At the faint and red end is HFF14Spo, which is a pair of gravitationally-lensed transients identified in the Hubble Frontier Fields [3]. At the bright end of the scale lie \lq Fast Blue Optical Transients\rq\ (FBOTs), which are characterized by their fast rise and decline timescales (Figure \ref{rodney}), but also by their extreme bolometric luminosities (in excess of 10$^{44}$ erg s$^{-1}$) and blue photometric colors [4--9]. 

Transients which are fast-evolving are often hard to detect, as they may fall within the gaps of survey cadences, and especially at the faint end, can easily escape the magnitude limits of existing surveys. Rapid photometric and spectroscopic followup of these events is fundamental to generating a better understanding of them. Monitoring their evolution across multiple wavelengths will provide stronger constraints on their spectral energy distributions, which can be used towards constraining models of their production. Though they are often difficult to detect, these events are fairly common in the nearby universe [4], reminding us that events of this class could be easily missed by wide field surveys as a result of their shallow depth and non-optimal cadence.

\hd{The big questions}

{\it What are the intrinsic properties of the fast transient population (color, evolution, etc.)?}

{\it What are the explosion mechanisms/progenitors of fast transients?}

{\it What are the nature of the rarest transients?}

The advent of deep, wide-field, multi-wavelength sky surveys has significantly increased the number of transients we can detect and follow up. However, this has also led to the discovery of a number of rare sub-classes of transients such as peculiar Type Ia SNe,  fast-evolving blue optical transients (FBOTs), off-axis gamma-ray bursts, Luminous Blue Variable outbursts (i.e., supernova ``imposters''), kilonovae, gravitationally-lensed SNe, and luminous red novae. These transients are either inherently rare, as they either have intrinsically low luminosity, are fast/slow evolving, or are just frequently missed by previous surveys due to their colors, nature, or evolution -- or due to the fact some previous surveys target bright galaxies to increase the rate of transient discovery. However, a number of these transients seem to favor lower luminosity galaxies, or nuclear regions and as such, these transients may have been missed even if they were bright, blue or long lived. These sources provide a unique insight into a wide range of physics associated with their progenitor origins, nature, explosion mechanisms, intrinsic rates, and  environments that are inaccessible to us by using more `standard' discoveries. 

\hd{What KWFI can do for this science case}

Rapid Target of Opportunity (ToO) capability.  KWFI would be able to observe rapidly evolving transients within hours of an alert.

Multi-band deep observations -- KWFI would be able to get deep photometry across the UV-optical spectrum for constraining models.

The blue sensitivity of KWFI can confirm candidates on the rise by luminosity increase/colour, which is hard to do at present. 

KWFI can perform candidate screening in the search fields to inform when to trigger spectroscopic follow-up resources.

\hd{References}

{\bf [1]} Armstrong et al. 2021, MNRAS, 507, 3125; 
{\bf [2]} Jacobson-Gal\'{a}n et al. 2021, arXiv210912136J;
{\bf [3]} Rodney et al. 2018, NatAs, 2, 324;
{\bf [4]} Drout et al. 2014, ApJ, 794, 23;
{\bf [5]} Arcavi et al. 2016, ApJ, 819, 35;
{\bf [6]} Tanaka et al. 2016, ApJ, 819, 5; 
{\bf [7]} Pursiainen et al. 2018, MNRAS, 481, 894; 
{\bf [8]} Prentice et al. 2018, ApJL, 865, 3;
{\bf [9]} Perley et al. 2019, MNRAS, 484, 1031

\clearpage

\hd{KWFI Science Case}

\begin{center}
  {\bf \Large Multiwavelength Very Fast Transients}
\end{center}

\hd{Contributing authors}

Jeff Cooke (Swinburne) \\
Jielai Zhang (Swinburne)\\
Katie Auchettl (UniMelb/UCSC)\\
Charlotte Angus (DARK/NBI)

\hd{Executive Summary}

\begin{figure}[!b]
\floatbox[{\capbeside\thisfloatsetup{capbesideposition={right,top},capbesidewidth=6.0cm}}]{figure}[\FBwidth]
{\caption{\small Approximate luminosity and time scale of optical fast transients [17]. Gray regions are slower-evolving transients (e.g., supernovae and novae). Colored regions indicate estimates for some fast transients. The arrow-shaped regions at the bottom denote that the luminosities are below the plot scale. Dashed lines indicate the distances reached as a function of time for continuous KWFI 30-second images, culminating in depths reached with stacked nightly images to the right. The gold-colored region is essentially unexplored to date. Given that the distributions of transients fill most of this phase space, it is highly likely that new phenomena will be discovered.}\label{FT_optical}}
{\includegraphics[width=11.0cm]{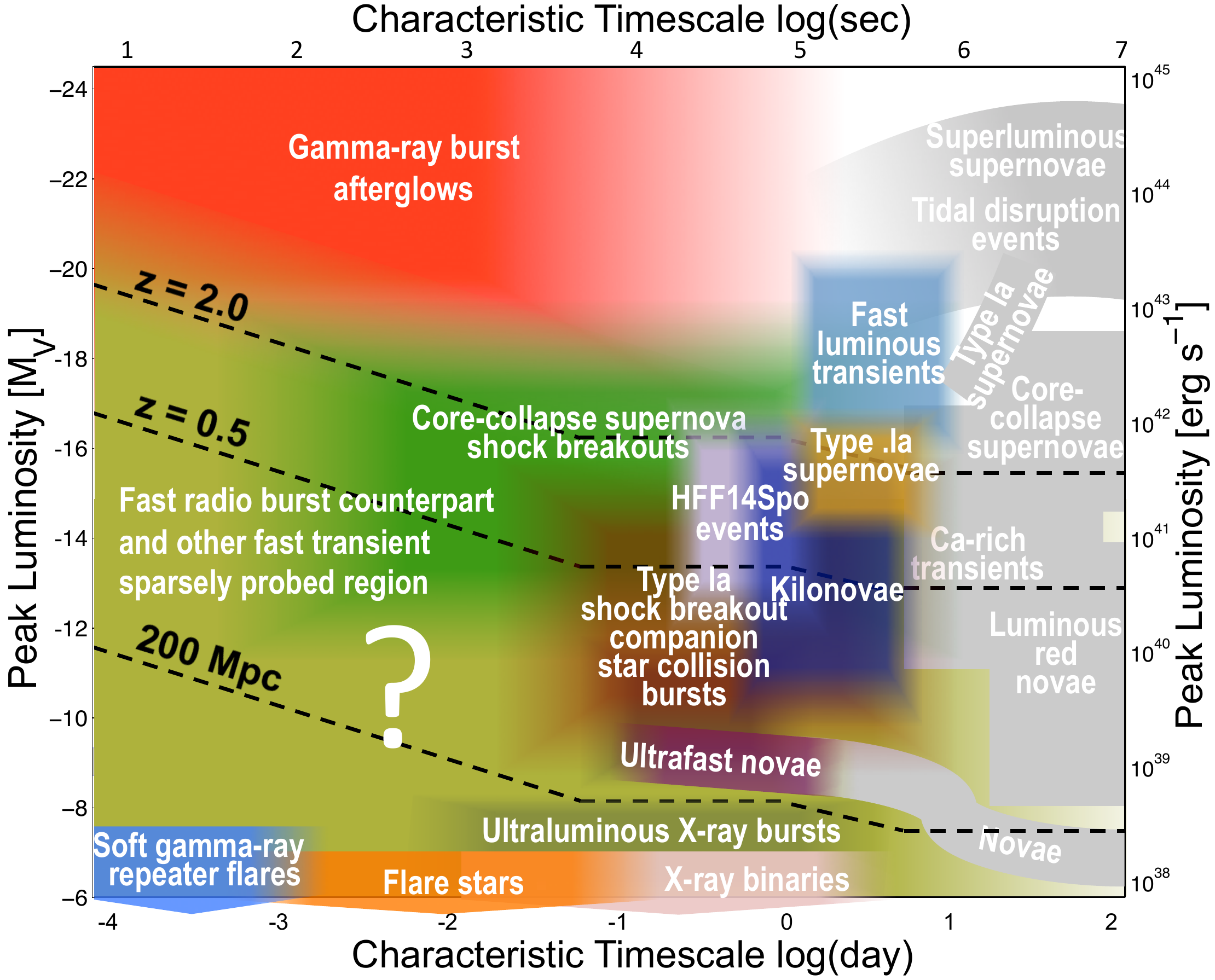}}
\end{figure}

\begin{figure}[!t]
\begin{center} 
\scalebox{0.64}[0.64]{\includegraphics{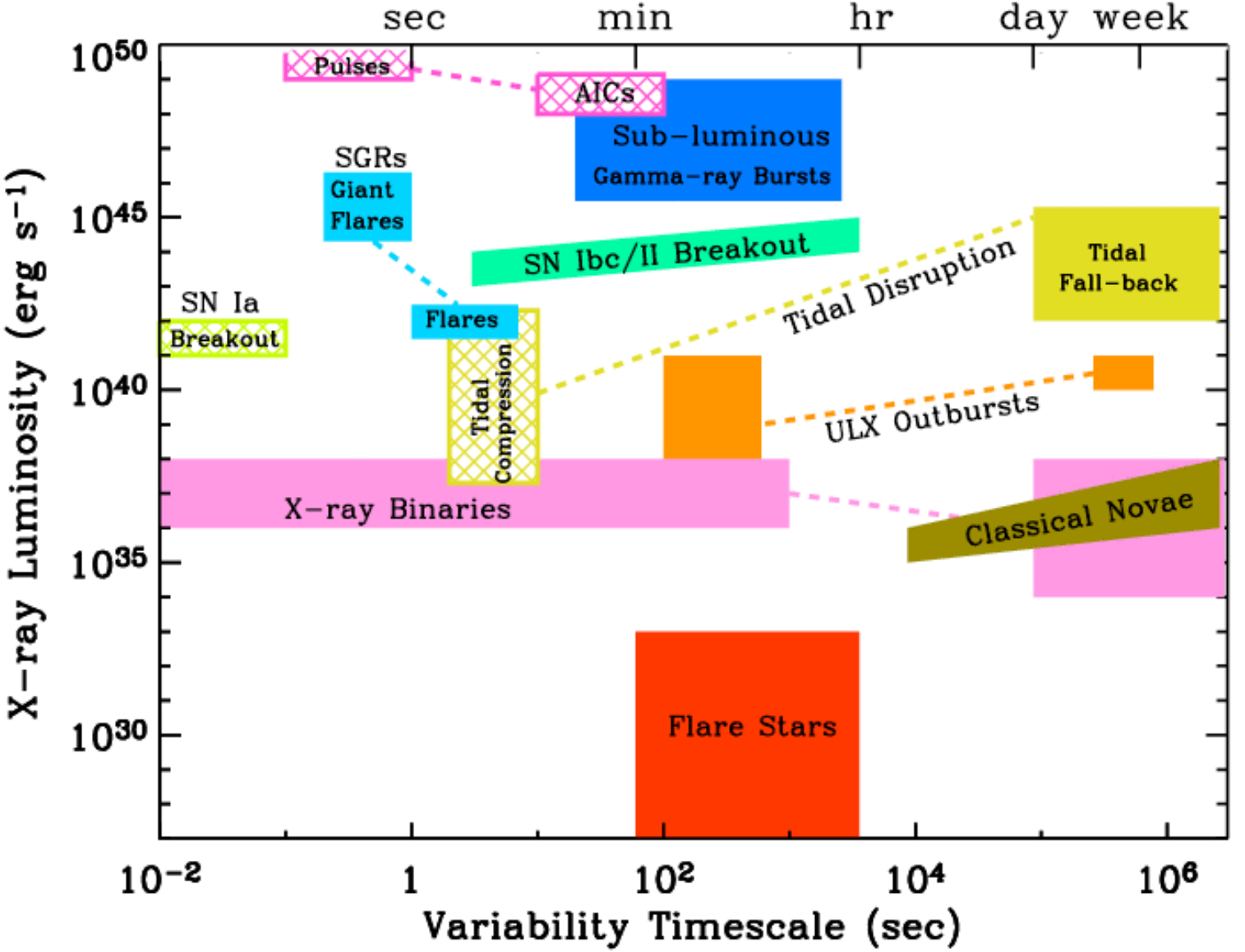}}
\scalebox{0.4}[0.38]{\includegraphics{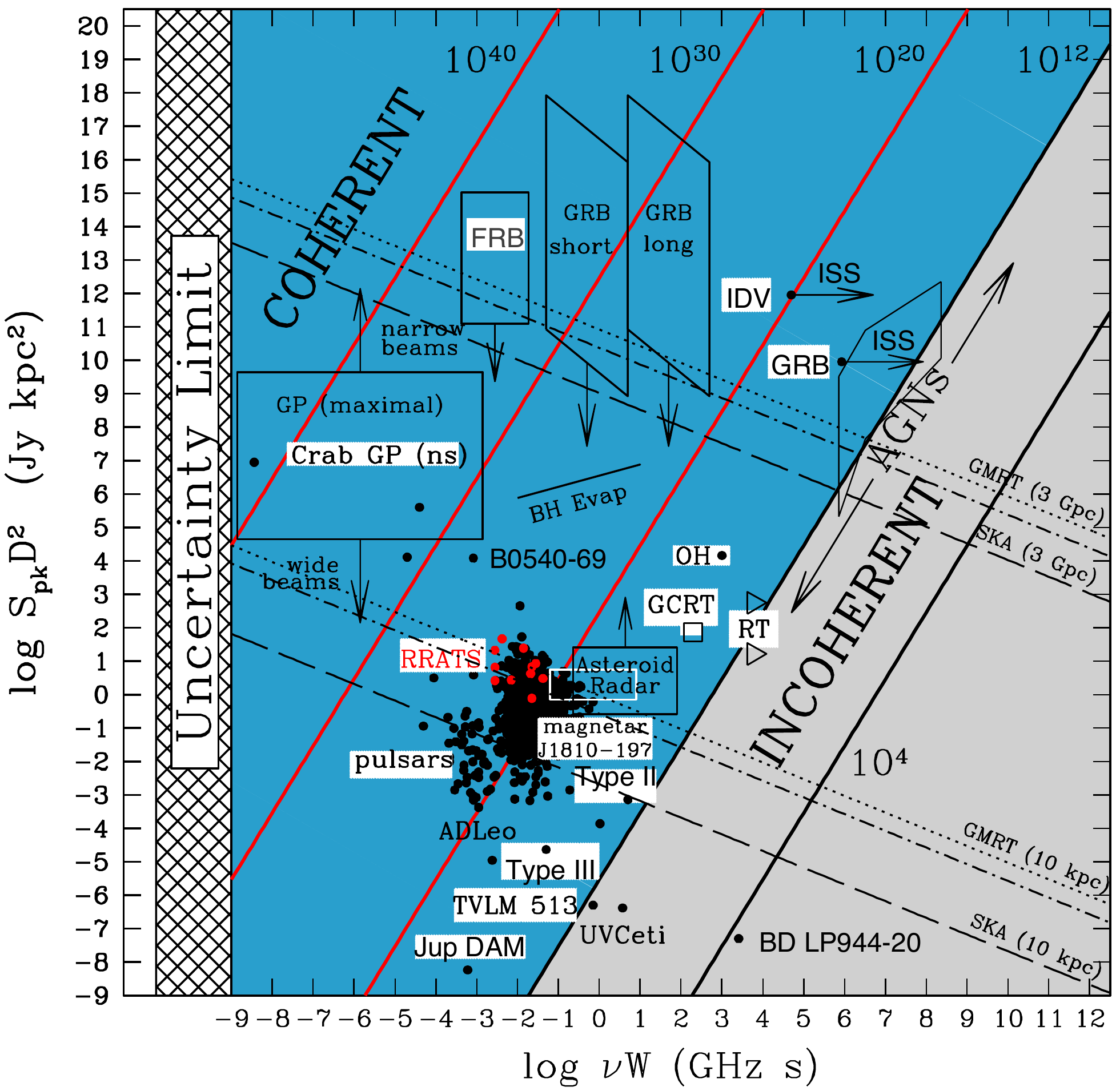}}
\caption{\small {\bf Left:} High-energy (X-ray) fast transients plotted similarly to Fig.\,\ref{FT_optical}, illustrating that fast transients fill the luminosity-timescale phase space when considering the distributions for each known class [18]. {\bf Right:} Low-energy (radio) fast transients [19]. The timescale as a function of observed frequency. For example, for observations at 1 GHz, the timescale is plotted in log seconds. These plots illustrate that fast transients occur at all wavelengths and fill the luminosity-timescale phase space in each (especially when considering their population distributions), leaving the fast optical regime open for exploration. Detections at other wavelengths can inform the time and location of transients to be explored in the deep KWFI images.}
\label{FT_xr}
\end{center} 
\end{figure}

Fast transients have been observed and theorized to occur on millisecond-to-hours timescales at all wavelengths (Figs.\,\ref{FT_optical} \&~\ref{FT_xr}), ranging from millisecond fast radio bursts (FRBs), off-axis gamma-ray bursts, X-ray and optical supernova shock breakouts, to kilonovae. Some fast transient classes are well-studied, a few have serendipitous observations, while many have had no observations to date. Very fast transients are rare and highly energetic, requiring fast-cadenced, blue imaging over wide fields and large cosmological volumes. KWFI provides the high sensitivity and wide-field of view needed to detect Galactic and distant extragalactic fast transients. Fast identification by KWFI image processing can inform and trigger rapid-response optical and other wavelength spectroscopy and imaging and, with a deployable secondary mirror, initiate deep Keck spectroscopy in minutes. When coordinated with wide-field telescopes at other wavelengths, KWFI will help acquire all possible information before the transients quickly fade away. 


\hd{Background}

Fast transients are found at all wavelengths and are rare. A number of fast transient classes have been observed and theorized to date (Figures\,\ref{FT_optical} \&~\ref{FT_xr}, including fast radio bursts (FRBs), gamma-ray bursts (GRBs), X-ray bursts, and their counterparts, flare stars, supernova shock breakouts, fast blue optical transients (FBOTs), Type Ia supernova ejecta collisions with companion stars, elusive off-axis GRBs, binary neutron star and neutron star--black hole kilonovae, soft gamma-ray repeater flares, blitzars, ultra-fast novae, and other sub-day and sub-minute bursts [e.g., 1--10]. These events probe extreme temperatures and densities and their detection provides insight into their explosion physics, their nature, their nucleosynthetic outputs, impacts on their environments, and a more complete understanding of the range of phenomena in the Universe. Fast transients impact other areas of research, such as providing a direct means to measuring the ionized baryon content of the Universe (FRBs) and understanding general relativity and compact matter (kilonovae). 

Exploration of fast transients has been ongoing with high-energy and radio facilities for decades. Previous fast transient searches have been in only one or two wavelength regime(s) or on single targets, with no deep, wide-field optical searches, leaving this parameter space largely unexplored. Progress in the optical has been slow, largely hampered by the availability of large-format detectors on large aperture telescopes and technological challenges with data processing, analysis, and fast identification of transients from the large number of sources in the deep, wide fields. However, recent progress in computational capabilities, data handling, data science analysis and visualization techniques has opened the door to optical fast transient science. Multiwavelength coordination enables acquisition of a wide range of fast transients simultaneously, a more complete understanding of the events, cross-matching between wavelength detections for deeper searches in all wavelength data, and an optimization of fast transient science programs. 

{\bf Progressing existing programs -} One approach to detect these very fast events is to have KWFI lead wide-field, multi-wavelength programs such as the Deeper, Wider, Faster (DWF) program. DWF [11--16] coordinates over 80 telescopes on all continents and in space at all wavelengths and particle detectors, to detect and follow up fast transients. Per observational run (6 consecutive half or full nights), DWF coordinates $\sim$10 major wide-field telescopes at all wavelengths (radio, millimeter, infrared, optical, ultraviolet, X-ray, gamma-ray, and particle detectors) to observe the same fields at the same time. Each telescope detects fast transients in real time and, for KWFI, fast processing will be done modeled after DWF pipelines for the Dark Energy Camera (DECam) and the Subaru Hyper-SuprimeCam. The KWFI pipeline will process the data {\it (in seconds)} on a local computing cluster, or on a supercomputer off site throughout the nights.  Fast transients are then identified within minutes of the light hitting the telescopes using machine learning and data visualization technology. The fast identifications enable rapid-response (within $\sim$2--15 min.) and ToO deep spectroscopy with program allocations on optical, infrared, radio and space-based high-energy telescopes. The option of a deployable secondary mirror for Keck will enable rapid deep spectroscopy in the optical and infrared using Keck instruments. Finally, long-term follow up with 1--2m-class telescopes worldwide is important (as is done with DWF) to classify transients and very early transient detections as some are associated with slower-evolving events, e.g., supernovae shock breakouts).

{\bf KWFI} is essential for deep optical fast transient and FRB counterpart detection. Its high sensitivity (m$_g$ $\sim$ 26 in 30\,s, m$_g$ $\sim$ 29 in nightly image stacks), wide field of view, and fast readout enables fast transient detection in large volumes to high redshift, while minimizing `off sky' time. The field of view of KWFI is well-matched to Parkes radio telescope and other wavelength wide-field telescopes. KWFI detections and fast data processing for `live' transient photometric evolution is crucial to inform deep spectroscopic follow up, rapid-response radio follow up, and rapid-response space-based high-energy imaging and detections.

\hd{The big questions}

{\it Big Question 1}

Fast transients fully populate the low- and high-energy luminosity-timescale phase space at different wavelength regimes (when considering their distributions) from months duration down to sub-second bursts. Do minute-to-sub-second transients exist in the optical? Is there a Universal limit to the duration of optical bursts and, if so, what is that limit? 

{\it Big Question 2}

What are the optical properties of the little-explored fast transients detected at other wavelengths? Optical detection can localize transients (e.g., high-energy and radio transients) and rapid-response deep spectra can study any optical emission and their host galaxy properties.

{\it Big Question 3}

What is the nature of fast radio bursts? Is there more than one population? Detecting counterparts in other wavelengths should resolve this problem (akin to the GRB mystery decades ago).

{\it Big Question 4}

What physics drives fast bursts? There appears to be a wide range of physics and progenitor systems. Optical detection and rapid-response spectroscopy can provide answers to the full range of physics. 

\hd{What KWFI can do for this science case}

Fast transients are relatively rare, thus very wide fields that probe large volumes is required. In addition, fast transients require very fast exposures and fast readout times (to minimize off-sky time) to detect and, ideally, monitor the evolution of the transients before they quickly fade away. Fast exposures are necessarily shallow due to the short shutter open time on sky. Thus, a large aperture telescope with a wide-field of view is essential to detect Galactic and extragalactic fast transients. Finally, fast bursts are highly energetic and typically very blue. Thus, high sensitivity at blue (u- and g-bands) is very beneficial.

This science requires the largest KWFI field of view possible to match the fields of view of the coordinated multi-wavelength facilities and to search the largest instantaneous cosmological volume. Radio transient search facilities have $\sim$1--30 deg$^2$ fields of view and high energy facilities have $0.5$ deg$^2$ to all-sky coverage. The most accessible and suitable radio and high-energy facilities have $\sim$1--3 deg$^2$ fields of view. Target fields for very fast transients is largely arbitrary, given the large volumes probed and their unknown locations. However, legacy fields, local galaxy clusters and known FRB repeater fields are good choices to maximize science. Corner CMOS chips on the KWFI focal plane would be ideal to monitor known FRB repeaters or other known bursts, while the full field is used to search for new FRBs and fast transients in the field. 

Various exposure strategies can be employed, but to date, a continuous stream of $\sim$20\,s images with exposure time similar to the readout time is efficient for maximum depth while minimizing off-sky time. Finally, this science needs fast readout time ($<$ 20\,s) and the filter exchange to be performed within the readout time for times when colors are needed. 

\hd{References}
{\bf [1]} Thornton et al.\ 2013, Science, 341, 53;
{\bf [2]} Petroff et al.\ 2015, MNRAS, 447, 246;
{\bf [3]} Berger 2014, ARAA, 52, 43;
{\bf [4]} Haisch, Strong \& Rodono 1991, ARAA, 29, 275;
{\bf [5]} Woosley \& Bloom 2006, ARAA, 44, 507;
{\bf [6]} Metzger 2019, LRR, 32, 1;
{\bf [7]} Falcke \& Rezzolla 2014, A\&A, 562, 137;
{\bf [8]} Drout et al.\ 2014, ApJ, 794, 23;
{\bf [9]} Cao et al.\ 2015, Nature, 521, 328;
{\bf [10]} Pian et al.\ 2006, Nature, 442, 1011;
{\bf [11]} Andreoni \& Cooke, 2019, IAUS, 339, 136;
{\bf [12]} Andreoni et al.\ 2017, PASA, 34. 37; 
{\bf [13]} Vohl et al.\ 2017, PASA, 34, 38;
{\bf [14]} Andreoni et al.\ 2020, MNRAS, 491, 5852; 
{\bf [15]} Webb et al.\ 2020, MNRAS, 498, 3077;
{\bf [16]} Webb et al.\ 2021, MNRAS, 506, 2089;
{\bf [17]} Cooke et al.\ 2021, {\it in prep.};
{\bf [18]} Soderberg, 2010, astro2010, 278;
{\bf [19]} Bhat, 2011, BASI, 38, 1

\clearpage

\hd{KWFI Science Case}

\begin{center}
  {\bf \Large Low-Redshift and Local Galaxies}
\end{center}

\hd{Contributing author}

R. Michael Rich (UCLA)

\hd{Executive Summary}

\begin{figure}[!h]
\begin{center} 
\scalebox{0.83}[0.83]{\includegraphics{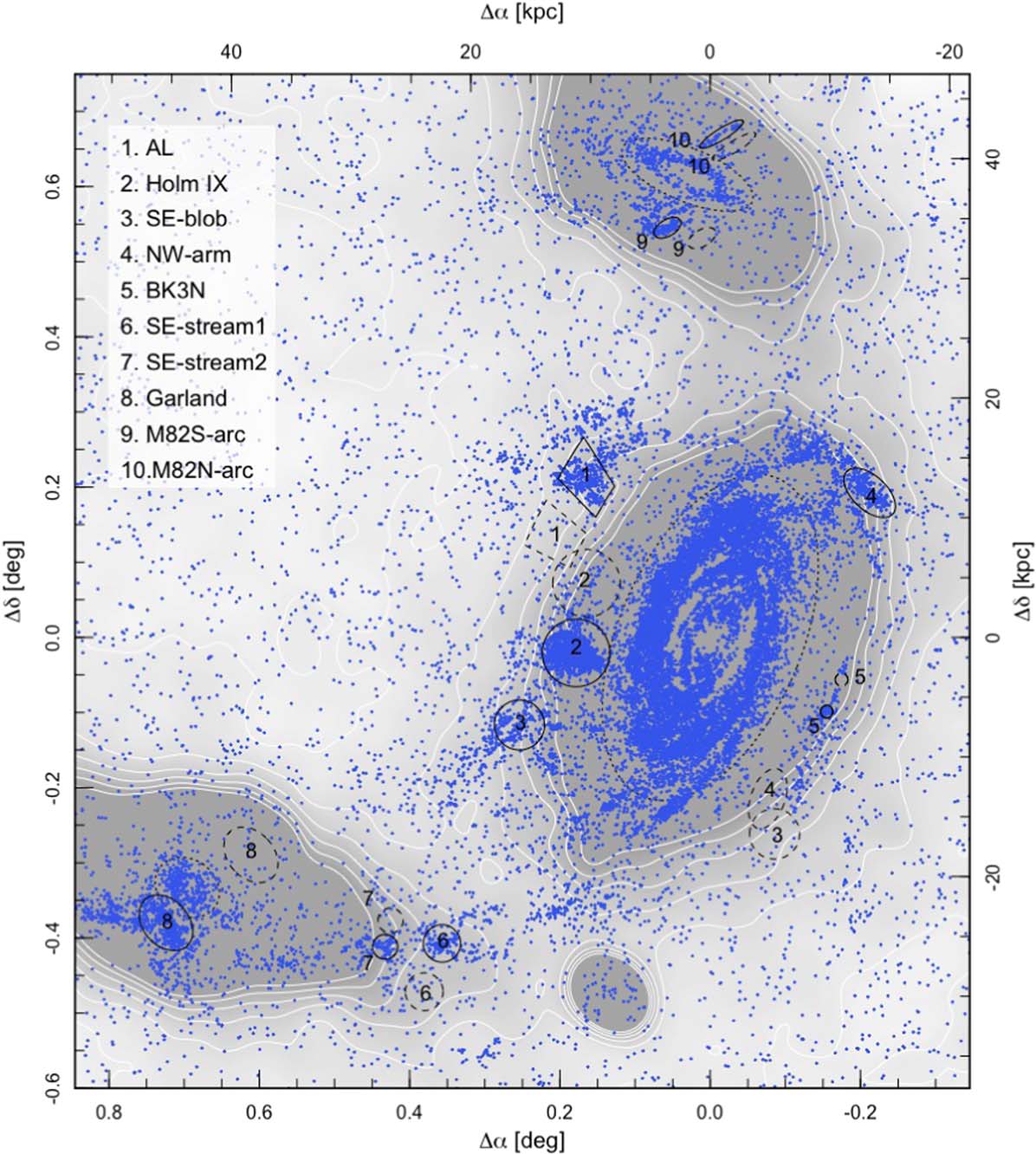}}
\caption{\small Spatial map of resolved stars in the core of the M81 Group [1] obtained using Hyper-SuprimeCam on the Subaru 8m telescope.  This kind of mapping is possible for the M81 system, even at a distance modulus of 27.8 corresponding to to 3.6 Mpc. The blue points represent the young MS, core He burning, and red supergiant stars, and they are plotted on top of a contour map of the red giant branch stellar density. The dotted lines show the R25 radii of M81, M82, and NGC 3077.  This image illustrates the power of ground-based imaging to study complex star formation histories on degree scales in the Local Volume.  Resolution of red giants will reveal and allow structural measurements of numerous ultra-faint dwarf galaxies that surround galaxies as close as 4--5 Mpc.}
\label{okamato}
\end{center} 
\end{figure}

Galaxies in the Local Volume (a radius of $\sim10$ Mpc) present a nearly complete census of the galaxy types and environments found in the low redshift Universe. KWFI will enable a novel study of these galaxies, powered by its UV sensitivity, optical system excellence, and Northern Hemisphere location.  Indeed, the Northern Hemisphere (not reached by LSST) places within reach the M31-M33 system, the M81 system ({\it only} observable from the North, and a range of other galaxies at $<$ 10 Mpc where some stellar populations can be resolved from the ground (Figure~\ref{okamato}). UV imaging allows searches for metal poor stars and precision photometric metallicity measurements for systems to 4--5 Mpc.  Wide field surveys will discover and characterize new dwarf galaxies in the Local Volume, extending the study of the structure of dark matter dominated systems beyond the Milky Way and M31.


\hd{The big questions}

{\it Big Question 1}

Can we explore the luminosity function of dwarf galaxies in the Local Volume and test the $\Lambda$CDM paradigm?

{\it Big Question 2}

What is the star formation history of galaxies and their satellites in the Local Volume?  

{\it Big Question 3}

Are most halo populations built from streams?

\hd{What KWFI can do for this science case}

While the scientific landscape will be challenging, KWFI will have extraordinary UV sensitivity and image quality, resulting in more precise stellar parameters and better sensitivity to metallicity (see also discussion in Milky Way Populations science case).  Searches for extremely metal poor stars will be possible (red giants) to 4 Mpc.  Reaching the fainter red giants and even the main-sequence turn-off, KWFI can discover and characterize new low luminosity dwarf galaxies, testing the missing satellites problem for a range of galaxies. 

KWFI requirements: UV sensitivity, excellent point spread function over a wide field are required. 

\hd{References}

{\bf [1]} Okamato, S., Arimoto, N. et al. 2019 ApJ, 884, 128

\clearpage

\hd{KWFI Science Case}

\begin{center}
  {\bf \Large The Low Surface Brightness Universe}
\end{center}

\hd{Contributing authors}

Cristina Mart{\'i}nez-Lombilla (UNSW, Sydney)\\
Sarah Brough (UNSW, Sydney)

\hd{Executive Summary}

The low surface brightness (LSB) Universe (detections fainter than $\sim$26~mag~arcsec$^{-2}$) is considered the last frontier of optical astronomy: objects with the lowest star densities, largely unseen by past wide field surveys and hardly studied by targeted observations. This almost unexplored field promises to deliver transformative insights into a wide range of phenomena from stellar outskirts to clusters of galaxies and requires wide-fields of view ($\sim$1 deg$^2$ and larger). KWFI can quickly reach the depth of current `Ultra Deep' surveys and will break new ground, reaching an unprecedented m $\sim$ 29, 5$\sigma$, in u g r filters in $\sim$5, $\sim$3, $\sim$13\,hrs respectively over its wide 1 deg diameter field. This makes KWFI ideal to detect low surface brightness structures and galaxies, such as ultra-diffuse galaxies, stellar streams, shells or intracluster light.

\hd{Background}

Significant progress has been made in the past decades with the construction of large telescopes and more sensitive detectors, together with technical advances in image reduction and treatment, telescope design and new observational strategies. In this sense, we are now starting to scratch the surface of all the interesting phenomena that are observable when the Universe is explored at such depths (see Knapen $\&$ Trujillo~2017, for a review). This point is demonstrated by the increasing interest in upcoming all sky deep surveys and the new generation of instruments, and their potential regarding Low Surface Brightness (LSB) science (e.g. Vera Rubin Observatory: Brough et al.~2020, Euclid: Borlaff et al.~2021). 

According to both theory (Martin et al.~2019) and observations (e.g. Dalcanton et al. 1997) the majority of galaxies actually reside in the LSB/dwarf regime, which means that our understanding of the physics of galaxy evolution is predicated on a small subset of the galaxy population. Studying the properties and stellar populations of LSB galaxies will be essential to complete the whole picture of how the Universe evolves. Deep imaging studies will also tackle historically open questions in astronomy, such as the missing satellite problem. LSB galaxies could be the perfect candidates to close the gap between simulations and observations (e.g. Moore et al.~1999). 

Models predict that all galaxies are surrounded by a network of ultra-LSB filaments and streams, relics of past merger events that shaped the galaxies we see today (e.g. Duc~2015). These very faint tidal features, which are undetectable in past wide-area surveys, are mainly produced by minor mergers and are a direct consequence of the aforementioned fact that low-mass galaxies far outnumber their massive counterparts (e.g. Kaviraj et al.~2010). The LSB Universe plays a crucial role in understanding the precise details of the hierarchical build-up of cosmological structure formation and the nature of dark matter. This faint regime will allow for a comprehensive analysis of the stellar properties in galaxy halos and the faint stellar structures around them to track back their mass assembly history and place constraints on the properties and nature of galaxy dark matter halos. 

In a larger scale framework, the formation and evolution of groups and clusters of galaxies can be also better understood with the contribution of LSB science. Deep observations have shown that there is an excess of light around the brightest cluster galaxies (BCGs) that extends up to a large radius. This light is from stars unbound from their host galaxies due to interactions and is known as the intracluster light (ICL; see Mihos~2016 and Montes~2022 for reviews). The ICL is a fossil record of the interaction a system has undergone, providing a holistic view of the past merger activity of the cluster (e.g. Merritt 1984). Simulations have shown that the ICL is a key in the evolution of galaxy clusters as when this component is considered, the growth rate of BCGs better agree with observations (e.g., Conroy et al.~2007; Contini et al.~2019). The ICL has also been proposed as a good luminous tracer of the dark matter distribution in clusters of galaxies (Montes $\&$ Trujillo 2019). However, the extremely faint nature of this light component results in a lack of reliable detections and very limited studies with very limited samples (1-20 clusters) on its nature (e.g., Mihos et al.~2005, Zibetti et al.~2005, Montes \& Trujillo 2014, Iodice et al.~2017, DeMaio et al.~2018,  Montes, Brough, et al.~2020). A detailed study of the ICL stellar populations at different mass and redshift ranges will allow us to infer the  properties of the galaxies from which the ICL accreted its stars and the main mechanisms on the formation of this component.

\hd{The big questions} 

{\it Big Question 1}

Do LSB observations agree with numerical predictions on the number and properties of dark matter substructures?

{\it Big Question 2}

What constraints do tidal features provide on the mass assembly history of their host galaxies? 

{\it Big Question 3}

Can we constrain the accretion history of clusters by understanding the nature of the intracluster light?

\hd{What KWFI can do for this science case}

\begin{figure}[!h]
\begin{center} 
\vspace{-0.2cm}
\scalebox{0.34}[0.335]{\includegraphics{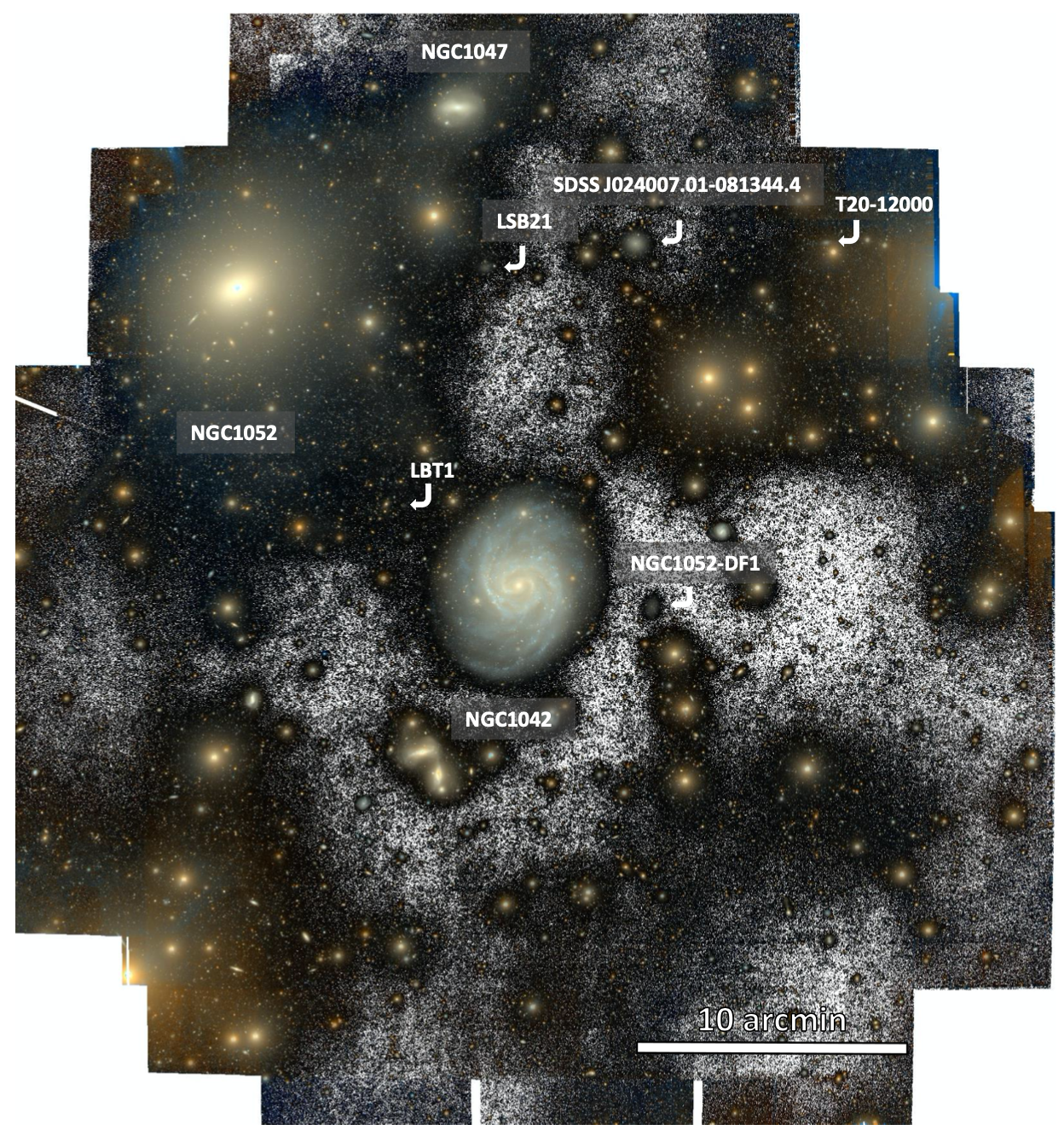}}
\scalebox{0.34}[0.335]{\includegraphics{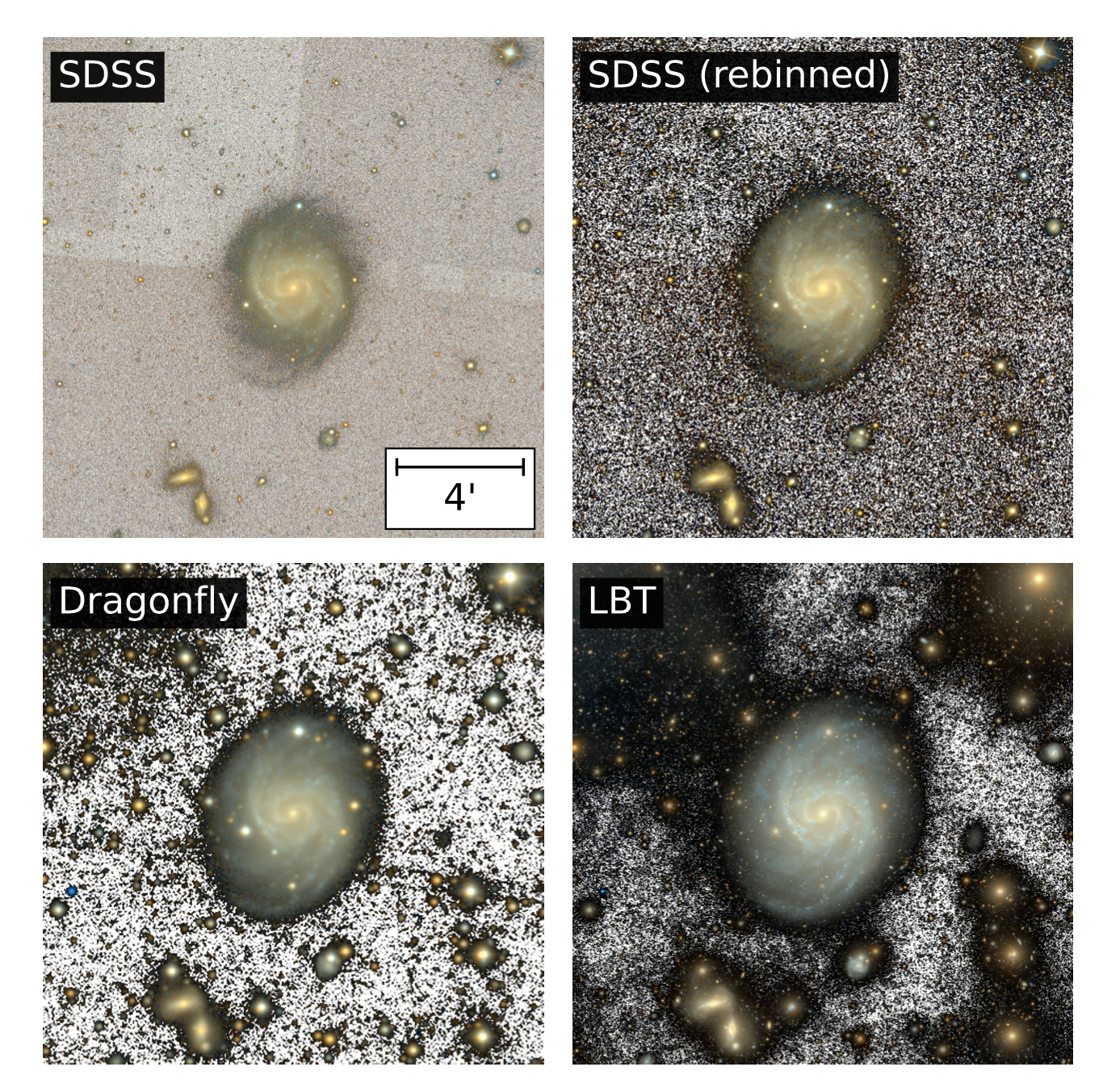}}
\vspace{-0.1cm}
\caption{\small \textbf{Left:} Color composite image around the galaxy NGC~1042 using r and r Sloan filters from Trujillo et al.~2021. The white pixels correspond to the sky region while the darker pixels are the brightest regions. North is up and east to the left. This image was obtained with the LBC camera (FOV 23'×25') of the 2×8.4m Large Binocular Telescope (LBT), with a total of 1.5 hours on source. This is part of the LBT Imaging of Galactic Halos and Tidal Structures (LIGHTS) survey where they reach a depth of $\sim$31~mag~arcsec$^{-2}$ (3$\sigma$ in 10''×10'' boxes). This program shows the potential of KWFI as in less than 1.5 hours on source will obtain equivalent images but with a FOV twice the size of this image ($\sim$5$\times$ the LBC area) and almost two times better spatial resolution. \textbf{Right:} Same image of the galaxy NGC~1042 at three different surface brightness limiting depths (3$\sigma$; 10''×10''): SDSS $\sim$26.9~mag~arcsec$^{-2}$ (r-band), Dragonfly $\sim$28~mag~arcsec$^{-2}$ (r-band) and LBT $\sim$30.5~mag~arcsec$^{-2}$ (r-band). }
\label{LBT}
\end{center} 
\end{figure}

The game-changing blue sensitivity over wide fields provided by the Keck Wide-Field Imager (KWFI), combined with the power of a 10~meter-class telescope, will allow to detect extremely low surface brightness features in large scale structures while analyzing a wide range of stellar populations in unprecedented detail. KWFI can quickly reach the goal depth of current ‘Ultra Deep’ surveys and will break new ground, reaching an unprecedented m~$\sim$~29~mag, 5$\sigma$, in u g r filters in $\sim$5, $\sim$3, $\sim$13~hrs respectively over its wide 1 degree diameter field. This makes KWFI ideal to detect low-surface brightness, ultra-diffuse, and ultra-faint galaxies, stellar streams and shells (Fig.~\ref{LBT}).

\begin{SCfigure}
\scalebox{0.33}[0.33]{\includegraphics{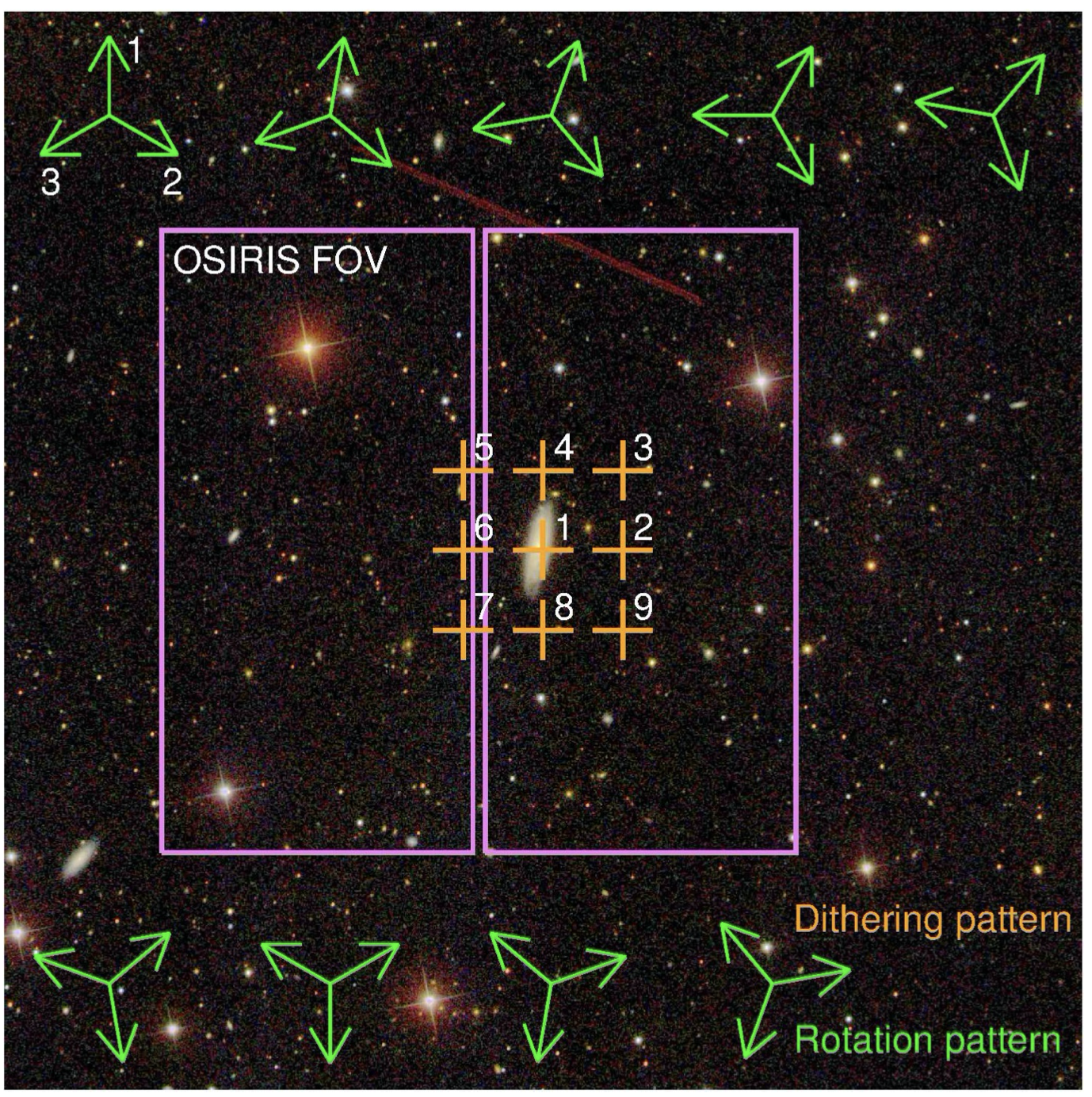}}
\caption{\small Observing strategy of Trujillo $\&$ Fliri~(2016) observing with the 10.4~m Gran Telescopio de Canarias (GTC) using the OSIRIS camera (FOV of 7.8'x8.5'). They spent 8.1 hours on source to reach surface brightness fainter than $\sim$31.5~mag~arcsec$^{-2}$ (3$\sigma$ in 10''×10'' boxes; r-band). This is perfectly achievable with KWFI and with a much larger FOV.}
\label{GTC}
\vspace{-0.3cm}
\end{SCfigure}

However, the surface brightness limit of an image is not only determined by photon statistics or by spatial resolution, but also by a long list of systematic errors such as internal reflections and scattering, which can be variable and/or difficult to control and account for (Slater et al~2009). These systematic errors can be reduced by using relatively simple imaging systems with fewer internal reflections (see e.g., Abraham $\&$ van Dokkum 2014). Certainly, this is the case of the very simple optical design of KWFI. In addition, the influence of ghosting can be accounted for by using optimal observation strategies. So far, the most convenient strategy is a combination of dithering and rotation patterns with a step size similar to or larger than the size of the main object under study (e.g., Trujillo $\&$ Fliri~2016; Fig.~\ref{GTC}). This strategy allows the target to never occupy exactly the same physical area of the CCDs and avoids the repetition of a similar orientation of the camera on the sky. It also helps with the estimation of an accurate background map around bright and extended galaxies. In LSB studies it is also crucial to model the contribution of the scattered light as it becomes a source of light that significantly contributes at brightnesses fainter than $\sim$28~mag~arcsec$^{-2}$ (Slater et al. 2009, Trujillo $\&$ Fliri~2016, Martinez-Lombilla $\&$ Knapen 2019) and it is not only related to the point sources (the stars) but also to the light of extended objects (Sandin~2014, 2015). To this aim, a camera rotator is an important component and is part of the design of KWFI.

KWFI will be powerful to map galaxy cluster ICL, stellar halos, and galaxy satellites to $\sim$10 Mpc, including their density and substructure. Deep, wide-field u-band is crucial for globular cluster selection, a key component to study stellar halos and properties of LSB galaxies (e.g., van Dokkum et al.~2018, 2019; Montes et al.~2020). As bright tracers of galaxy halo chemodynamics and accretion histories, KWFI will compile an unprecedented census in all galaxy types and environments beyond our local group. The high blue sensitivity together with the great spatial resolution of KWFI, allow age and metallicity estimates of the stellar populations in other LSB structures such as tidal features or ICL. H$\alpha$ narrowband imaging at these depths provide details on the star forming regions and star formation rates and provide unprecedented details on the study of the outer regions of nearby galaxy disks (see Martinez-Lombilla et al.~2019).

\begin{figure}[!h]
\begin{center} 
\vspace{-0.2cm}
\scalebox{0.45}[0.47]{\includegraphics{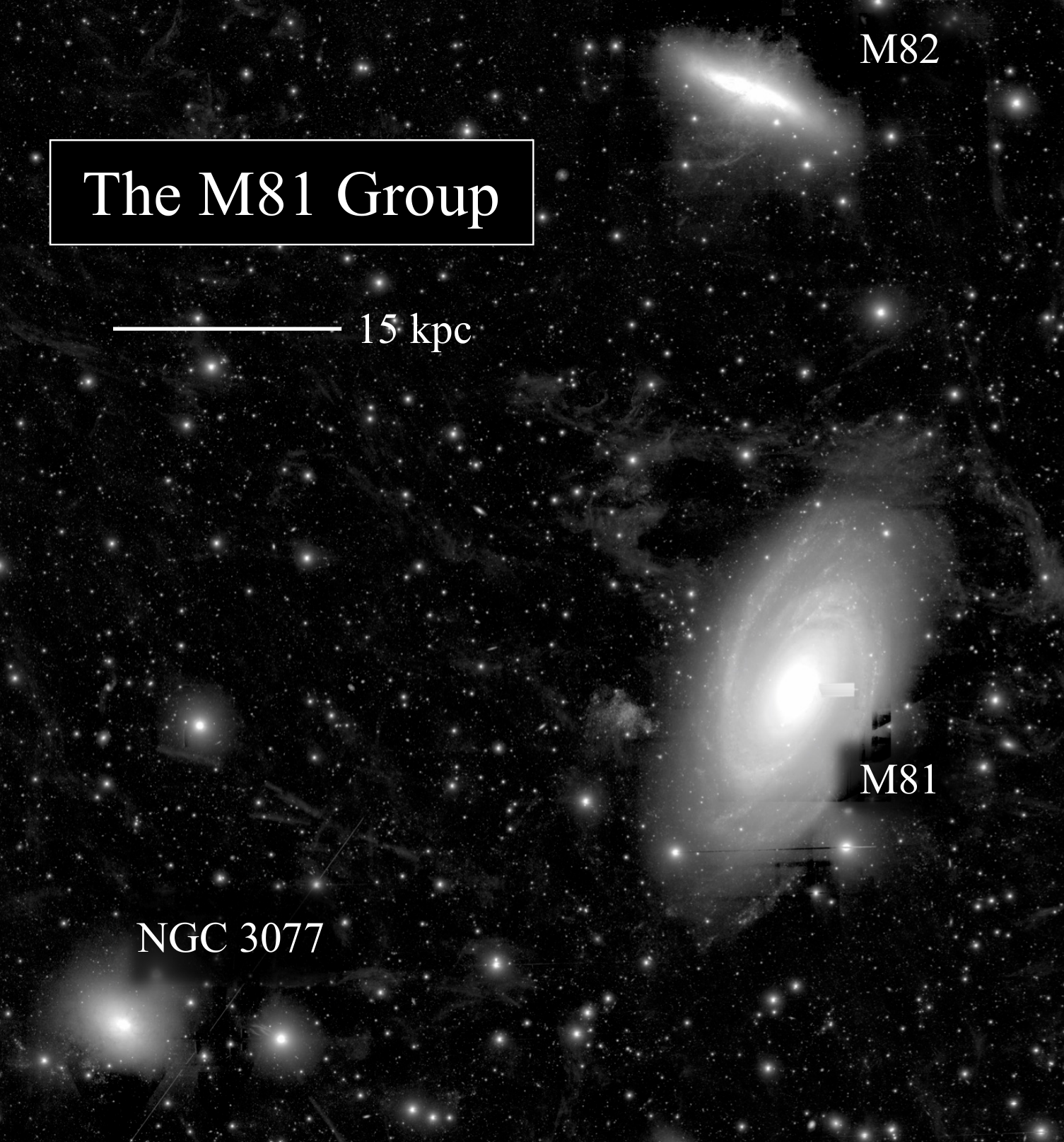}}
\scalebox{1.0}[1.0]{\includegraphics{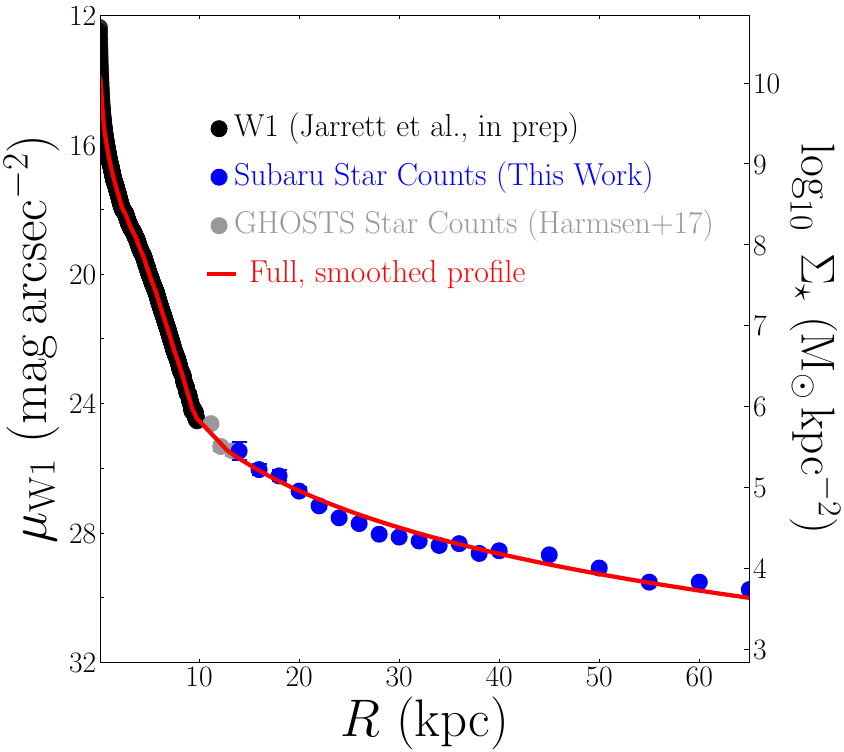}}
\vspace{-0.1cm}
\caption{\small \textbf{Left:}  A deep, wide-field ($\sim$65 $\times$ 75 kpc) g-band mosaic of the M81 Group, taken with Subaru HSC. A logarithmic stretch was used. The three primary interacting group members are labeled (M81, M82, and NGC 3077). The visible dark patches around the three galaxies, as well as bright stars, are the result of chip bleeds. \textbf{Right:} Near-infrared Surface Brightness profile along the minor axis of M81, combining WISE W1 probing the galaxy interior, with the outer resolved star profile obtained from the Subaru HSC imaging. Corresponding stellar mass density is shown on the right axis. Star counts have been converted to W1 using our adopted fiducial isochrone model (10 Gyr, [Fe/H] = $-$1.2). Black points show the W1 measurements, while blue points show Subaru star counts, and the three gray points show inner GHOSTS measurements. A smooth, integrated profile is fit to the total profile and shown in red, for visual effect.}
\label{M81}
\end{center} 
\end{figure}

\begin{figure}[!h]
\begin{center} 
\vspace{-0.2cm}
\scalebox{0.50}[0.50]{\includegraphics{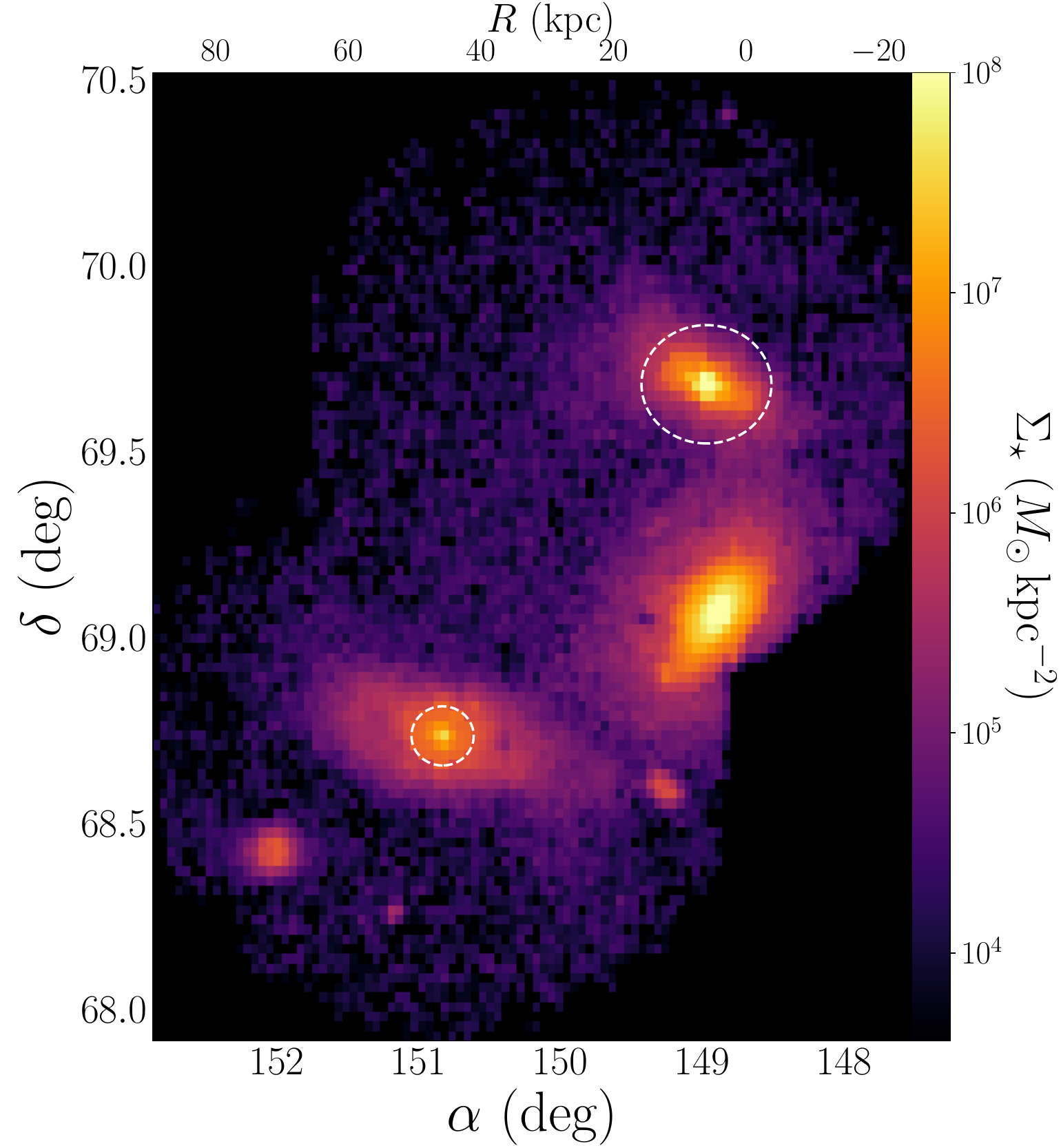}}
\scalebox{0.50}[0.50]{\includegraphics{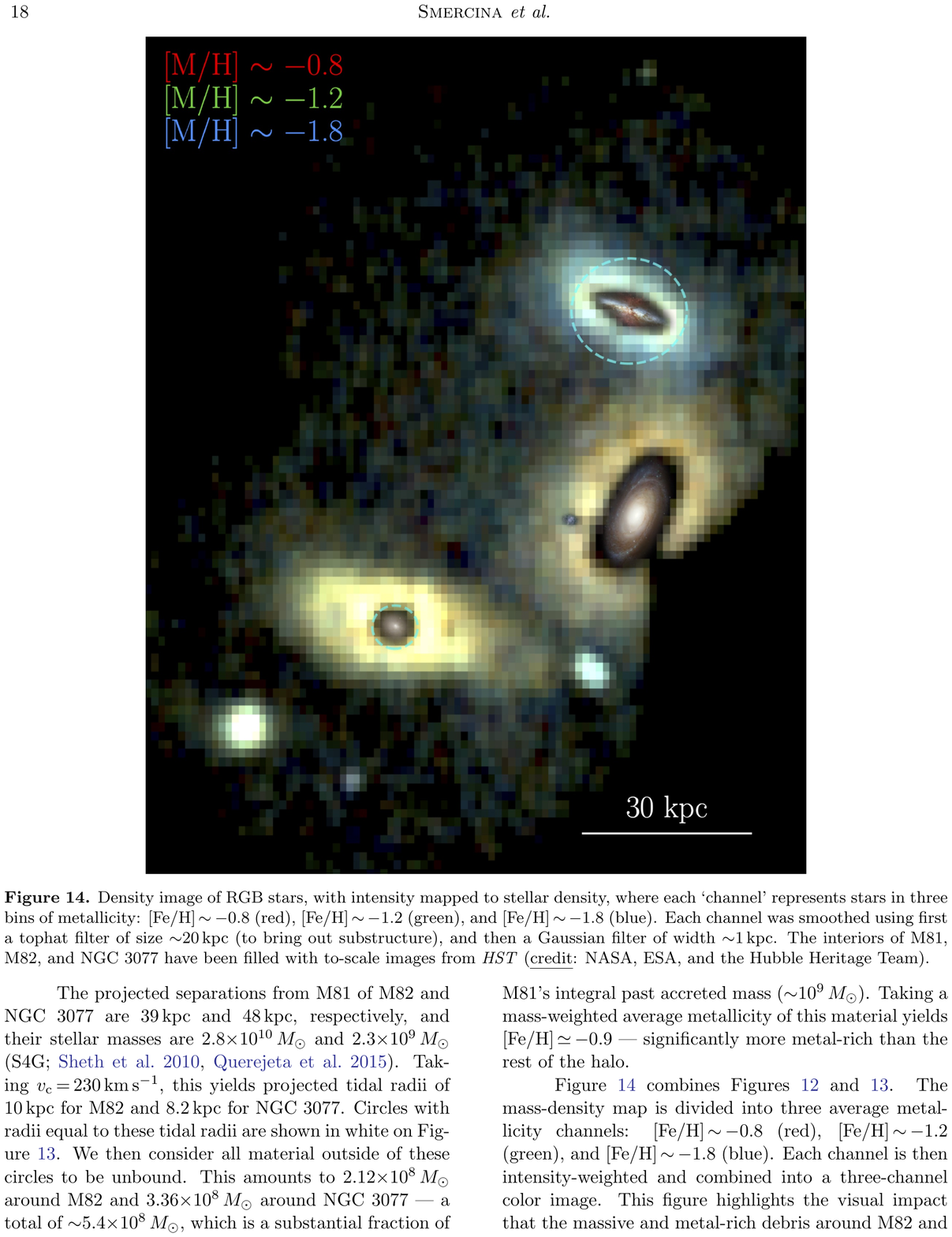}}
\vspace{-0.1cm}
\caption{\small {\bf Left:} Stellar mass density map of the M81 Group. The map has been logarithmically scaled and color-coded according to the right bar. Density was calculated for each $\sim$1~kpc$^2$ pixel and converted to stellar mass. The interior regions of M81, M82, and NGC~3077, where the data were too crowded to detect individual stars with Subaru were filled in using calibrated Ks images from the 2MASS Large Galaxy Atlas (Jarrett et al.~2003) re-binned to $\sim$1~kpc physical resolution. The final map was lightly smoothed with a 0.5~kpc Gaussian kernel and spans an impressive four orders of magnitude in mass density. White dashed circles show the estimated tidal radii of M82 and NGC~3077. All material outside of these circles is counted as unbound, to estimate the total current accreted mass of M81.  {\bf Right:} Density image of RGB stars, with intensity mapped to stellar density, where each ‘channel’ represents stars in three bins of metallicity: [Fe/H] $\sim$ $-$0.8 (red), [Fe/H] $\sim$ $-$1.2 (green), and [Fe/H] $\sim$ $-$1.8 (blue). Each channel was smoothed using first a tophat filter of size $\sim$20 kpc (to bring out substructure), and then a Gaussian filter of width $\sim$1 kpc. The interiors of M81, M82, and NGC 3077 have been filled with scaled HST images (credit: NASA, ESA, and the Hubble Heritage Team).}
\label{M81_2}
\end{center} 
\end{figure}

The above considerations mandate a large field of view. Currently, most targets will `fit' in $\sim$0.8 deg$^2$ (but see below on regions around the targets), with some larger targets requiring tiling (e.g., 10+ deg$^2$ for the Virgo cluster). Examples include nearby galaxies such as M101 ($\sim$42 $\times$ 42 arcmin$^2$ for the galaxy and $\sim$1 deg$^2$ for the group), M31 (a few deg$^2$), NGC 2043 ($\sim$30 $\times$ 20 arcmin$^2$) or galaxy groups such as M81 (a few deg$^2$; Fig.\,\ref{M81} \&~\ref{M81_2}) or NGC 1023 (a few deg$^2$ with some interesting individual galaxies). Each system requires space around it to study known and new substructures. Thus, a larger field of view will help discover outer features around galaxies that have not been uncovered to date, especially as KWFI will reach deeper flux limits than has been previously possible, and in the blue. Finally, star-counting studies in the outskirts of nearby galaxies such as those above and M33 are examples of very interesting LSB science, as it allows the determination of ages and metallicities of the stellar populations in those external regions, as well as deep H-alpha narrowband imaging to explore the extended disks and tidal features in these systems.

\clearpage

\hd{KWFI Science Case}

\begin{center}
  {\bf \Large Globular Clusters, Compact Elliptical Galaxies and Ultra Diffuse Galaxies}
\end{center}

\hd{Contributing authors}

Duncan Forbes (Swinburne)\\
Jonah Gannon (Swinburne)

\hd{Executive Summary}

Galaxy halos and their surroundings contain important chemodynamics and accretion histories. KWFI will compile an unprecedented census in all galaxy types and environments beyond our local group. KWFI will map stellar halos and ultra diffuse galaxies to 10s of Mpc, including their density, metallicity, and substructure. KWFI is ideal to detect low-surface brightness galaxy halos and streams but also ultra-diffuse, and ultra-faint galaxies. These halos also contain globular clusters and compact elliptical galaxies. 

\bigskip

\hd{Background}

Galaxy halos hold a record of their assembly histories and star formation in diffuse components and satellites. Deep u-band enables resolved star isolation and metallicity sensitivity over redder optical bands. KWFI enables mapping of resolved stellar halos and galaxy satellites to $\sim$10 Mpc, including their density, metallicity, and substructure. Globular clusters (GCs) are bright tracers of halo chemodynamics in galaxies beyond the Local Group. Deep, wide-field $u$-band is crucial for reducing GC photometric sample contamination. Keck spectroscopic follow-up will reach to $\sim$30 Mpc and GC systems probed photometrically with KWFI can reach the luminosity function turnover mag out to $\sim$100~Mpc. This capability would allow for an unprecedented census (numbers, spatial density profiles, colors as proxies for metallicities) of GCs in a variety of galaxy types and environments, and unique constraints on their host accretion histories. KWFI will be a powerful detector of compact elliptical galaxies (cEs), as well as low-surface brightness targets, including ultra-diffuse galaxies (UDGs), ultra-faint dwarfs, stellar streams and shells. KWFI can efficiently map galaxy outer regions for low level star formation and to improve SEDs, as well as the distribution of dust and escaping radiation.

\begin{figure}[!h]
\begin{center} 
\scalebox{0.48}[0.48]{\includegraphics{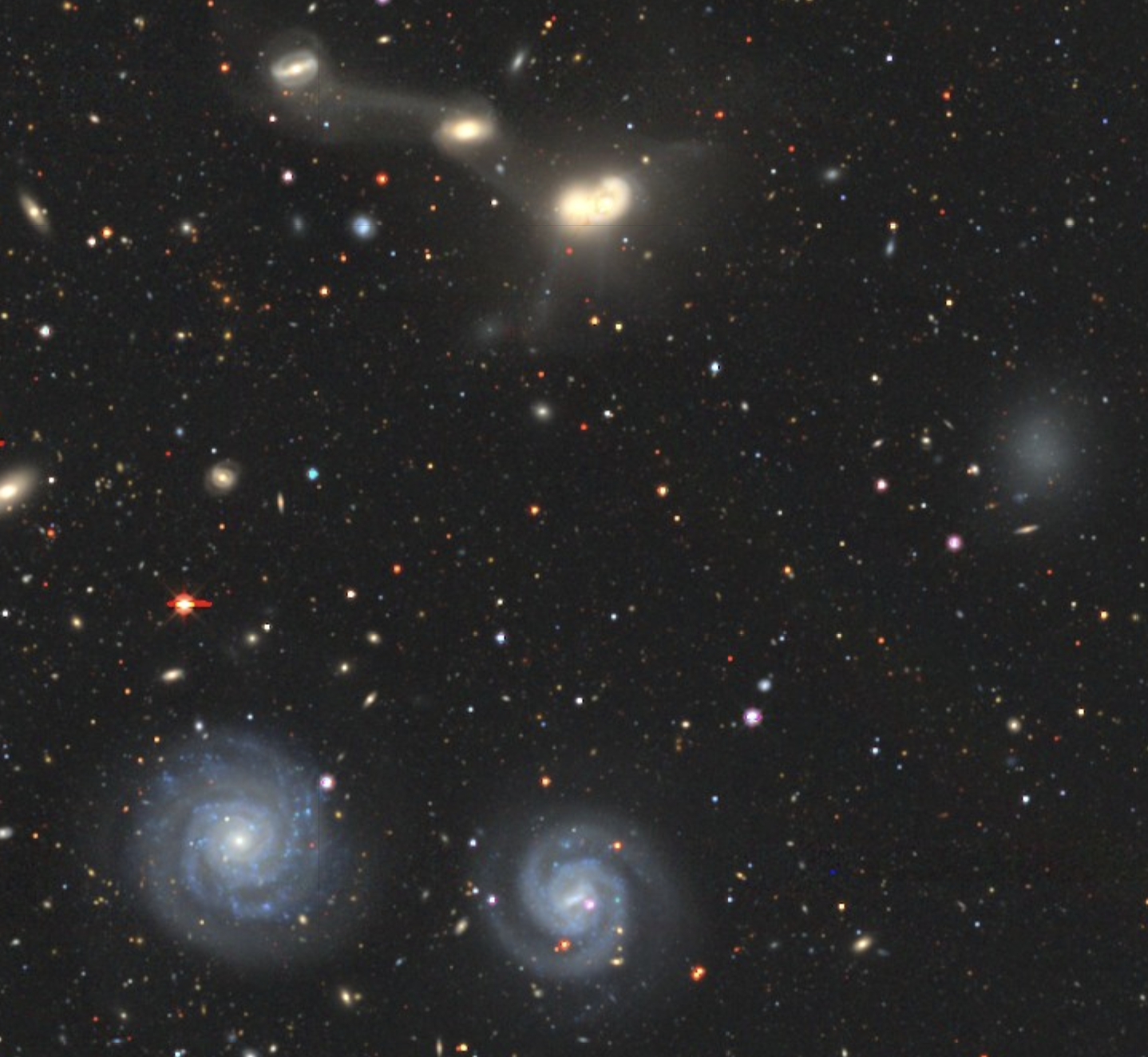}}
\caption{\small Image of an Ultra-Diffuse galaxy (middle, far right) along with several high surface brightness galaxies. Interacting galaxies (near the top) show tidal tails and stellar streams, and well as faint sources.}
\label{fig1}
\end{center} 
\end{figure}

\hd{The big questions}

{\it What is the origin of outer halo globular clusters?}

Galaxies are thought to have formed in two phases: from in-situ formation and ex-situ or accreted material. In the most massive galaxies the fraction of accreted mass is predicted to dominate over in-situ formed mass. It is difficult observationally to test this prediction using galaxy field stars as they are well mixed. However, GCs are robust to mergers and long-lived. By identifying outer halo GCs one can hope to determine the fraction of accreted versus in-situ mass. These observations can be done efficiently with the wide field u-band capability of KWFI. Follow-up spectroscopy of GCs within 30 Mpc may reveal different mean ages to those in-situ formed GCs in galaxy central regions.

{\it What is the relative frequency of compact elliptical galaxies that are stripped remnants versus an extension of the low mass elliptical galaxy family?}

Only a few hundred compact elliptical galaxies (cEs) have been cataloged. It appears that most are the compact remnant of a larger stripped galaxy. However, a few examples are found in isolated environments too far from any nearby galaxy to have been stripped. These are good candidates for low mass ellipticals that are simply an extension of the elliptical galaxy family to the lowest masses. Wide area imaging is required to identify further examples of cEs in dense environs but also particularly those outside of groups/clusters. Once large samples have been gathered their properties can be contrasted to better determine their true origin, lending insight into the process of galaxy formation. \\

{\it What is the abundance and properties of field ultra diffuse galaxies?}

UDGs are relatively rare class of galaxy, with most found in group and cluster environments and a small fraction in the field. In groups and clusters, UDGs are typically red and quiescent, while those in the field have been found to be both blue and star forming or red and quiescent. Recently, simulations have provided an explanation for red and quiescent field UDGs, showing that these field UDGs can form as `backsplash' galaxies on unbound or eccentric galaxy group and cluster orbits (Benavides et al.~2021). However, these simulations expect more UDGs in the field than observed and  require deep, wide-field searches to test the predictions. 

Ultra diffuse galaxies (UDGs) are very rare in the field but they are important clues to the formation and evolution of UDGs more generally. The ability to take deep imaging with a large field of view (1 $\times$ 1 degree) over a large photometric baseline (u-band to z-band), the Keck Wide Field Imager is uniquely poised to build a statistical sample of UDGs in the field. Its unprecedented ability to deeply image in u through z bands will allow photometric redshifts, providing far quicker environmental confirmation than is currently available (a single spectroscopic redshift for a UDG can take many hours of 10m-class telescope time). This large photometric coverage will also allow SED fitting which has proven to be an effective tool to distinguish between UDG formation mechanisms in small samples (e.g., Pandya et al.~2018). In addition, the unparalleled ability of KWFI to deeply image in the u-band will allow an investigation of star formation in UDGs. While other instruments (e.g., LSST, Euclid, HSC) can provide small parts of this science, they lack the u-band sensitivity of KWFI and so lack the accuracy and discriminatory power that its wide photometric baseline will provide. Ultimately, KWFI is the only instrument currently proposed that will be able to build an accurate, statistical sample of UDGs in field environs.

\hd{KWFI requirements}

Deep, wide-field ($\sim$1 deg) u-band imaging enables GC and galaxy selection and detection.

Full ugriz broadband imaging over the wide field.

\clearpage

\hd{KWFI Science Case}

\begin{center}
  {\bf \Large Supernova Remnants and Light Echos}
\end{center}

\hd{Contributing author}

Armin Rest (Space Telescope Institute)

\hd{Executive Summary}

Tycho Brahe's observations of a supernova (SN) in 1572 challenged the teachings of Aristotle that the celestial realm was unchanging. We have discovered a way to see the same light that Tycho saw 440 years ago by observing SN light that only now reaches Earth after bouncing off dust filaments. These light echoes (LEs) give us a unique opportunity in astronomy: direct observation of the cause (the explosion) as well as the effect (the expanded remnant) of the same astronomical event. Furthermore, multiple LEs allow us to see the same explosion from different directions, providing the only way to directly map asymmetry. This is valuable because modern theoretical work suggests that asymmetry may be a critical ingredient in the SN explosion mechanism. In addition, the spectroscopic time series of these ancient events can be recovered with LEs. The main goal is to combine these approaches to LE follow up and obtain high S/N spectroscopic time series from multiple directions for as many Galactic SNe as possible. So far, we have discovered LEs from two Galactic SNe (Tycho and Cas A), but with KFWI, we have the depth and field of view to not only significantly expand the sample of light echoes around Cas A and Tycho, but also to finally discover light echoes around some of the other Galactic SNRs, e.g., Kepler (SN 1604), the Crab Nebula (SN 1054), 3C 58, W49B, and Kes 75.

\bigskip


\begin{figure}[!h]
\begin{center} 
\scalebox{1.52}[1.52]{\includegraphics{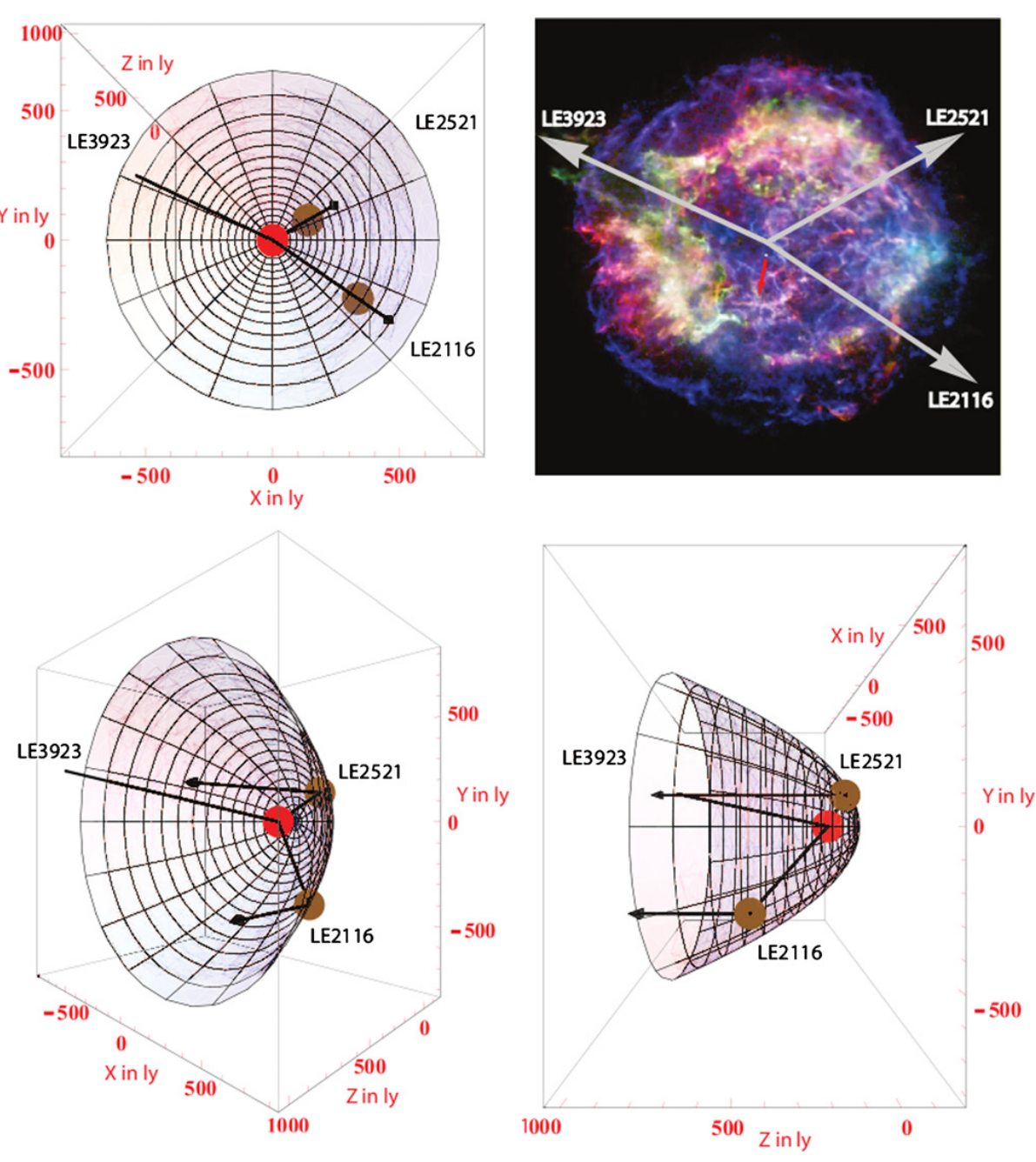}}
\caption{\small Cas A light echo (LE) observations and geometry [3]. North is toward the positive y-axis (up), east is toward the negative x-axis (left), and the positive z-axis points toward the observer with the origin at the SNR. The SN to LE cone apex distance, in light years, is half the interval of time since the SN explosion. The red and brown circles indicate the SN and scattering dust, respectively. The black lines show the path of the light scattering from the LE-producing dust concentrations. The scattering dust of LE3923 is more than 2000 ly in front of Cas A, much farther than any other scattering dust; thus, we show only part of the light path and do not include its dust location. The top left panel shows a Chandra X-ray image [5], with the projected light path from SN to scattering dust overplotted (gray arrows). The red arrow indicates the X-ray compact object and its apparent motion. In this false-color image, red corresponds to low-energy X-rays around the Fe L complex ($\sim$1 keV and below), green to mid-energy X-rays around the Si K blend ($\sim$2 keV), and blue to high-energy X-rays in the 4–6 keV continuum band between the Ca K and Fe K blends.
}
\label{cas}
\end{center} 
\end{figure}

\hd{The big questions}

{\bf Spectroscopic classification of Ancient SNe:} Current transient searches are providing a flood of extragalactic SNe, and this will accelerate in the era of LSST. These increased numbers provide valuable new information about the statistics of diverse SN types and they reveal exceedingly rare events, but connecting these to the underlying physical parameters is fraught with uncertainty and sometimes impossible. Centuries-old Galactic SN remnants, on the other hand, provide our most detailed and direct measurements of the ejected mass, kinetic energy, element stratification, nucleosynthetic yield, explosion geometry, spatially resolved shock fronts, the surrounding environment, and the nature of any resulting compact remnant ---  but the original SN explosions themselves were not subjected to the scrutiny of spectroscopic analysis. LEs are the only way to bridge this gap. Combining the physical diagnostics of a collection of nearby remnants with LE spectra of the corresponding historical SN event allows us to make solid connections between the underlying physics and observed cosmic explosions.

{\bf Probing SNe in 3D:} Since a single LE probes a hemisphere of the explosion, LEs from different directions can give 3D information about the SN. We applied this technique first to LEs of Cas~A, which revealed outflow velocities of \about 4000~\kms\ for LEs from different directions (see Fig.~\ref{cas}; [2,3]). X-ray and optical data of Cas~A show Fe-rich material in the same direction as the LE and blue-shifted velocities, which also traces a direction opposite the known motion of the resulting neutron star. This suggests that the mechanism which gave the neutron star its kick also affected the outer layers of the SN and may suggest a diversity in viewing angles and outflow velocities observed in extragalactic SNe. Cas A appears to be the first instance where the asymmetry in the SN remnant can be directly associated with asymmetry observed in the spectrum of the explosion itself [2,3].

\begin{figure}[!h]
\begin{center} 
\scalebox{1.7}[1.7]{\includegraphics{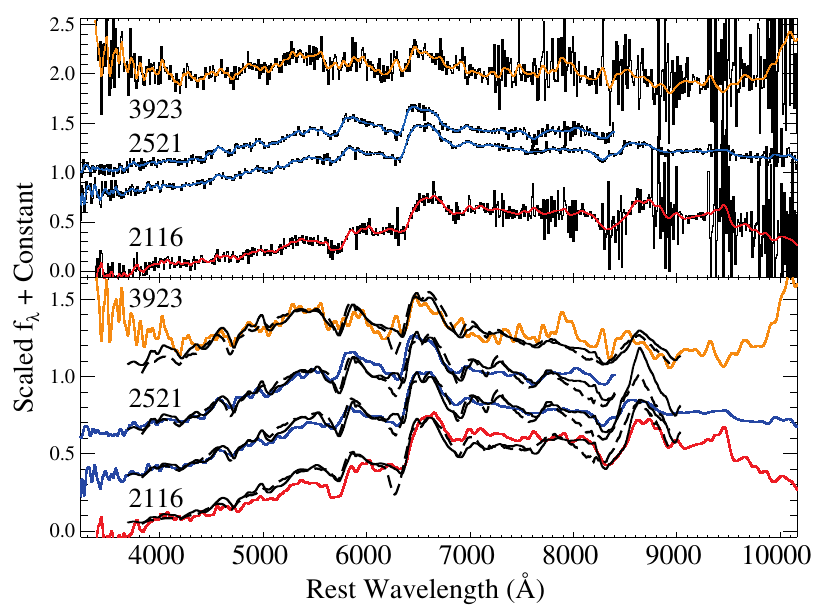}}
\caption{\small Top panel: Observed LE spectra binned to 10\AA\ pixel$^{-1}$ (black curves) [3]. The colored curves are the variance-weighted smoothed spectra with the red, blue, and orange curves corresponding to LEs LE2116, LE2521, and LE3923, respectively. The LE2521 spectrum with the shorter wavelength range is from the MMT; the remaining spectra are from Keck. Bottom panel: variance-weighted smoothed LE spectra (colored curves) compared to their corresponding light-curve weighted integrated SN 1993J (solid black curves) and SN 2003bg spectra (dashed black curves). The LE spectra are equivalent to the spectra in the top panel. The SNe 1993J and 2003bg spectra have been processed to reflect the dust scattering and reddening associated with each LE (see the text for details). The spectra of Cas A are very similar to both SNe 1993J and 2003bg from all three directions probed; however, there are some significant differences.}
\label{spec}
\end{center} 
\end{figure}

{\bf LE Spectroscopic Time Series of SNe:}
A rarely exploited opportunity provided by SN LEs is the ability to resolve a bright LE across a very thin dust sheet and observe the evolution of the SN explosion as it appears in scattered light [1,3]. 
We performed this measurement for a Tycho LE with follow up designed to sample the LE emission in the direction it propagates (Fig.~\ref{spec}). By extracting small, resolved regions of LEs, we obtain spectra that probe different epochs of the SN, with a time resolution of \about1~week. However, this requires a bright LE and optimal dust geometry, which is rare and requires wide area search and rapid follow up.

The main goal is to combine these approaches to LE follow up and obtain high S/N spectroscopic time series from multiple directions for as many Galactic SNe as possible. These observations will allow us to connect the SNR with the original SN explosion, and address open questions about intrinsic asymmetry and viewing angle dependence in SNe. In particular, there is evidence that the variations in SN ejecta velocity are due to orientation effects [6]: an off-center explosion might cause different ejecta velocities for different lines of sight [7]. 

\hd{What KWFI can do for this science case}

Light echo surface brightness is proportional to 1/t$^2$, where t is the time since the light of the event reached the Earth. Therefore the light echoes for older SNe are in general fainter than for young SNe. With KWFI, we have the depth and FOV to hunt for light echoes of the older galactic SNRs. With the new PS1 dust maps [8], we are now able to identify regions of the sky which have dust structures at distances for which we can expect light echoes (within the uncertainties of age and distance to the source event). This allows us to target specific regions of the sky, which are still large, but much smaller than the full search annulus. Large fields of view are necessary for faint LE searches.

\hd{References}

{\bf [1]} Rest et al. 2005, Nature, 438, 1132;
{\bf [2]} Rest et al. 2008, ApJ, 681, 81;
{\bf [3]} Rest et al. 2011, ApJ, 732, 3;
{\bf [4]} Rest et al. 2008, ApJ, 680, 1137;
{\bf [5]} Hwang et al. 2004, ApJ, 615, 117;
{\bf [8]} Schlafly et al. 2014, ApJ, 789, 15

\clearpage

\hd{KWFI Science Case}

\begin{center}
  {\bf \Large Milky Way Star Formation and Feedback}
\end{center}

\hd{Contributing authors}

Josh Walawender (W. M. Keck Observatory)\\
John Bally (University of Colorado at Boulder)


\bigskip

\hd{Background}

Traditionally, “initial conditions” in the formation of star-forming molecular clouds are thought to be key agents. Two classes of theory have been advanced: “Competitive accretion” in which cores compete for gas from a common reservoir (e.g. Bate \& Bonnell - the ”British school”), and “turbulent collapse” (e.g. McKee, Tan, and Krumholz - the “Berkeley school”) in which cores evolve in isolation. In either scenario, the core-mass-function (CMF) is thought to map directly into the stellar initial mass function (IMF).

An alternative hypothesis is that N-body dynamics combined with feedback dominates the establishment of stellar masses and the shape of the IMF. Dynamics and feedback erase the effects of initial conditions. Feedback from forming stars stirs the parent cloud, generates turbulence, disrupts, and blows-out gas, thereby limiting the efficiency of clump formation.  Gravitational collapse of turbulent clumps produces non-hierarchical clusters of star-forming cores. Most cores form non-hierarchical multiple systems of stars. N-body dynamics of cores and forming stars results in ejections. Ejection of  stars from their birth-sites sets  final stellar masses.  

As stars form and grow in mass, ever-more powerful feedback mechanisms operate. In order of increasing momentum and energy input these mechanisms are: protostellar outflows, soft, non-ionizing ultraviolet (UV) flux (FUV with $\lambda > $912 \AA ), ionizing UV (EUV with $\lambda < $912 \AA ), stellar winds, radiation pressure, and supernova explosions. Finally, in only the most massive “super-star-cluster” forming environments and galactic nuclei there is black-hole feedback. Note that the least energetic and earliest rungs of this ladder (outflows, FUV, EUV) are most directly coupled to parent clouds. While higher rungs (stellar winds, supernovae, and BHs) are poorly coupled to star forming clouds; these mechanisms couple more to the general interstellar medium (ISM) or intergalactic medium (IGM).

\hd{The big questions}

{\it What determines the final masses of stars?  }

{\it What determines the stellar initial mass function (IMF)? }

\hd{What KWFI can do for this science case}

\begin{figure}[!h]
\begin{center} 
\scalebox{0.44}[0.44]{\includegraphics{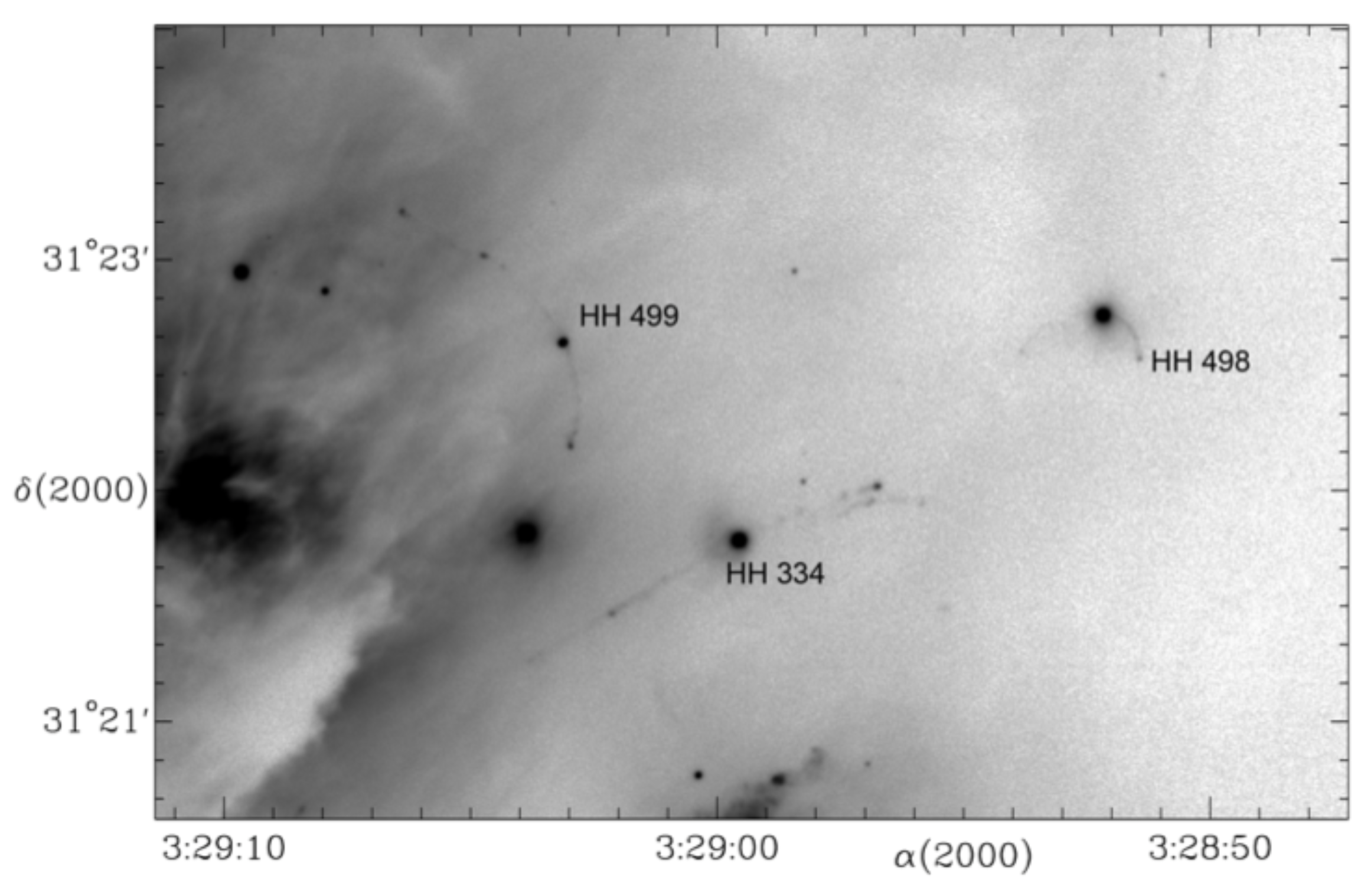}}
\caption{\small An H$\alpha$ $+$ [\textsc{Sii}] image of the HH 334, 498, and 499 bent jets [2,3]}
\label{bent_jets}
\end{center} 
\end{figure}

Emission lines trace shock-excited material. Thus narrowband imaging is a necessary step in the quantitative measurement and mapping of feedback impacts and the characterization of feedback mechanisms (Figure~\ref{bent_jets}).  

Narrowband interference filters isolating critical nebular emission lines such as \ha{} (6563 \AA), \hb{} (4861 \AA), \sii{} (6716 \& 6731 \AA), [OI] (6300 \& 6363 \AA ), [OII] (3726 \& 3729 \AA ), and [OIII] (4959 \& 5007 \AA) are the best tracers of the various feedback mechanisms. Narrowband (<100 \AA{} bandwidth) filters isolate these lines from background continuum -- both from the sky and astrophysical backgrounds.

While \ha{} and \sii{} are classic tracers of shocks, the [OI], [OII], and [OIII] images give all ionization stages of O in shocks as well as photo-ionized regions. [OII] seems to be complimentary to, and reveals different morphologies from, \ha{} and \hb{} images [1]. While [OIII] traces the hardest (fastest) shocks and [OI] is an excellent tracer of photodissociation regions.

Imaging provides the morphology of outflow systems. This is critical in the identification of the source star and the spatial extent of the outflow. Multi-band morphologies (such as comparing \ha{}, \sii{}, and [OII]) distinguishes shocks from photo-ionization regions. In addition, any apparent symmetries of the shocks around the source stars provide fossil records of the underlying N-body dynamics of the source stars.

The photometry of emission lines provides a number of diagnostics: The \ha{} to \hb{} ratio provides a pixel by pixel map of the extinction to the radiating medium. This map can then be used to recover the the intrinsic \ha{} flux. Source morphology combined with emission measure yields the electron density ($n_e = (EM/L)^{0.5}$).

Multi-epoch imaging provides proper motions and thus velocities. These velocities, combined with densities from the emission measure, give momentum injection rates into the media being impacted by shocks and thus quantifies stellar feedback directly.

Equipped with narrowband filters, KWFI will provide characterization of feedback from stars of all ages (proto, main-sequence, post-main sequence). Using $\sim$80 angstrom bandpass filters will mean that quite a few galaxies will be within reach. Ultra-deep surveys of nearby galaxies such as M33, M31, M101 will probe feedback on kpc sales. Ultra-deep images of systems such as M82 and M83 will probe the “galactic ecologies” of these systems, and show how matter can be expelled from galaxies into the IGM and accreted from the IGM onto galaxies.    

\hd{References}

{\bf [1]} Bally \& Reipurth 2018, RNAAS, 2, 46; 
{\bf [2]} Walawender et al. 2005, ApJ, 129, 2308;
{\bf [3]} Walawender et al. 2005, ApJ, 130, 1795

\clearpage

\hd{KWFI Science Case}

\begin{center}
  {\bf \Large Milky Way Stellar Populations and Metallicities}
\end{center}

\hd{Contributing author}

R. Michael Rich (UCLA)

\hd{Executive Summary}

\begin{figure}[!h]
\begin{center} 
\scalebox{0.57}[0.57]{\includegraphics{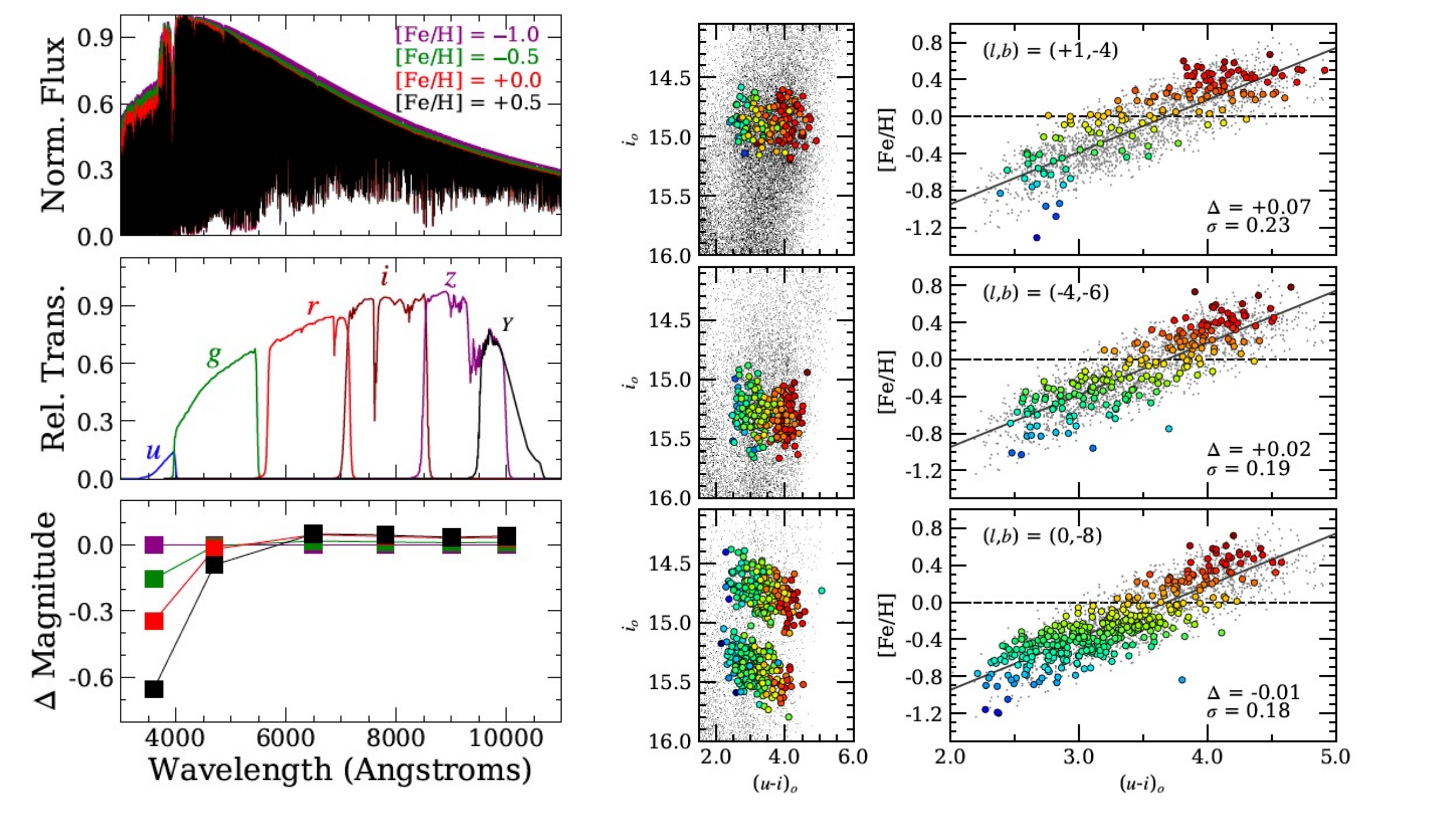}}
\caption{\small UV magnitudes are powerful in constraining metallicity over the full range, even at relatively high metallicity. (Left panel) Iron line absorption as a function of metallicity for a typical red clump star; notice the very high sensitivity of the u band [2]. (Right) Correlation between $(u-i)_0$ and [Fe/H] (spectroscopic) for red clump stars in 3 bulge fields [3]. The reddening corrected color gives a metallicity to a 0.2 dex precision. It is expected that dereddened u band photometry will be useful in measuring metallicities across a wide range of stellar populations. Although the Gaia mission will stop operating on the 2020s, if excellent astrometry can be attained, techniques such as proper motion cleaning etc.\ can be applied to the data. These techniques are applicable throughout the Local Volume, especially in M31.}
\label{metals}
\end{center} 
\end{figure}

The ultraviolet (UV) has long been employed as a metallicity sensitive photometric magnitude [1] (Figure~\ref{metals}). KWFI can be used to obtain measurements in the UV that can constrain stellar metallicities in the Galactic bulge and thick disk, even in regions of high extinction. If narrow band filters can be supported, a deep search for ultra-metal poor stars can be undertaken in the Galactic bulge, thick disk, and halo/dwarf spheroidal galaxies.

\bigskip


\hd{The big questions}

{\it Where are the first generations of stars in the Milky Way stellar populations?}

{\it Can complete measurements of the metallicity distribution in a stellar population be used to constrain models of chemical evolution and metal enrichment?}

{\it Can age-metallicity relationships and star formation histories be elucidated in Milky Way populations?}

{\it Can UV photometry be used to search for extremely faint substructure in the Milky  Way halo, e.g., streams (also using astrometry)?  What fraction of the halo, bulge and thick disk is comprised of substructures?}

Searches for extremely metal poor stars can be contemplated if a narrow-band Ca HK filter can be supported. Proper motion membership and UV photometry can be used to explore the metallicity distributions of e.g., dwarf spheroidal galaxies to faint magnitudes below the main sequence turnoff. Each 1 mag fainter results in a factor of $\sim10$ more stars in the sample (this is due to the rising luminosity function for all red giants and dwarfs), which is important for systems where the total luminosity is small. The scientific landscape is competitive (e.g., the PRISTINE survey [4] on the CFHT). However KWFI will have the best UV sensitivity and widest field of view of any ground-based detector. For PRISTINE, photometric precision has proven to be a limiting factor in the discovery of extremely metal poor stars, especially for dense fields in the Milky Way bulge; this would be no issue for KWFI.

\hd{What KWFI can do for this science case}

Extraordinary ultraviolet sensitivity combined with appropriate filter selections can transform KWFI into a search engine for metal poor stars across the Milky Way and the Local Volume.  Within the Milky Way, it will be possible to contemplate deep searches even in ultra-faint dwarf galaxies.  

Top level science requirements: UV sensitivity, excellent PSF, wide field, broadband and narrow-band filters.

\hd{References}

\noindent{\bf [1]} Eggen, O.J., Lynden-Bell, D., \& Sandage, A.R. 1962, ApJ 136, 748
\noindent{\bf [2]} Johnson, C., Rich, R.M. et al. 2022 MNRAS, {\it in press}
\noindent{\bf [3]} Johnson, C., Rich, R.M. et al. 2020  MNRAS, 499, 2357 
\noindent{\bf [4]} Arentsen, A., et al. 2020 MNRAS 496, 4964

\clearpage

\hd{KWFI Science Case}

\begin{center}
  {\bf \Large Solar System Science}
\end{center}

\hd{Contributing authors}

Bryce Bolin (Caltech)\\
Richard Wainscoat (University of Hawaii)

\hd{Executive Summary}

With careful design, KWFI will become a uniquely powerful instrument with capabilities and sensitivity for solar system science well exceeding Hyper-SuprimeCam on Subaru and the Rubin Observatory. The wide-field capability and high sensitivity of KWFI will enable deep images over large areas very quickly, including blue imaging. KWFI can uniquely address and help resolve a number of problems, as described below, regarding main belt and active asteroids, trans-Neptunian objects, comets, and Near-Earth objects.


\hd{Important considerations in KWFI design and capability}
\begin{itemize}
    \item Solar system objects move, so increasing exposure time is not an option to go deeper. Keck has a larger aperture than Subaru or the Rubin observatory, so it is capable of seeing fainter solar system objects than either of these telescopes. It will become the most sensitive telescope in the world for Solar System surveys.
    \item Nearly all solar system science cases will benefit from as large a field-of-view as possible.
    \item The optics should be designed such that wavelengths in the range 400--830 nm are parfocal, with any chromatic aberration minimized. This will allow use of a wide filter such as the Pan-STARRS {\it w\/} filter, which is similar to the CFHT {\it gri\/} filter, and spans 400--830 nm. A wide filter such as this can produce as much as a factor 3 increase in sensitivity compared to a narrower filter such as {\it r\/}. This would be a capability unique to KWFI. Subaru and Rubin do not have wide filters. Together with Keck's larger aperture, the increase in sensitivity compared to Rubin can be as much as a factor 5.
    \item An atmospheric dispersion corrector is absolutely essential to complement the use of wide filters.
    \item The optics should produce seeing limited images across the entire field.
    \item Fast readout is essential --- solar system exposures will typically be no longer than 45--60 seconds. A 5 second readout is desirable; readout should be no longer than 10 seconds.
    \item We need to know exactly when the shutter opens and closes
    \item Filter change should be relatively quick (HSC takes 30 minutes)
\end{itemize}

\hd{The big questions}

{\bf Main Belt Asteroids/Trojans (2--5 AU from the Sun):}
\vspace{-0.1cm}

{\it What is the size distribution of 100-m scale Main Belt asteroids, the feeder population for NEOs?}

KWFI can probe the size distribution of 100 m-scale Main Belt asteroids, 1,000s of $\sim$100 m-scale Main Belt asteroids per $\sim$1 deg$^2$ KWFI pointing (down to r $\sim$ 25.7 in a 100\,s r-band exposure).

{\it Is the rotation distribution of sub-km-scale Main Belt asteroids and km-scale Trojan asteroids more processed by non-destructive collisions compared to larger asteroids?}

KWFI will study the effect of non-catastrophic collisions on the rotations of sub-km Main Belt asteroids using light curves of sub-km Main Belt asteroids. KWFI can obtain 20--30 $\times$ SNR $\sim$ 10--15 time-series photometric data points in r band of 1,000s of sub-km scale asteroids.

{\bf Trans-Neptunian objects (30--100 AU+ from the Sun):}
\vspace{-0.1cm}

{\it How common are the binarity and elongated shape of 10 km-scale Arrakoth-like objects in the trans-Neptunian object population?}

KWFI would obtain light curves of 10 km-scale trans-Neptunian asteroids like Arrokoth and provide rough constraints on shape and binarity. 

KWFI can perform wide-field searches for distant trans-Neptunian objects such as detached objects/Planet 9. It can detect $\sim$100 km-scale objects out to 100 AU in deep-stack g and r band imaging.

{\bf Active asteroids and comets (0.5--10 AU from the Sun):}
\vspace{-0.1cm}

{\it How prevalent is the activity of asteroids in the Main Belt/Trojan Swarms? How common are observations of low levels of comet-like activity in Main Belt asteroids}

KWFI can provide SNR $\sim$ 10--15 g and r band photometry of 1,000s of Main Belt asteroids down to g and r $\sim$ 26 and search for activity.

{\it What are the dust properties and activity-driving mechanisms of active asteroids in the Main Belt?}

KWFI can provide imaging of low surface brightness features associated with individual Main Belt asteroids becoming active illuminating the properties of the dusty coma and its activity-driving mechanism. 

{\it What is the turn-on distance/onset of activity of comets and Centaurs beyond the frost line?}

KWFI can probe the activity of distant comets/Centaurs past 10 AU with deep g/r imaging.

{\it Do near-Sun comets like C/2012 S1 ISON and rock comets like 3200 Phaethon contain activity and/or fragments?}

KWFI can detect fragments as small as 50 m at $\sim$1.5AU from the Earth in g and r band imaging. A wide 1 sq. field of view is crucial for covering debris field of fragmenting bodies.

{\bf Near-Earth objects (0.5 - 1.3 AU from the Sun):}
\vspace{-0.1cm}

KWFI enables detection of NEOs as small as $\sim$1 m-scale several lunar distances or more from the Earth and characterization of the spectral type and rotation of 1-m scale NEOs and mini-moons like 2020 CD3.

We know very little about NEOs that spend most of their time inside Earth's orbit. This region can harbor dangerous objects, because objects move more slowly and spend more time near aphelion (i.e., they spend more time near Earth's orbit). The major opposition surveys such as Pan-STARRS and Catalina cannot find many of these objects because they are seldom (or never) in the opposition region of the sky. We can search for these at low solar elongation, but this is difficult, because of the phase angle (the NEO is only partially illuminated), brighter sky (due to zodiacal light and high airmass), and poorer seeing at high airmass. All these factors are helped by a highly sensitive instrument such as KWFI. KWFI will be a powerful tool to search for NEOs at small solar elongation and help better assess the risk from these objects.

It is well-known that trailing losses make detection of fast-moving objects much more difficult. The large increase in sensitivity relative to the competing 8-m telescopes will help us to discover more fast-moving objects, Fast-moving objects may include small Near-Earth objects, mini-moons of Earth (objects temporarily captured into Earth orbit), and interstellar objects. KWFI will help us to better understand the size-number distribution of small NEOs.

{\bf Other planets in our Solar System}

Subaru/HSC is presently being used to search for additional planets in our solar system. KWFI could would be more sensitive and would have greater accessibility to the Keck community.

\hd{Other KWFI considerations}

The low solar elongation observations described require dark time, and must avoid the Galactic plane. The crescent moon cannot be in the part of the sky where the observations are being acquired. Creative scheduling --- perhaps including queue scheduling --- would be helpful for science cases such as this, which can only use part of a night.

Use of wide filters such as {\it w\/} generally require dark time. Crescent moon can be tolerated, but more than 60\% lunar illumination should be avoided. A moon avoidance angle will need to be established.

\end{document}